\documentclass[12pt]{article}
\usepackage{mathrsfs}
\makeatletter \@addtoreset{equation}{section} \makeatother
\renewcommand{\theequation}{\thesection.\arabic{equation}}
\newcommand{\ba}{\begin{array}}
\newcommand{\ea}{\end{array}}
\newcommand{\beq}{\begin{equation}}
\newcommand{\eeq}{\end{equation}}
\newcommand{\bea}{\begin{eqnarray}}
\newcommand{\eea}{\end{eqnarray}}

\def\bce{\begin{center}}
\def\ece{\end{center}}

\def\nonu{\nonumber}

\def\pa{\partial}
\def\al{\alpha}
\def\be{\beta}

\def\de{\delta}

\def\la{\lambda}

\def\eps6{{\displaystyle \mathop{\epsilon}^{6}}{}}
\def\g6{{\displaystyle \mathop{g}^{6}}{}}

\def\nab6{{\displaystyle \mathop{\nabla}^{6}}{}}

\def\0{{\sst{(0)}}}
\def\1{{\sst{(1)}}}
\def\2{{\sst{(2)}}}
\def\3{{\sst{(3)}}}
\def\4{{\sst{(4)}}}
\def\5{{\sst{(5)}}}
\def\6{{\sst{(6)}}}
\def\7{{\sst{(7)}}}
\def\8{{\sst{(8)}}}

\def\ba{\begin{array}}
\def\ea{\end{array}}
\def\beq{\begin{equation}}
\def\eeq{\end{equation}}
\def\be{\begin{equation}}
\def\ee{\end{equation}}

\def\la{\lambda}
\def\eps{\epsilon}

\def\ba{\begin{array}}
\def\ea{\end{array}}
\def\beq{\begin{equation}}
\def\eeq{\end{equation}}
\def\be{\begin{equation}}
\def\ee{\end{equation}}

\def\la{\lambda}
\def\eps{\epsilon}

\def\eps6{{\displaystyle \mathop{\epsilon}^{6}}{}}

\def\nab6{{\displaystyle \mathop{\nabla}^{6}}{}}

\newcommand{\bean}{\begin{eqnarray*}}
\newcommand{\eean}{\end{eqnarray*}}

\begin{document}
\thispagestyle{empty} \addtocounter{page}{-1}
   \begin{flushright}
\end{flushright}

\vspace*{1.3cm}
  
\centerline{ \Large \bf
A Deformed Supersymmetric $w_{1+\infty}$ Symmetry
}
\vspace*{0.3cm}
\centerline{ \Large \bf
in the Celestial Conformal Field Theory } 
\vspace*{1.5cm}
\centerline{ {\bf  Changhyun Ahn}
} 
\vspace*{1.0cm} 
\centerline{\it 
 Department of Physics, Kyungpook National University, Taegu
41566, Korea} 
\vspace*{0.5cm}
\centerline{\tt ahn@knu.ac.kr
} 
\vskip2cm

\centerline{\bf Abstract}
\vspace*{0.5cm}

By using the $K$-free complex bosons and
the $K$-free complex fermions,
we construct the ${\cal N}=2$ supersymmetric $W_{\infty}^{K,K}$
algebra which is the matrix generalization of
previous ${\cal N}=2$ supersymmetric $W_{\infty}$ algebra.
By twisting
this ${\cal N}=2$ supersymmetric $W_{\infty}^{K,K}$
algebra, we obtain
the ${\cal N}=1$ supersymmetric $W_{\infty}^{K}$
algebra  which is the matrix generalization of
known ${\cal N}=1$ supersymmetric
topological $W_{\infty}$ algebra.
From this two-dimensional symmetry algebra,
we propose the operator product expansion (OPE)
between the soft graviton and gravitino (as a first example),
at nonzero deformation parameter,
in the supersymmetric
Einstein-Yang-Mills theory 
explicitly. Other six OPEs
between the graviton, gravitino, gluon and gluino
can be determined completely.
At vanishing deformation parameter, we reproduce
the known result of
Fotopoulos, Stieberger, Taylor and Zhu on the above OPEs
via celestial holography.


\baselineskip=18pt
\newpage
\renewcommand{\theequation}
{\arabic{section}\mbox{.}\arabic{equation}}

\tableofcontents

\section{ Introduction}

Recently, the celestial holography
has been proposed by the fact that there exists
a duality between the gravitational scattering in asymptotically
flat spacetimes  and  the conformal field theory which lives
on the celestial sphere. See the review papers
\cite{PPR,Aneeshetal,Pasterski,Raclariu} on the
celestial holography.
By using the low energy scattering problems, the symmetry
algebra of the conformal field
theory for flat space (a celestial conformal field theory)  
has been studied in \cite{GHPS}.
Moreover, in \cite{Strominger},
the group of symmetries on the celestial sphere satisfies
the wedge subalgebra of $w_{1+\infty}$ algebra \cite{Bakas}.
This implies that we should understand
the structures behind these findings thoroughly 
in order to check the above duality.
In \cite{Ahn2111},
the supersymmetric $w_{1+\infty}$ algebra has been identified with
the corresponding soft current algebra in the supersymmetric
Einstein-Yang-Mills theory.
The relevant works on the celestial holography in the various directions
can be found in \cite{
Pasterski:2022jzc,Pasterski:2022joy,Costello:2022wso,
Pasterski:2022lsl,Freidel:2021ytz,Krishnan:2021nqo,Agarwal:2021ais,
Crawley:2021auj,Bu:2021avc,
Freidel:2021dfs,Chandrasekaran:2021vyu,MRSV,Ball:2021tmb,
Giribet:2021zie,Adamo:2021zpw,Campiglia:2021oqz,Adamo:2021lrv,
Jiang:2021csc,Gupta:2021cwo,Guevara:2021tvr,Donnay:2021wrk,
Pano:2021ewd,Jiang:2021ovh,HPS,Banerjee:2021dlm,
Sharma:2021gcz,Hu:2021lrx}.
See \cite{PPR}
for more complete literatures. 

The higher-spin extension of the Virasoro algebra 
has been found by Zamolodchikov \cite{Zamol}.
The so-called $W_3$ algebra consists of the spin-$2$ stress energy
tensor and the spin-$3$ current.
Subsequently, this $W_3$ algebra is generalized to the $W_N$ algebra
\cite{FZ,FL}
which is generated by the spin-$2$ stress energy tensor and the
higher-spin currents of each spin, $s=3, 4, \cdots, N$. 
It is also possible to construct a linear $W_{\infty}$ algebra
\cite{PRS1990,PRS1990-1}
generated by the currents with spins $s=2, 3, 4, \cdots, \infty$.
A simple contraction of the $W_{\infty}$ algebra leads to
the 
$w_{\infty}$ algebra \cite{Bakas}.
Moreover, 
the 
$W_{1+\infty}$ algebra \cite{PRS1990-2} contains
all spins $s=1, 2, 3, \cdots, \infty$.
The ${\cal N}=2$ supersymmetric $W_{\infty}$ algebra \cite{BPRSS}
whose bosonic sector is given by
$W_{\infty}$ and $W_{1+\infty}$
is obtained.

Bakas and Kiritsis \cite{BK} have found the 
$W_{\infty}^K$ algebra which is an $U(K)$-matrix generalization of
$W_{\infty}$ algebra. For each current of spin $s$, there are
$K^2$ multicomponent generators.
Odake and Sano \cite{OS}
also have constructed 
the
$W_{1+\infty}^L$ algebra which is an $U(L)$-matrix extension of $W_{1+\infty}$ algebra.
There exist $L^2$ multicomponent generators for each current of spin
$s$.
Furthermore, the 
supersymmetric $W_{\infty}^{K,L}$ algebra, whose bosonic sector
is given by $W_{\infty}^K$ and $W_{1+\infty}^L$,
has been studied in \cite{Odake}.

In this paper, by taking the condition $K=L$,
we construct
the ${\cal N}=2$ supersymmetric $W_{\infty}^{K,K}$ algebra
with $U(K) \times U(K)$ symmetry.
Due to the above condition $K=L$,
we can multiply the generators in the fermionic currents.
The ${\cal N}=2$ supersymmetry is reduced to the
${\cal N}=1$ supersymmetry by topological twisting \cite{Witten,EY,PRSS}.
Then 
the ${\cal N}=1$ supersymmetric $W_{\infty}^{K}$ algebra
with $U(K)$ symmetry is obtained.
That is, we obtain the matrix
generalization of \cite{PRSS}.
The seven commutator relations between the bosonic and fermionic
currents can be written in terms of the various structure constants and
the deformation parameter. 
By considering the vanishing limit of this deformation parameter,
we reproduce the previous result of \cite{FSTZ}.
We propose that the OPEs between the graviton, gravitino, gluon and gluino
in the supersymmetric Einstein-Yang-Mills theory can be determined
from the above two-dimensional symmetry algebra
\footnote{\label{firstfootnote} We are focusing on the soft currents
  where the celestial operators have the specific
  conformal dimensions for the bosonic and fermionic fields.
  For the former $\Delta = 1, 0, -1, \cdots $
  and for the latter $\Delta=\frac{1}{2}, -\frac{1}{2},
  -\frac{3}{2}, \cdots $. See also \cite{GHPS,Jiang:2021ovh}.}.

For the nonvanishing deformation parameter,
the commutators contain the possible terms in the right hand sides.
In general, the structure constants depend on the two modes of the
commutator and the weights. Among the weights, the weights $h_1$ and $h_2$
appearing in the
left hand side of the commutator  are given. On the other hand,
the weight $h$ appearing in the right hand side vary from its lowest
value to the highest value depending on the previous weights
$h_1$ and $h_2$. The weight $h$ plays the role of dummy variable
in the summation of the right hand side of the commutator.
We would like to determine the OPEs between
the above soft currents in the supersymmetric Einstein-Yang-Mills theory
by looking at the two dimensional symmetry algebra characterized by
seven commutators. Then the question is how we can determine the OPEs
in the soft currents which will eventually
lead to the commutators we have found in two dimensional conformal field
theory after performing
the appropriate contour integrals.

As mentioned before, the structure constants
consists of mode dependent part and mode independent part and they
do depend on the above three weights dependence.
We should figure out how the mode dependent part can be read off
from the relevant OPE between the soft currents because
the mode independent part can be multiplied into this inside of the
dummy variable weight $h$. From the experience of the various
contour integrals \cite{GHPS}, we expect that there should be
$h$ dependence in the OPE when we consider the case of
the nonzero deformation
parameter. Once we have obtained the correct OPEs
which produces the mode dependent part of the commutators, then
it is straightforward to determine the full OPEs
by multiplying the weights dependence parts and summing over
above dummy variable $h$ within the possible range.

In section $2$,
we obtain the ${\cal N}=2$ supersymmetric
$W_{\infty}^{K,K}$ algebra after reviewing
the supersymmetric $W_{\infty}^{K,L}$ algebra.
In section $3$, we determine the ${\cal N}=1$ supersymmetric
$W_{\infty}^{K}$ algebra. The free field realization is given.
At the vanishing deformation parameter,
the previous result \cite{Ahn2111} is reproduced.
vWe also present the seven commutator relations in terms of
the corresponding OPEs.
Finally, we propose its realization in the
${\cal N}=1$ supersymmetric Einstein-Yang-Mills theory.
In section $4$, we summarize what we have
obtained in this paper. 
In Appendices, we provide some details in sections $2$ and $3$.

\section{ The ${\cal N}=2$ supersymmetric
$W_{\infty}^{K,K}$ algebra with $U(K)\times U(K)$
symmetry}

\subsection{The supersymmetric
$W_{\infty}^{K,L}$ algebra with $U(K)\times U(L)$
symmetry: Review}

The nontrivial (anti)commutator relations of $W_{\infty}^{K,L}$ algebra
\cite{Odake} (See also \cite{AKK})
are given by
\bea
\big[(W^{\bar{\alpha} {\beta}}_{\mathrm{F},h_1})_m,(W^{\bar{\gamma}
{\delta}}_{\mathrm{F},h_2})_n\big] 
\!&=& \! 
\sum_{h=-1}^{h_1+h_2-3} \,
\frac{\la^h}{2} \, p_{\mathrm{F}}^{h_1,h_2, h}(m,n)
\, \Bigg[ 
\delta^{\bar{\gamma}
{\beta}}  \, (W^{\bar{\alpha} {\delta}}_{\mathrm{F},h_1+h_2-2-h})_{m+n}
\nonu \\
\! & + \!& (-1)^{h}  \delta^{\bar{\alpha} {\delta}}
(W^{\bar{\gamma} {\beta}}_{\mathrm{F},h_1+h_2-2-h})_{m+n} \Bigg]
+ c_{W_{\mathrm{F},h_1}}(m) 
\delta^{\bar{\alpha} {\delta}} \delta^{{\beta} \bar{\gamma}}
\delta^{h_1 h_2} \la^{2(h_1-2)}\delta_{m+n}\ ,
\nonu\\
\big[(W^{\bar{a} {b}}_{\mathrm{B},h_1})_m,(W^{\bar{c} {d}}_{\mathrm{B},h_2})_n\big] 
\!& = &\!
\sum_{h=-1}^{h_1+h_2-4} \,\frac{\la^h}{2} \,
p_{\mathrm{B}}^{h_1,h_2, h}(m,n) \,
\Bigg[ 
\delta^{\bar{c} {b}}  \, (W^{\bar{a} {d}}_{\mathrm{B},h_1+h_2-2-h})_{m+n}
\nonu \\
\!& + \!& (-1)^{h} 
\delta^{\bar{a} {d}}  (W^{\bar{c} {b}}_{\mathrm{B},h_1+h_2-2-h} )_{m+n}
\Bigg]
+ c_{W_{\mathrm{B},h_1}}(m) 
\delta^{\bar{a} {d}}  \delta^{{b} \bar{c}}
\delta^{h_1 h_2} \la^{2(h_1-2)} \delta_{m+n}\ ,
\nonu\\
\big[(W^{\bar{\alpha} {\beta}}_{\mathrm{F},h_1})_m,(Q^{\bar{a} {\gamma}}_{h_2+\frac{1}{2}})_r\big] 
\!&= &\!
\sum_{h=-1}^{h_1+h_2-3}\, \la^h \, q_{\mathrm{F}}^{h_1,h_2+\frac{1}{2}, h}(m,r) 
\,\delta^{\bar{\alpha} {\gamma}} 
\, (Q^{\bar{a} {\beta}}_{h_1+h_2-\frac{3}{2}-h})_{m+r}\ ,
\nonu\\
\big[(W^{\bar{\alpha} {\beta}}_{\mathrm{F},h_1})_m,(\bar{Q}^{{a} \bar{\gamma}}_
{h_2+\frac{1}{2}})_r\big] 
\!&= &\!
\sum_{h=-1}^{h_1+h_2-3}\, \la^h \, (-1)^h
\,q_{\mathrm{F}}^{h_1,h_2+\frac{1}{2}, h}(m,r) \, 
\delta^{ \beta \bar{\gamma}} 
\, (\bar{Q}^{a \bar{\alpha}}_{h_1+h_2-\frac{3}{2}-h})_{m+r}\ ,
\nonu\\
\big[(W^{\bar{a} {b}}_{\mathrm{B},h_1})_m,(Q^{\bar{c} {\alpha}}_{h_2+\frac{1}{2}})_r \big] 
\!&=&\!
\sum_{h=-1 }^{h_1+h_2-3}\, \la^h
\,q_{\mathrm{B}}^{h_1,h_2+\frac{1}{2},h}(m,r)\,
\delta^{\bar{c} {b}} 
\, (Q^{\bar{a} {\alpha}}_{h_1+h_2-\frac{3}{2}-h})_{m+r}\ ,
\nonu\\
\big[(W^{\bar{a} {b}}_{\mathrm{B},h_1})_m,(\bar{Q}^{c \bar{\alpha}}_{h_2+\frac{1}{2}})_r\big] 
\!&=&\!
\sum_{h=-1 }^{h_1+h_2-3}\, \la^h \, (-1)^{h}
\,q_{\mathrm{B}}^{h_1,h_2+\frac{1}{2},h}(m,r)\,
\delta^{\bar{a} c} 
\, (\bar{Q}^{{b}\bar{\alpha}}_{h_1+h_2-\frac{3}{2}-h})_{m+r}\ ,
\nonu\\
\{(Q^{\bar{a} {\alpha}}_{h_1+\frac{1}{2}})_r,(\bar{Q}^{{b} \bar{\beta}}_{
h_2+\frac{1}{2}})_s\} 
\!&=&\!
\sum_{h=0}^{h_1+h_2-1} \,
\la^h
\, o_{\mathrm{F}}^{h_1+\frac{1}{2},h_2+\frac{1}{2},h}(r,s) \,
\delta^{\bar{a} {b}} \,
(W^{\bar{\beta} \alpha}_{\mathrm{F},h_1+h_2-h})_{r+s}
\nonu \\
& + & \sum_{h=0}^{h_1+h_2-2} \,
\la^h\,
o_{\mathrm{B}}^{h_1+\frac{1}{2},h_2+\frac{1}{2},h}(r,s) \, \delta^{\alpha \bar{\beta}}\,
(W^{\bar{a} {b}}_{\mathrm{B},h_1+h_2-h})_{r+s} 
\nonu \\
\!& + \!& c_{Q\bar{Q}_{h_1+\frac{1}{2}}}(r)  \,
\delta^{\bar{a} {b}} \, \delta^{{\alpha} \bar{\beta}}\,
\delta^{h_1 h_2} \, \la^{2(h_1+\frac{1}{2}-1)}\,
\delta_{r+s} \ .
\label{seven}
\eea
The bosonic $W_{\infty}^K$ subalgebra corresponding to the second
equation of
(\ref{seven}) is generated by the $U(K)$-adjoint
$W^{\bar{a} {b}}_{\mathrm{B},h}$ with
an integer weight $h=2,3, \cdots, \infty$.
The subscript $B$ stands for the bilinear of complex free bosons
in next subsection.
The fundamental index $a, b,  \cdots $ of $U(K)$ runs over
$a, b,  \cdots = 1, 2, \cdots, K$
and the antifundamental index $\bar{a}, \bar{b}, \cdots $
of $U(K)$
runs over $\bar{a}, \bar{b}, \cdots  =1,2, \cdots, K$.
The bosonic $W_{1+\infty}^L$ subalgebra corresponding to
the first equation of (\ref{seven}) is generated by the $U(L)$-adjoint
$W^{\bar{\alpha} {\beta}}_{\mathrm{F},h}$ with an integer
weight $h=1,2, \cdots, \infty$. Note the presence of weight-$1$ current.
The subscript $F$ stands for the bilinear of complex free fermions
in next subsection.
The fundamental index $\al, \beta,  \cdots $ of $U(L)$ runs over
$\al, \beta,  \cdots  = 1, 2, \cdots, L$
and the antifundamental index $\bar{\al}, \bar{\beta},  \cdots $
of $U(L)$
runs over $\bar{\al}, \bar{\beta},  \cdots =1,2, \cdots, L$.
There are also the bifundamental $Q^{\bar{a} {\alpha}}_{h+\frac{1}{2}}$
and
the bifundamental $\bar{Q}^{{b} \bar{\beta}}_{
h+\frac{1}{2}}$ under the $U(K) \times U(L)$
with the half-integer
weight $h+\frac{1}{2}= \frac{3}{2}, \frac{5}{2},
\cdots $ for the remaining (anti)commutator relations.
Note that the lower and upper limits for the dummy variable $h$
in (\ref{seven}) can be determined by the fact that
i) the maximum weight for the current in the right hand side
is equal to the sum of two weights in
the left hand side minus one and
ii)
the minimum weight for the current
in the right hand side
is equal to $2, 1$ or $\frac{3}{2}$
as above. 

The $\la$ is a deformation parameter \footnote{
This corresponds to the parameter $q$ in \cite{Odake} and is nothing to
do with the one in the higher spin algebra in \cite{AKK}.}
and the central terms in (\ref{seven}) except the
$\la$-dependent factors are given by
\bea
c_{W_{\mathrm{F},h}}(m)&=&
N\, k \,\frac{2^{2(h-3)}\, (h-1)!)\, (h-1)!}{(2h-3)!!\,(2h-1)!!}
\, \prod_{j=1-h}^{h-1}(m+j)\ ,
\nonu\\
c_{W_{\mathrm{B},h}}(m)&=&
N\, k \,\frac{2^{2(h-3)}(h-2)!\,h!}{(2h-3)!!\,(2h-1)!!}\,
\prod_{j=1-h}^{h-1}(m+j)\ ,
\nonu\\
c_{Q\bar{Q}_{h}}(r)\!&=\!&
N\, k \,\frac{2^{2(h-\frac{3}{2})}(h-\frac{3}{2})!\,(h-\frac{1}{2})!}
{(2h-2)!!\,(2h-2)!!}\,
\prod_{j=\frac{1}{2}-h}^{h-\frac{3}{2}}(r+j+\frac{1}{2})\, .
\label{centralterms}
\eea
There exists an overall factor $N$ which is related to the
number of free complex bosons (or fermions).
The $k$ is the level of $\hat{SU}(L)$ and the corresponding
weight-one current is given by $\frac{4 \, \la}{\sqrt{L\, N}}\,
W^{\bar{\alpha} {\beta}}_{\mathrm{F},1}\,
\de_{\beta \bar{\al}}$ with $c_{W_{\mathrm{F},1}}(m)=\frac{N\, m}{16}$.
By introducing the $k$ copies of the free field realization,
we construct the general level $k$ realization \cite{Odake} because
$W_{\infty}^{K,L}$ algebra is linear.
The Virasoro central charge is given by $c=N k (2K+L)$
from  $c_{W_{\mathrm{F},2}}(m)=\frac{1}{12}\, N\, k\, m(m^2-1)$ and
$c_{W_{\mathrm{B},2}}(m)=\frac{1}{6} \, N\, k \, m(m^2-1)$ from
(\ref{centralterms}) and
the Sugawara stress energy tensor is
given by
$W^{\bar{a} {b}}_{\mathrm{B},2}\,\de_{b \bar{a}}
-
W^{\bar{\alpha} {\beta}}_{\mathrm{F},2}\,\de_{\beta \bar{\al}}$.

The mode-dependent structure constants appearing in (\ref{seven})
are described as follows:
\bea
p_{\mathrm{F}}^{h_1,h_2, h}(m,n)
\! &
\equiv \! &\frac{1}{2(h+1)!}\,\phi^{h_1,h_2}_{h}(0,\textstyle{-\frac{1}{2}})
\,N^{h_1,h_2}_{h}(m,n) \ ,
\nonu\\
p_{\mathrm{B}}^{h_1,h_2, h}(m,n)
\! &
\equiv \! &\frac{1}{2(h+1)!}\,\phi^{h_1,h_2}_{h}(0,0)
\,N^{h_1,h_2}_{h}(m,n) \ ,
\nonu\\
q_{\mathrm{F}}^{h_1,h_2, h}(m,r) 
\! &
\equiv \! &\frac{(-1)^h}{4(h+2)!}\Bigg[
(h_1-1)\,\phi^{h_1,h_2+\frac{1}{2}}_{h+1}(0,0)
\nonu \\
\!& - \!& (h_1-h-3)\,\phi^{h_1,h_2+\frac{1}{2}}_{h+1}(0,\textstyle{-\frac{1}{2}})
\Bigg]
\,N^{h_1,h_2}_{h}(m,r) \ ,
\nonu\\
q_{\mathrm{B}}^{h_1,h_2, h}(m,r) 
\! &
\equiv \! &\frac{-1}{4(h+2)!}\Bigg[
(h_1-h-2)\,\phi^{h_1,h_2+\frac{1}{2}}_{h+1}(0,0)-(h_1)\,\phi^{h_1,h_2+\frac{1}{2}}_{h+1}(0,\textstyle{-\frac{1}{2}})
\Bigg]
\,N^{h_1,h_2}_{h}(m,r) \ ,
\nonu\\
o_{\mathrm{F}}^{h_1,h_2,h}(r,s) 
\! &
\equiv \! &\frac{4(-1)^h}{h!}\Bigg[
(h_1+h_2-1-h)\,\phi^{h_1+\frac{1}{2},h_2+\frac{1}{2}}_{h}(
\textstyle{\frac{1}{2}},\textstyle{-\frac{1}{4}})
\nonu \\
\!& - \!&
(h_1+h_2-\frac{3}{2}-h)\,\phi^{h_1+\frac{1}{2},h_2+\frac{1}{2}}_{h+1}(
\textstyle{\frac{1}{2}},\textstyle{-\frac{1}{4}})
\Bigg]\,  N^{h_1,h_2}_{h-1}(r,s)\ ,
\nonu\\
o_{\mathrm{B}}^{h_1,h_2,h}(r,s) 
\!&
\equiv \!&-\frac{4}{h!}\Bigg[
(h_1+h_2-2-h)\,\phi^{h_1+\frac{1}{2},h_2+\frac{1}{2}}_{h}(
\textstyle{\frac{1}{2}},\textstyle{-\frac{1}{4}})
\nonu \\
\!& - \!&
(h_1+h_2-\frac{3}{2}-h)\,\phi^{h_1+\frac{1}{2},h_2+\frac{1}{2}}_{h+1}(\textstyle{\frac{1}{2}},\textstyle{-\frac{1}{4}})
\Bigg]
\,  N^{h_1,h_2}_{h-1}(r,s)\ .
\label{modedependence}
\eea
The structure constants are polynomials in the modes.
The modes $m,n, \cdots$ are integers and the modes
$r,s, \cdots$ are half-integers.
We introduce
the following quantities
\bea
N^{h_1,h_2}_{h}(m,n)
\!&
\equiv \!&
\sum_{l=0 }^{h+1}(-1)^l
\left(\begin{array}{c}
h+1 \\  l \\
\end{array}\right)
[h_1-1+m]_{h+1-l}[h_1-1-m]_l
\nonu \\
\!& \times \!& [h_2-1+n]_l [h_2-1-n]_{h+1-l} \ ,
\nonu\\
\phi^{h_1,h_2}_{h}(x,y)
\!&
\equiv \!&
{}_4 F_3
\Bigg[
\begin{array}{c}
-\frac{1}{2}-x-2y, \frac{3}{2}-x+2y, -\frac{h+1}{2}+x,
-\frac{h}{2} +x \\
-h_1+\frac{3}{2},-h_2+\frac{3}{2},h_1+h_2-h-\frac{3}{2}
\end{array} ; 1
\Bigg]\ .
\label{Nphi}
\eea
We use the falling Pochhammer symbol
$[a]_n \equiv a(a-1) \cdots (a-n+1)$ in (\ref{Nphi}) and
we use the binomial coefficients for parentheses.
Moreover, the generalized hypergeometric function, 
with four upper arguments $a_i$,  three lower arguments $b_i$
and variable $z$, is 
defined as the series
\bea
{}_4 F_3
\Bigg[
\begin{array}{c}
a_1, a_2, a_3, a_4 \\
b_1,b_2,b_3
\end{array} ; z
\Bigg] 
=
\sum_{n=0 }^{\infty}
\frac{(a_1)_n (a_2)_n (a_3)_n (a_4)_n}
{(b_1)_n (b_2)_n (b_3)_n}
\frac{z^n}{n!}\,,
\label{defF43}
\eea
where the rising Pochhammer symbol
$(a)_n \equiv a(a+1) \cdots (a+n-1)$ is used in (\ref{defF43})
\footnote{\label{typos}
In the right hand side of (\ref{Nphi}),
the $h_1$ and $h_2$ play the role of
$(i+2)$ (or $(i+\frac{3}{2})$) and
$(j+2)$ (or $(j+\frac{3}{2})$) of \cite{Odake}.
The $h$ corresponds to their $r$. Due to the typos in \cite{BPRSS}
(See also the
footnote $2$ of \cite{PRSS}), the structure constants in \cite{Odake}
are different from the ones in \cite{BPRSS}. For example,
the structure constant $a^{i \, \al}_l(m,r)$ of \cite{BPRSS} is
given by our 
$(-1)^h \, q_B^{h_1,h_2,h}(m,r)$ with
the identification $h_1=i+2, h_2=\al+\frac{3}{2},h=l-1$ and
the structure constant $\tilde{a}^{i \, \al}_l(m,r)$ of \cite{BPRSS}
is given by our $(-1)^h \, q_F^{h_1,h_2,h}(m,r)$
with the identification $h_1=i+2, h_2=\al+\frac{3}{2},h=l-1$.}.

In this paper,
by acting the generators of $U(K)$ (or $U(L)$)
with the contractions of the (anti)fundamental indices
on the (anti)commutator relations in (\ref{seven}),
we will determine the ${\cal N}=2$ supersymmetric
$W_{\infty}^{K,K}$ algebra and
the ${\cal N}=1$ supersymmetric
$W_{\infty}^{K}$ algebra.
Then the fermionic currents will not have any (anti)fundamental indices. 
 
\subsection{Free field realization: Review}

The $W_{\infty}^{K,L}$ algebra with level $k=1$
is realized by $K$-free complex bosons of weight-$1$
($\bar{\pa}\, \phi^{\bar{i},a}$
and $\bar{\pa}\, \bar{\phi}^{i,\bar{a}}$) and
$L$-free complex fermions of weight-$\frac{1}{2}$
($\psi^{\bar{i},\al}$ and $\bar{\psi}^{i, \bar{\al}}$). The index $i$ is the
fundamental index of $U(N)$
and the index $\bar{i}$ is the antifundamental index of $U(N)$.
Their operator product expansions in the antiholomorphic sector
are
\bea
\bar{\pa} \, \bar{\phi}^{i,\bar{a}}(\bar{z}) \, \bar{\pa} \,
\phi^{\bar{j},b}(\bar{w}) \!& = \!& \frac{1}{(\bar{z}-\bar{w})^2}\,
\de^{i \bar{j}}
\, \de^{\bar{a}  b} + \cdots \ ,
\nonu \\
\bar{\psi}^{i,\bar{\al}}(\bar{z}) \, \psi^{\bar{j},\beta}(\bar{w})
\! & = \! & \frac{1}{(\bar{z}-\bar{w})}\,
\de^{i \bar{j}} \, \de^{\bar{\al} \beta}+ \cdots \ .
\label{OPE}
\eea
The $U(N)$-singlet currents of  $W_{\infty}^{K,L}$ algebra
are described by the bilinears of these free fields as follows
(See also \cite{EGR}):
\bea
W^{\bar{\alpha} \beta}_{\mathrm{F},h}
&
=& n_{W_{\mathrm{F},h}}
\sum_{l=0}^{h-1} \sum_{i,\, \bar{\imath}=1}^N
\delta_{i,\bar{\imath}}(-1)^l
\left(\begin{array}{c}
h-1 \\  l \\
\end{array}\right)^2
\,(\,\bar{\partial}^{h-l-1}\bar{\psi}^{i,\bar{\alpha}} \,
\bar{\partial}^l\psi^{\bar{\imath},\beta}\,) \, ,
\nonu \\
W^{\bar{a} b}_{\mathrm{B},h}
&
=& n_{W_{\mathrm{B},h}}
\sum_{l=0}^{h-2}\sum_{i,\, \bar{\imath}=1}^N 
\delta_{i,\bar{\imath}}\frac{(-1)^l}{(h-1)}
\left(\begin{array}{c}
h-1 \\  l \\
\end{array}\right)
\left(\begin{array}{c}
h-1 \\  l+1 \\
\end{array}\right)
\,(\,\bar{\partial}^{h-l-1}\bar{\phi}^{i,\bar{a}}  \,
\bar{\partial}^{l+1}\phi^{\bar{\imath},b} \,)\, , \nonu \\
Q^{\bar{ a} \alpha}_{h+\frac{1}{2}} 
&
=& n_{Q_{h+\frac{1}{2}}} \sum_{l=0}^{h-1}\sum_{i,\, \bar{\imath}=1}^N 
\delta_{i,\bar{\imath}}(-1)^l
\left(\begin{array}{c}
h-1 \\  l \\
\end{array}\right)
\left(\begin{array}{c}
h \\  l \\
\end{array}\right)
\,(\,\bar{\partial}^{h-l}\bar{\phi}^{i,\bar{a}}
\bar{\partial}^l
\psi^{\bar{\imath},\alpha}\,)\, , \nonu \\
\bar{Q}^{a \bar{\alpha}}_{h+\frac{1}{2}} 
&=& n_{\bar{Q}_{h+\frac{1}{2}}} \sum_{l=0}^{h-1} \sum_{i,\, \bar{\imath}=1}^N
\delta_{i,\bar{\imath}}(-1)^{h-1+l}
\left(\begin{array}{c}
h-1 \\  l \\
\end{array}\right)
\left(\begin{array}{c}
h \\  l \\
\end{array}\right)
\,(\,\bar{\partial}^{h-l}\phi^{\bar{\imath},a}
\bar{\partial}^l  \bar{\psi}^{i,\bar{\alpha}}\,)\, .
\label{WWQQ}
\eea
The $l=0$ cases of single summations
correspond to the lowest weights, $1,2, \frac{3}{2}$ and
$\frac{3}{2}$. 
The normalizations are given by
\bea
n_{W_{\mathrm{F},h}} \!& = \!&  \frac{2^{h-3}(h-1)!}{(2h-3)!!}\,\la^{h-2} \ ,
\qquad
n_{W_{\mathrm{B},h}}=\frac{2^{h-3}\,h!}{(2h-3)!!}\,\la^{h-2} \ ,
\nonu \\
n_{Q_{h+\frac{1}{2}}} \!& = \!&
\frac{2^{h-\frac{1}{2}}h!}{(2h-1)!!}\,\la^{h-1}
= n_{\bar{Q}_{h+\frac{1}{2}}} \ .
\label{nor}
\eea

Then we can check that the modes of (\ref{WWQQ}) satisfy
the previous (anti)commutator relations
(\ref{seven}) by using the mode expansion for the normal
ordering between the free fields in the conformal field theory.
After using the
(anti)commutator relations corresponding to (\ref{OPE}),
the left hand sides of (\ref{seven}) contain
the quadratic free fields having the explicit modes
(where the coefficients depend on the weights, the modes and the
dummy variable from the infinite sum)
and the central terms.
Similarly, the right hand sides of (\ref{seven}) contain
the quadratic free fields
and the central terms in the presence of the nontrivial structure
constants (\ref{modedependence}).
For several low values of the weights and the modes, we can check
the several nontrivial identities.
Alternatively, after using the Thielemans package \cite{Thielemans}
for low values of the weights,
simplifying the right hand sides of the OPEs between the currents and
rewriting down them in terms of the (anti)commutator relations
with the help of the explicit formula in \cite{CFT,Blumenhagenetal},
the previous algebra (\ref{seven}) can be
checked also explicitly \footnote{This paragraph is based on
the discussion with S. Odake some years ago.}.

\subsection{ The ${\cal N}=2$ supersymmetric
$W_{\infty}^{K,K}$ algebra with $U(K)\times U(K)$
symmetry}

Let us consider the case of $K=L$.
Then the number of (anti)fundamental indices is the same.
From the decomposition of $U(K+K) = U(K) \oplus U(K) \oplus
({\bf K},\bar{\bf K}) \oplus (\bar{\bf K}, {\bf K})$,
the generators consists of $ t^{\hat{A}}_{\al \bar{\beta}}$,
$ t^{\hat{A}}_{a \bar{b}}$, $ t^{\hat{A}}_{\alpha \bar{a}}$ and
$ t^{\hat{A}}_{a \bar{\alpha}}$ in addition to $\delta_{\al
\bar{\beta}}$, $\delta_{a \bar{b}}$, $\delta_{\alpha
\bar{a}}$ and $\delta_{a\bar{\alpha}}$
with $\hat{A} = 1, 2, \cdots, (K^2-1)$.

By multiplying the generators into the four kinds of currents,
we obtain four kinds of singlets and adjoints of $U(K)$
as follows:
\bea
W_{\mathrm{F},h} \!& \equiv
\! & W^{\bar{\alpha} \beta}_{\mathrm{F},h}\, \delta_{\beta
\bar{\alpha}} \ , \qquad
W^{\hat{A}}_{\mathrm{F},h} \equiv
W^{\bar{\alpha} \beta}_{\mathrm{F},h}\, t^{\hat{A}}_{\beta \bar{\alpha}}\ ,
\nonu \\
W_{\mathrm{B},h} \! & \equiv \! & W^{\bar{a} b}_{\mathrm{B},h}\,
\delta_{b \bar{a}} \ , \qquad
W^{\hat{A}}_{\mathrm{B},h} \equiv
W^{\bar{a} b}_{\mathrm{B},h}\, t^{\hat{A}}_{b \bar{a}} \ ,
\nonu \\
Q_{h+\frac{1}{2}} \!& \equiv
\! & Q^{\bar{a} \alpha}_{h+\frac{1}{2}}\, \delta_{\alpha
  \bar{a}} \ , \qquad
Q^{\hat{A}}_{h+\frac{1}{2}} \equiv
Q^{\bar{a} \alpha}_{h+\frac{1}{2}}\, t^{\hat{A}}_{\alpha \bar{a}}\ ,
\nonu \\
\bar{Q}_{h+\frac{1}{2}} \!& \equiv
\! & \bar{Q}^{a \bar{\alpha} }_{h+\frac{1}{2}}\, \delta_{a
\bar{\alpha}} \ , \qquad
\bar{Q}^{\hat{A}}_{h+\frac{1}{2}} \equiv
\bar{Q}^{a \bar{\alpha} }_{h+\frac{1}{2}}\, t^{\hat{A}}_{a \bar{\alpha}}\ ,
\qquad \hat{A} = 1, 2, \cdots, (K^2-1) \ .
\label{singletadjoint}
\eea

Now we would like to rewrite down (\ref{seven}) in terms of
(\ref{singletadjoint}) after multiplying the various generators
\footnote{From now on, we do not have to distinguish the two (anti)
fundamental indices.}.

\subsubsection{ The $W_{1+\infty}^K$ algebra}

Now we can multiply the generators into the first equation of (\ref{seven})
and the three commutator relations can be obtained as follows:
\bea
\big[(W_{\mathrm{F},h_1})_m,(W_{\mathrm{F},h_2})_n\big] 
\!&=& \!
\sum^{h_1+h_2-3}_{h= 0, \mbox{\footnotesize even}} \, \la^h\,
p_{\mathrm{F}}^{h_1,h_2, h}(m,n)
\, (   W_{\mathrm{F},h_1+h _2-2-h} )_{m+n}\nonu \\
\!& + \!&
K\, c_{W_{\mathrm{F},h_1}} \,
\delta^{h_1 h_2}\,\la^{2(h_1-2)}\,\delta_{m+n}\ ,
\nonu \\
\big[(W_{\mathrm{F},h_1})_m,(W^{\hat{A}}_{\mathrm{F},h_2})_n\big] 
\!&=& \!
\sum^{h_1+h_2-3}_{h= 0, \mbox{\footnotesize even}} \, \la^h\,
p_{\mathrm{F}}^{h_1,h_2, h}(m,n)
\, (   W^{\hat{A}}_{\mathrm{F},h_1+h_2-2-h} )_{m+n}\ ,
\nonu \\
\big[(W^{\hat{A}}_{\mathrm{F},h_1})_m,(W^{\hat{B}}_{\mathrm{F},h_2})_n\big] 
\!&=& \!
-\sum^{h_1+h_2-3}_{h= -1,
\mbox{\footnotesize odd}} \, \la^h\,
p_{\mathrm{F}}^{h_1,h_2, h}(m,n)
\, \frac{i}{2}\, f^{\hat{A} \hat{B}  \hat{C}} \, (   W^{\hat{C}}_{\mathrm{F},h_1+h_2-2-h} )_{m+n}
\nonu \\
\!& + \!&
c_{W_{\mathrm{F},h_1}} \,
\delta^{\hat{A} \hat{B}}\, \delta^{h_1 h_2}\,\la^{2(h_1-2)}\,\delta_{m+n}
\nonu \\
\!&+\!& \sum^{h_1+h_2-3}_{h= 0, \mbox{\footnotesize even}} \, \la^h\,
p_{\mathrm{F}}^{h_1,h_2, h}(m,n)
\, \Big[ \frac{1}{2}\, d^{\hat{A} \hat{B} \hat{C}} \,
(   W^{\hat{C}}_{\mathrm{F},h_1+h_2-2-h} )_{m+n} \nonu \\
\!& + \!& \frac{1}{K}\, \delta^{\hat{A} \hat{B}}\,
(   W_{\mathrm{F},h_1+h_2-2-h} )_{m+n} \Big]\ .
\label{OSalgebra}
\eea
This $W_{1+\infty}^K$ algebra (or $\hat{SU}(K)_k$
$W_{1+\infty}$ algebra) was found in \cite{OS}.
The first equation of (\ref{OSalgebra}) generated by
the singlet current of $U(K)$ is $W_{1+\infty}$ algebra
and its extension with the adjoint of $U(K)$
appears in the remaining equations.
In the last equation of (\ref{OSalgebra}),
the identity for the product of two generators
$t^{\hat{A}}\, t^{\hat{B}}= \frac{1}{K}\, \de^{\hat{A} \hat{B}} \,
{\bf 1}_K + \frac{1}{2}\, (i\, f + d)^{\hat{A} \hat{B} \hat{C}}\, t^{\hat{C}}
$ is used.
Note that the weights in the right hand side
appear in even or odd integers. Of course,
by taking the contractions of the currents with the vanishing
$\la$ limit, the first equation reduces to
the $w_{1+\infty}$ algebra (which is the weight-$1$ extension
of $w_{\infty}$ algebra \cite{Bakas}) as shown in \cite{Ahn2111}
\footnote{
\label{decoupling}
  By redefining the currents of weights $2, 3, 4$ nonlinearly,
  we can decouple the weight-$1$ current from other currents. That is,
  there are no singular OPEs between the weight-$1$ current and others.
  The OPEs between the above currents of weights $2,3,4$ do not contain
  the weight-$1$ current at the poles in the right hand side. The lowest
  pole we are considering contains the weight-$4$ current.
  For the poles having the higher spin current of spins, $5, 6, \cdots $,
  we need to find out the corresponding redefined currents step by step.
  We expect that this will be true
for higher weights. See also the relevant paper \cite{PRSplb}.}.

\subsubsection{The $W_{\infty}^K$ algebra}

The second equation of (\ref{seven}) can be rewritten as
\bea
\big[(W_{\mathrm{B},h_1})_m,(W_{\mathrm{B},h_2})_n\big] 
\!&=& \!
\sum^{h_1+h_2-4}_{h= 0, \mbox{\footnotesize even}} \, \la^h\,
p_{\mathrm{B}}^{h_1,h_2, h}(m,n)
\, (   W_{\mathrm{B},h_1+h_2-2-h} )_{m+n} \nonu \\
\!& + \!&
K\, c_{W_{\mathrm{B},h_1}} \,
\delta^{h_1 h_2}\,\la^{2(h_1-2)}\,\delta_{m+n}\ ,
\nonu \\
\big[(W_{\mathrm{B},h_1})_m,(W^{\hat{A}}_{\mathrm{B},h_2})_n\big] 
\!&=& \!
\sum^{h_1+h_2-4}_{h= 0, \mbox{\footnotesize even}} \, \la^h\,
p_{\mathrm{B}}^{h_1,h_2, h}(m,n)
\, (   W^{\hat{A}}_{\mathrm{B},h_1+h_2-2-h} )_{m+n}\ ,
\nonu \\
\big[(W^{\hat{A}}_{\mathrm{B},h_1})_m,(W^{\hat{B}}_{\mathrm{B},h_2})_n\big] 
\!&=& \!
-\sum^{h_1+h_2-4}_{h= -1, \mbox{\footnotesize odd}} \,
\la^h\, p_{\mathrm{B}}^{h_1,h_2, h}(m,n)
\, \frac{i}{2}\, f^{\hat{A} \hat{B}  \hat{C}} \,
(   W^{\hat{C}}_{\mathrm{B},h_1+h_2-2-h} )_{m+n}
\nonu \\
\!& + \!&
c_{W_{\mathrm{B},h_1}} \,
\delta^{\hat{A} \hat{B}}\,
\delta^{h_1 h_2}\,\la^{2(h_1-2)}\,\delta_{m+n}
\nonu \\
\!&+\!& \sum^{h_1+h_2-4}_{h= 0, \mbox{\footnotesize even}} \, \la^h\,
p_{\mathrm{B}}^{h_1,h_2, h}(m,n)
\, \Bigg[ \frac{1}{2}\, d^{\hat{A} \hat{B} \hat{C}} \,
(   W^{\hat{C}}_{\mathrm{B},h_1+h_2-2-h} )_{m+n} \nonu \\
\!& + \!& \frac{1}{K}\, \delta^{\hat{A} \hat{B}}\,
(   W_{\mathrm{B},h_1+h_2-2-h} )_{m+n} \Bigg]\ .
\label{BKalgebra}
\eea
This $W_{\infty}^K$ algebra was found in \cite{BK}.
The first equation of (\ref{BKalgebra}) generated by
the singlet current of $U(K)$ is $W_{\infty}$ algebra
and its extension with the adjoint of $U(K)$
appears in the remaining equations.
The algebraic structure of (\ref{BKalgebra}) looks similar to
the one of (\ref{OSalgebra}). The upper bound of dummy
variable $h$ and the structure constants are different from
each other.
By taking the contractions of the currents with the vanishing
$\la$ limit, the first equation reduces to
the $w_{\infty}$ algebra \cite{Bakas}.

\subsubsection{The commutators between the bosonic and fermionic
currents}

In this case, we have four commutator relations after
multiplying the generators into the third equation of (\ref{seven})
\bea
\big[(W_{\mathrm{F},h_1})_m,(Q_{h_2+\frac{1}{2}})_r\big] 
\!&= &\!
\sum^{h_1+h_2-3}_{h=-1}\, \la^h \, q_{\mathrm{F}}^{h_1,h_2+\frac{1}{2}, h}(m,r) 
\, (Q_{h_1+h_2-\frac{3}{2}-h})_{m+r}\ ,
\nonu \\
\big[(W_{\mathrm{F},h_1})_m,(Q^{\hat{A}}_{h_2+\frac{1}{2}})_r\big] 
\!&= &\!
\sum^{h_1+h_2-3}_{h=-1}\, \la^h \, q_{\mathrm{F}}^{h_1,h_2+\frac{1}{2}, h}(m,r) 
\, (Q^{\hat{A}}_{h_1+h_2-\frac{3}{2}-h})_{m+r}\ ,
\nonu\\
\big[(W^{\hat{A}}_{\mathrm{F},h_1})_m,(Q_{h_2+\frac{1}{2}})_r\big] 
\!&= &\!
\sum^{h_1+h_2-3}_{h=-1}\, \la^h \, q_{\mathrm{F}}^{h_1,h_2+\frac{1}{2}, h}(m,r) 
\, (Q^{\hat{A}}_{h_1+h_2-\frac{3}{2}-h})_{m+r}\ ,
\nonu \\
\big[(W^{\hat{A}}_{\mathrm{F},h_1})_m,(Q^{\hat{B}}_{h_2+\frac{1}{2}})_r\big] 
\!&=& \!
\sum^{h_1+h_2-3}_{h= -1} \, \la^h\, q_{\mathrm{F}}^{h_1,h_2+\frac{1}{2}, h}(m,r)
\, \frac{i}{2}\, f^{\hat{A} \hat{B}  \hat{C}} \, (
Q^{\hat{C}}_{h_1+h_2-\frac{3}{2}-h} )_{m+r}
\nonu \\
\!&+\!& \sum^{h_1+h_2-3}_{h= -1} \, \la^h\, q_{\mathrm{F}}^{h_1,h_2+\frac{1}{2}, h}(m,r)
\, \Bigg[ \frac{1}{2}\, d^{\hat{A} \hat{B} \hat{C}} \,
(   Q^{\hat{C}}_{h_1+h_2-\frac{3}{2}-h} )_{m+r} \nonu \\
\!& + \!& \frac{1}{K}\, \delta^{\hat{A} \hat{B}}\,
(   Q_{h_1+h_2-\frac{3}{2}-h} )_{m+r} \Bigg]\ .
\label{WFQ}
\eea

\subsubsection{The commutators between the other bosonic and fermionic
currents}

From the fifth equation of (\ref{seven}), the following four commutator
relations can be obtained by multiplying the generators 
\bea
\big[(W_{\mathrm{B},h_1})_m,(Q_{h_2+\frac{1}{2}})_r\big] 
\!&= &\!
\sum^{h_1+h_2-3}_{h=-1}\, \la^h \, q_{\mathrm{B}}^{h_1,h_2+\frac{1}{2}, h}(m,r) 
\, (Q_{h_1+h_2-\frac{3}{2}-h})_{m+r}\ ,
\nonu \\
\big[(W_{\mathrm{B},h_1})_m,(Q^{\hat{A}}_{h_2+\frac{1}{2}})_r\big] 
\!&= &\!
\sum^{h_1+h_2-3}_{h=-1}\, \la^h \, q_{\mathrm{B}}^{h_1,h_2+\frac{1}{2}, h}(m,r) 
\, (Q^{\hat{A}}_{h_1+h_2-\frac{3}{2}-h})_{m+r}\ ,
\nonu\\
\big[(W^{\hat{A}}_{\mathrm{B},h_1})_m,(Q_{h_2+\frac{1}{2}})_r\big] 
\!&= &\!
\sum^{h_1+h_2-3}_{h=-1}\, \la^h \, q_{\mathrm{B}}^{h_1,h_2+\frac{1}{2}, h}(m,r) 
\, (Q^{\hat{A}}_{h_1+h_2-\frac{3}{2}-h})_{m+r}\ ,
\nonu \\
\big[(W^{\hat{A}}_{\mathrm{B},h_1})_m,(Q^{\hat{B}}_{h_2+\frac{1}{2}})_r\big] 
\!&=& \!
-\sum^{h_1+h_2-3}_{h= -1} \, \la^h\, q_{\mathrm{B}}^{h_1,h_2+\frac{1}{2}, h}(m,r)
\, \frac{i}{2}\, f^{\hat{A} \hat{B}  \hat{C}} \,
(   Q^{\hat{C}}_{h_1+h_2-\frac{3}{2}-h} )_{m+r}
\nonu \\
\!&+\!& \sum^{h_1+h_2-3}_{h= -1} \, \la^h\, q_{\mathrm{B}}^{h_1,h_2+\frac{1}{2}, h}(m,r)
\, \Bigg[ \frac{1}{2}\, d^{\hat{A} \hat{B} \hat{C}} \,
(   Q^{\hat{C}}_{h_1+h_2-\frac{3}{2}-h} )_{m+r} \nonu \\
\!& + \!& \frac{1}{K}\, \delta^{\hat{A} \hat{B}}\,
(   Q_{h_1+h_2-\frac{3}{2}-h} )_{m+r} \Bigg]\ .
\label{WBQ}
\eea
We present here (\ref{WFQ}) and (\ref{WBQ})
which are necessary to describe the discussion of
next section and
the remaining
(anti)commutator relations are presented in Appendix $A$.

\subsection{Free field realization}

From (\ref{singletadjoint}) and (\ref{WWQQ}) with (\ref{nor}),
we obtain the following free field realization 
\bea
W_{\mathrm{F},h}
&
=& n_{W_{\mathrm{F},h}}
\sum_{l=0}^{h-1} \sum_{i,\, \bar{\imath}=1}^N
\delta_{i,\bar{\imath}}(-1)^l
\left(\begin{array}{c}
h-1 \\  l \\
\end{array}\right)^2
\,(\,\bar{\partial}^{h-l-1}\bar{\psi}^{i,\bar{\alpha}} \,
\de_{\beta \bar{\al}}\,
\bar{\partial}^l\psi^{\bar{\imath},\beta}\,) \, ,
\nonu \\
W^{\hat{A}}_{\mathrm{F},h}
&
=& n_{W_{\mathrm{F},h}}
\sum_{l=0}^{h-1} \sum_{i,\, \bar{\imath}=1}^N
\delta_{i,\bar{\imath}}(-1)^l
\left(\begin{array}{c}
h-1 \\  l \\
\end{array}\right)^2
\,(\,\bar{\partial}^{h-l-1}\bar{\psi}^{i,\bar{\alpha}} \,
t^{\hat{A}}_{\beta \bar{\al}}\,
\bar{\partial}^l\psi^{\bar{\imath},\beta}\,) \, ,
\nonu \\
W_{\mathrm{B},h}
&
=& n_{W_{\mathrm{B},h}}
\sum_{l=0}^{h-2}\sum_{i,\, \bar{\imath}=1}^N 
\delta_{i,\bar{\imath}}\frac{(-1)^l}{(h-1)}
\left(\begin{array}{c}
h-1 \\  l \\
\end{array}\right)
\left(\begin{array}{c}
h-1 \\  l+1 \\
\end{array}\right)
\,(\,\bar{\partial}^{h-l-1}\bar{\phi}^{i,\bar{a}}  \,
\de_{b \bar{a}}\,
\bar{\partial}^{l+1}\phi^{\bar{\imath},b} \,)\, , \nonu \\
W^{\hat{A}}_{\mathrm{B},h}
&
=& n_{W_{\mathrm{B},h}}
\sum_{l=0}^{h-2}\sum_{i,\, \bar{\imath}=1}^N 
\delta_{i,\bar{\imath}}\frac{(-1)^l}{(h-1)}
\left(\begin{array}{c}
h-1 \\  l \\
\end{array}\right)
\left(\begin{array}{c}
h-1 \\  l+1 \\
\end{array}\right)
\,(\,\bar{\partial}^{h-l-1}\bar{\phi}^{i,\bar{a}}  \,
t^{\hat{A}}_{b \bar{a}}\,
\bar{\partial}^{l+1}\phi^{\bar{\imath},b} \,)\, , \nonu \\
Q_{h+\frac{1}{2}} 
&
=& n_{Q_{h+\frac{1}{2}}} \sum_{l=0}^{h-1}\sum_{i,\, \bar{\imath}=1}^N 
\delta_{i,\bar{\imath}}(-1)^l
\left(\begin{array}{c}
h-1 \\  l \\
\end{array}\right)
\left(\begin{array}{c}
h \\  l \\
\end{array}\right)
\,(\,\bar{\partial}^{h-l}\bar{\phi}^{i,\bar{a}}\,
\de_{\al \bar{a}}\,
\bar{\partial}^l
\psi^{\bar{\imath},\alpha}\,)\, , \nonu \\
Q^{\hat{A}}_{h+\frac{1}{2}} 
&
=& n_{Q_{h+\frac{1}{2}}} \sum_{l=0}^{h-1}\sum_{i,\, \bar{\imath}=1}^N 
\delta_{i,\bar{\imath}}(-1)^l
\left(\begin{array}{c}
h-1 \\  l \\
\end{array}\right)
\left(\begin{array}{c}
h \\  l \\
\end{array}\right)
\,(\,\bar{\partial}^{h-l}\bar{\phi}^{i,\bar{a}}
\,
t^{\hat{A}}_{\al \bar{a}} \,
\bar{\partial}^l
\psi^{\bar{\imath},\alpha}\,)\, , \nonu \\
\bar{Q}_{h+\frac{1}{2}} 
&=& n_{\bar{Q}_{h+\frac{1}{2}}} \sum_{l=0}^{h-1} \sum_{i,\, \bar{\imath}=1}^N
\delta_{i,\bar{\imath}}(-1)^{h-1+l}
\left(\begin{array}{c}
h-1 \\  l \\
\end{array}\right)
\left(\begin{array}{c}
h \\  l \\
\end{array}\right)
\,(\,\bar{\partial}^{h-l}\phi^{\bar{\imath},a}\,
\de_{a\bar{\al}}\,
\bar{\partial}^l  \bar{\psi}^{i,\bar{\alpha}}\,)\, ,
\nonu \\
\bar{Q}^{\hat{A}}_{h+\frac{1}{2}} 
&=& n_{\bar{Q}_{h+\frac{1}{2}}} \sum_{l=0}^{h-1} \sum_{i,\, \bar{\imath}=1}^N
\delta_{i,\bar{\imath}}(-1)^{h-1+l}
\left(\begin{array}{c}
h-1 \\  l \\
\end{array}\right)
\left(\begin{array}{c}
h \\  l \\
\end{array}\right)
\,(\,\bar{\partial}^{h-l}\phi^{\bar{\imath},a}\,
t^{\hat{A}}_{a \bar{\al}}\,
\bar{\partial}^l  \bar{\psi}^{i,\bar{\alpha}}\,)\, .
\label{WWQQ1}
\eea
The number $N$ which is contracted in
(\ref{WWQQ1}) does not play any role of our discussion.
In this basis, the singlet and adjoint property of
$U(K)$ is clear.
It is evident that the above
free field realizations satisfy (\ref{WFQ}) and (\ref{WBQ})
and Appendices (\ref{app1}), (\ref{app2}) and (\ref{app3}).
Except the last two of (\ref{WWQQ1}),
the remaining ones will be used in next section.

\subsection{ The existence of ${\cal N}=2$ supersymmetric
$w_{\infty}^{K,K}$ algebra with $U(K)\times U(K)$
symmetry?}

By taking the simple rescalings
\bea
W_{h} \!& \rightarrow \!& \la \, W_h \ , \qquad W_{h}^{\hat{A}} \rightarrow
\la \, W_{h}^{\hat{A}}\ , \nonu \\
Q_{h+\frac{1}{2}} \!& \rightarrow \!& \la \,
Q_{h+\frac{1}{2}} \ , 
\qquad
Q^{\hat{A}}_{h+\frac{1}{2}} \rightarrow \la\,
Q^{\hat{A}}_{h+\frac{1}{2}} \ ,
\label{rescaling}
\eea
and
putting the $\la$ to vanish,
we obtain the following commutator relations 
\bea
\big[(W^{\hat{A}}_{\mathrm{F},h_1})_m,(W^{\hat{B}}_{\mathrm{F},h_2})_n\big] 
\!&= & \!
-
\, \frac{i}{4}\, f^{\hat{A} \hat{B}  \hat{C}} \,
(   W^{\hat{C}}_{\mathrm{F},h_1+h_2-1} )_{m+n} \ ,
\nonu \\
\big[(W^{\hat{A}}_{\mathrm{B},h_1})_m,(W^{\hat{B}}_{\mathrm{B},h_2})_n\big] 
\!&= & \!
-
\, \frac{i}{4}\, f^{\hat{A} \hat{B}  \hat{C}} \,
(   W^{\hat{C}}_{\mathrm{B},h_1+h_2-1} )_{m+n} \ ,
\nonu \\
\big[(W_{\mathrm{F},h_1})_m,(Q_{h_2+\frac{1}{2}})_r\big] 
\!&= &\!
-\frac{1}{4}
\, (Q_{h_1+h_2-\frac{1}{2}})_{m+r}\ ,
\nonu \\
\big[(W_{\mathrm{F},h_1})_m,(Q^{\hat{A}}_{h_2+\frac{1}{2}})_r\big] 
\!&= &\!
-\frac{1}{4}
\, (Q^{\hat{A}}_{h_1+h_2-\frac{1}{2}})_{m+r}\ ,
\nonu\\
\big[(W^{\hat{A}}_{\mathrm{F},h_1})_m,(Q_{h_2+\frac{1}{2}})_r\big] 
\!&= &\!
-\frac{1}{4}
\, (Q^{\hat{A}}_{h_1+h_2-\frac{1}{2}})_{m+r}\ ,
\nonu \\
\big[(W^{\hat{A}}_{\mathrm{F},h_1})_m,(Q^{\hat{B}}_{h_2+\frac{1}{2}})_r\big] 
\!&=& \!
- \frac{i}{8}\, f^{\hat{A} \hat{B}  \hat{C}} \, (
Q^{\hat{C}}_{h_1+h_2-\frac{1}{2}} )_{m+r}
- \frac{1}{4}
\, \Bigg[ \frac{1}{2}\, d^{\hat{A} \hat{B} \hat{C}} \,
(   Q^{\hat{C}}_{h_1+h_2-\frac{1}{2}} )_{m+r} \nonu \\
\!& + \!& \frac{1}{K}\, \delta^{\hat{A} \hat{B}}\,
(   Q_{h_1+h_2-\frac{1}{2}} )_{m+r} \Bigg] \ ,
\nonu \\
\big[(W_{\mathrm{B},h_1})_m,(Q_{h_2+\frac{1}{2}})_r\big] 
\!&= &\!
\frac{1}{4}
\, (Q_{h_1+h_2-\frac{1}{2}})_{m+r}\ ,
\nonu \\
\big[(W_{\mathrm{B},h_1})_m,(Q^{\hat{A}}_{h_2+\frac{1}{2}})_r\big] 
\!&= &\!
\frac{1}{4}
\, (Q^{\hat{A}}_{h_1+h_2-\frac{1}{2}})_{m+r}\ ,
\nonu \\
\big[(W^{\hat{A}}_{\mathrm{B},h_1})_m,(Q_{h_2+\frac{1}{2}})_r\big] 
\!&= &\!
\frac{1}{4}
\, (Q^{\hat{A}}_{h_1+h_2-\frac{1}{2}})_{m+r}\ ,
\nonu \\
\big[(W^{\hat{A}}_{\mathrm{B},h_1})_m,(Q^{\hat{B}}_{h_2+\frac{1}{2}})_r\big] 
\!&=& \!
 -\frac{i}{8}\, f^{\hat{A} \hat{B}  \hat{C}} \, (
Q^{\hat{C}}_{h_1+h_2-\frac{1}{2}} )_{m+r}
+ \frac{1}{4}
\, \Bigg[ \frac{1}{2}\, d^{\hat{A} \hat{B} \hat{C}} \,
(   Q^{\hat{C}}_{h_1+h_2-\frac{1}{2}} )_{m+r} \nonu \\
\!& + \!& \frac{1}{K}\, \delta^{\hat{A} \hat{B}}\,
(   Q_{h_1+h_2-\frac{1}{2}} )_{m+r} \Bigg] \ .
\label{n2reduced}
\eea
There are no mode dependent terms in the right hand side.
In the OPE language,
the $\frac{1}{\la}$ term in the first order pole
in the original OPEs survives.
Other reduced commutator relations similar to
(\ref{n2reduced}) appear in Appendix $A$.
Once we keep  the commutators in the bosonic singlet currents, then
we have some $\frac{1}{\la}$ dependence in other commutators.


\section{  The
${\cal N}=1$
supersymmetric
$W_{\infty}^{K}$
algebra with $U(K)$
symmetry }

\subsection{ The
${\cal N}=1$
supersymmetric
$W_{\infty}^{K}$
algebra with $U(K)$
symmetry}

\subsubsection{ The commutators between the bosonic currents}

In \cite{PRSS}, the bosonic current of weight $h$
is given by the linear
combination of $W_{B,h}$, $W_{F,h}$, $\bar{\pa}\, W_{B,h-1}$ and
$\bar{\pa}\, W_{F,h-1}$.
In terms of their modes with correct deformation parameter $\la$
(the power of $\la$ should be equal to $(h-2)$),
we obtain the following $U(K)$-singlet and $U(K)$-adjoint currents
together with (\ref{singletadjoint})
\bea
(W_{h})_m & \equiv & (W_{\mathrm{B},h})_m +  (W_{\mathrm{F},h})_m
+ \la \, \frac{2(h-2)(m+(h-2)+1)}{2(h-2)+1}\, (W_{\mathrm{B},h-1})_m
\nonu \\
\!& - \!& \la \,
\frac{(2(h-2)+2)(m+(h-2)+1)}{2(h-2)+1}\, (W_{\mathrm{F},h-1})_m \ ,
\nonu \\
(W^{\hat{A}}_{h})_m & \equiv & (W^{\hat{A}}_{\mathrm{B},h})_m +
(W^{\hat{A}}_{\mathrm{F},h})_m
+ \la \, \frac{2(h-2)(m+(h-2)+1)}{2(h-2)+1}\, (W^{\hat{A}}_{\mathrm{B},h-1})_m
\nonu \\
\!& - \!& \la \, \frac{(2(h-2)+2)(m+(h-2)+1)}{2(h-2)+1}\,
(W^{\hat{A}}_{\mathrm{F},h-1})_m \ .
\label{WWhat}
\eea
According to the construction of (\ref{WWhat}), the lowest value for
the weight $h$ of the bosonic current $W_{B,h}$ is given by $h=2$.
When $h=2$, the third term of each
expression vanishes due to the coefficient.
The $U(K)$-adjoint current is new and is the matrix generalization of
\cite{PRSS}. The mode $m$ dependence in the coefficients
appears in the third and
fourth terms when we write down  the mode of
derivative of current
in terms of mode of the current itself \footnote{In principle
  \cite{PRSplb}, we can add
  the weight-$1$ currents
  by reversing
  the procedure in the footnote \ref{decoupling} and obtain
the
$W_{1+\infty}^{K}$
algebra and by contractions the corresponding
$w_{1+\infty}^{K}$
algebra can be obtained).}.

Then we can determine their commutator relations explicitly
by using both (\ref{OSalgebra}) and (\ref{BKalgebra})
as follows:
\bea
\big[(W_{h_1})_m,(W_{h_2})_n\big] 
\!&=& \!
\sum^{h_1+h_2-4}_{h= 0} \, \la^h\, q^{h_1, h_2, h}(m,n)\, (W_{h_1+h_2-2-h})_{m+n}
+c_{W,h_1}(m)
\ ,
\nonu \\
\big[(W_{h_1})_m,(W^{\hat{A}}_{h_2})_n\big] 
\!&=& \!
\sum^{h_1+h_2-4}_{h= 0} \, \la^h\, q^{h_1, h_2, h}(m,n)\,
(W^{\hat{A}}_{h_1+h_2-2-h})_{m+n} \ ,
\nonu \\
\big[(W^{\hat{A}}_{h_1})_m,(W^{\hat{B}}_{h_2})_n\big] 
\!&=& \!
-\sum^{h_1+h_2-4}_{h=-1} \, \la^h\, \tilde{q}^{h_1,h_2, h}(m,n)
\, \frac{i}{2}\, f^{\hat{A} \hat{B}  \hat{C}} \,
(   W^{\hat{C}}_{h_1+h_2-2-h} )_{m+n}
\nonu \\
\!& + \!&
\frac{1}{K}\, \de^{\hat{A}\hat{B}}\, c_{W,h_1}(m)
\nonu \\
\!&+\!& \sum^{h_1+h_2-4}_{h= 0} \, \la^h\, q^{h_1,h_2, h}(m,n)
\, \Bigg[ \frac{1}{2}\, d^{\hat{A} \hat{B} \hat{C}} \,
(   W^{\hat{C}}_{h_1+h_2-2-h} )_{m+n} \nonu \\
\!& + \!& \frac{1}{K}\, \delta^{\hat{A} \hat{B}}\,
(   W_{h_1+h_2-2-h} )_{m+n} \Bigg] \ .
\label{WWn=1}
\eea
Again,
the last two relations in (\ref{WWn=1}) are new.
Compared to (\ref{BKalgebra}), the algebraic structure looks similar,
but the structure constants are different from each other
and the range for
the dummy variable $h$ in the right hand sides is different.
It is claimed in \cite{PRSS}
that the algebra in the first relation of
(\ref{WWn=1}) is isomorphic to the $W_{\infty}$ algebra
\cite{PRS1990,PRS1990-1} \footnote{
  It is known in \cite{PRSS} that
  the diagonal $W_{\infty}$ algebra from the current
  $W_h$ is generated
  in $W_{\infty}$ algebra (generated by
  $W_{B,h}$) and $W_{1+\infty}$ algebra (generated by $W_{F,h}$). }.
Here the central terms appearing in the
first and the last equations of (\ref{WWn=1}) are given by the following
expression
\bea
c_{W,h_1}(m) \!& = \!& K \Bigg[ c_{W_{B,h_1}}(m) \, \de^{h_1 h_2}\,
  \la^{2(h_1-2)}
\nonu \\
\!&+ \!& \la\, \frac{2(h_2-2)(n+(h_2-2)+1)}{2(h_2-2)+1}\,
 c_{W_{B,h_1}}(m) \, \de^{h_1, h_2-1}\, \la^{2(h_1-2)}\, \nonu \\
\!& +\!& \la\,
\frac{2(h_1-2)(m+(h_1-2)+1)}{2(h_1-2)+1}\,
 c_{W_{B,h_1-1}}(m) \, \de^{h_1-1, h_2}\, \la^{2(h_1-1-2)}\,
\nonu \\
\!& + \!& \la^2\,
\frac{2(h_1-2)(m+(h_1-2)+1)}{2(h_1-2)+1}\,
\frac{2(h_2-2)(n+(h_2-2)+1)}{2(h_2-2)+1}\,
\nonu \\
\!& \times \!&  c_{W_{B,h_1-1}}(m) \, \de^{h_1-1, h_2-1}\, \la^{2(h_1-1-2)}\,
\nonu \\
\!& + \!&  c_{W_{F,h_1}}(m) \, \de^{h_1 h_2}\, \la^{2(h_1-2)}\,
\nonu \\
\!&- \!& \la \, \frac{(2(h_2-2)+2)(n+(h_2-2)+1)}{2(h_2-2)+1}\,
 c_{W_{F,h_1}}(m) \, \de^{h_1, h_2-1}\, \la^{2(h_1-2)}\,
 \nonu \\
\!& -\!& \la\,
\frac{(2(h_1-2)+2)(m+(h_1-2)+1)}{2(h_1-2)+1}\,
 c_{W_{F,h_1-1}}(m) \, \de^{h_1-1, h_2}\, \la^{2(h_1-1-2)}\,
\nonu \\
\!& + \!& \la^2\,
\frac{(2(h_1-2)+2)(m+(h_1-2)+1)}{2(h_1-2)+1}\,
\frac{(2(h_2-2)+2)(n+(h_2-2)+1)}{2(h_2-2)+1}\,
\nonu \\
\!& \times \!&  c_{W_{F,h_1-1}}(m) \, \de^{h_1-1, h_2-1}\, \la^{2(h_1-1-2)}\,
\Bigg] \, \de_{m+n} \ .
\label{cen}
\eea
We can check that the above expression (\ref{cen})
vanishes (topological property)
by using the Kronecker delta conditions properly.
The second and the sixth, the third and the seventh, and the
remaining ones can be combined as the independent terms.
We introduce the following structure constants
\bea
q^{h_1, h_2, h}(m,n) \!& \equiv \!&
 q_{\mathrm{B}}^{h_1,h_2-\frac{1}{2}, h}(m,n+\frac{1}{2})
\nonu \\
\!& + \!&  \frac{2(h_1-2)(m+(h_1-2)+1)}{2(h_1-2)+1}\,
q_{\mathrm{B}}^{h_1-1,h_2-\frac{1}{2}, h-1}(m,n+\frac{1}{2})
\nonu \\
\!&+\!& q_{\mathrm{F}}^{h_1,h_2-\frac{1}{2}, h}(m,n+\frac{1}{2})
\nonu \\
\! & -\! & 
\frac{(2(h_1-2)+2)(m+(h_1-2)+1)}{2(h_1-2)+1}\,
q_{\mathrm{F}}^{h_1-1,h_2-\frac{1}{2}, h-1}(m,n+\frac{1}{2}) \ ,
\nonu \\
\tilde{q}^{h_1, h_2, h}(m,n) & \equiv &
 q_{\mathrm{B}}^{h_1,h_2-\frac{1}{2}, h}(m,n+\frac{1}{2})
\nonu \\
\!& + \!&  \frac{2(h_1-2)(m+(h_1-2)+1)}{2(h_1-2)+1}\,
q_{\mathrm{B}}^{h_1-1,h_2-\frac{1}{2}, h-1}(m,n+\frac{1} {2})
\nonu \\
\!&-\!& q_{\mathrm{F}}^{h_1,h_2-\frac{1}{2}, h}(m,n+\frac{1}{2})
\nonu \\
\!& + \!& 
\frac{(2(h_1-2)+2)(m+(h_1-2)+1)}{2(h_1-2)+1}\,
q_{\mathrm{F}}^{h_1-1,h_2-\frac{1}{2}, h-1}(m,n+\frac{1}{2}) \ .
\label{qqtilde}
\eea
The first relation of (\ref{qqtilde}) was found in \cite{PRSS}
with our convention and is more natural in the commutator
relation between the bosonic current and the fermionic
current in next subsection.
That is the reason why there are
shifts in the weight $h_2$ and the mode $n$.
The second and the fourth terms  have the explicit
mode $m$ dependence due to the derivative terms as we explained
before.
Note that there are precise relations between
the structure constants $(p_B,p_F)$
which appear in the equations (\ref{OSalgebra}) and
(\ref{BKalgebra})
and the structure constants $(q_F, q_B)$ which appear
in the equations (\ref{WFQ}) and (\ref{WBQ})
at each term in the commutator relations.
Their relations will appear later.

\subsubsection{ The commutators between the bosonic currents
and the fermionic currents}

Now we obtain the commutator relations including the
fermionic currents.
By using (\ref{WFQ}), (\ref{WBQ}) and (\ref{singletadjoint}),
the following relations satisfy
\bea
\big[(W_{h_1})_m,(Q_{h_2+\frac{1}{2}})_r\big] 
\!&=& \!
\sum^{h_1+h_2-3}_{h=-1} \, \la^h\, q^{h_1, h_2+1, h}(m,r-\frac{1}{2})
\, (Q_{h_1+h_2-\frac{3}{2}-h})_{m+r} \ ,
\nonu \\
\big[(W_{h_1})_m,(Q^{\hat{A}}_{h_2+\frac{1}{2}})_r\big] 
\!&=& \!
\sum^{h_1+h_2-3}_{h=-1} \, \la^h\, q^{h_1, h_2+1, h}(m,r-\frac{1}{2})\,
(Q^{\hat{A}}_{h_1+h_2-\frac{3}{2}-h})_{m+r} \ ,
\nonu \\
\big[(W^{\hat{A}}_{h_1})_m,(Q_{h_2+\frac{1}{2}})_r\big] 
\!&=& \!
\sum^{h_1+h_2-3}_{h=-1} \, \la^h\, q^{h_1, h_2+1, h}(m,r-\frac{1}{2})\,
(Q^{\hat{A}}_{h_1+h_2-\frac{3}{2}-h})_{m+r} \ ,
\nonu \\
\big[(W^{\hat{A}}_{h_1})_m,(Q^{\hat{B}}_{h_2+\frac{1}{2}})_r\big] 
\!&=& \!
\sum^{h_1+h_2-3}_{h= -1} \, \la^h\, \hat{q}^{h_1,h_2+\frac{1}{2}, h}(m,r)
\, \frac{i}{2}\, f^{\hat{A} \hat{B}  \hat{C}} \,
(   Q^{\hat{C}}_{h_1+h_2-\frac{3}{2}-h} )_{m+r}
\nonu \\
\!&+\!& \sum^{h_1+h_2-3}_{h= -1} \, \la^h\,
\check{q}^{h_1,h_2+\frac{1}{2}, h}(m,r)
\, \Bigg[ \frac{1}{2}\, d^{\hat{A} \hat{B} \hat{C}} \,
(   Q^{\hat{C}}_{h_1+h_2-\frac{3}{2}-h} )_{m+r} \nonu \\
\!& + \!& \frac{1}{K}\, \delta^{\hat{A} \hat{B}}\,
(   Q_{h_1+h_2-\frac{3}{2}-h} )_{m+r} \Bigg]
\nonu \\
\!& + \!& 
\sum^{h_1+h_2-3}_{h= -1} \, \la^h\, \dot{q}^{h_1, h_2+\frac{1}{2}, h}(m,r)
\, \frac{i}{2}\, f^{\hat{A} \hat{B}  \hat{C}} \,
(   Q^{\hat{C}}_{h_1-1+h_2-\frac{3}{2}-h} )_{m+r}
\nonu \\
\!&+\!& \sum^{h_1+h_2-3}_{h= -1} \, \la^h\, \bar{q}^{h_1, h_2+\frac{1}{2}, h}(m,r)
\, \Bigg[ \frac{1}{2}\, d^{\hat{A} \hat{B} \hat{C}} \,
(   Q^{\hat{C}}_{h_1-1+h_2-\frac{3}{2}-h} )_{m+r} \nonu \\
\!& + \!& \frac{1}{K}\, \delta^{\hat{A} \hat{B}}\,
(   Q_{h_1-1+h_2-\frac{3}{2}-h} )_{m+r} \Bigg]\ .
\label{WQn=1}
\eea
Again, the first relation of (\ref{WQn=1}) was found in \cite{PRSS}.
The remaining ones are the matrix generalization.
In the last equation of (\ref{WQn=1}), the following
structure constants are introduced by collecting
each contribution 
\bea
\hat{q}^{h_1, h_2, h}(m,r) \! & \equiv \! & -q_{\mathrm{B}}^{h_1,h_2, h}(m,r)
+ q_{\mathrm{F}}^{h_1,h_2, h}(m,r) \ ,
\nonu  \\
\check{q}^{h_1, h_2, h}(m,r) \! & \equiv \! & q_{\mathrm{B}}^{h_1,h_2, h}(m,r)
+ q_{\mathrm{F}}^{h_1,h_2, h}(m,r) \ ,
\nonu  \\
\dot{q}^{h_1, h_2, h}(m,r) \! & \equiv \! &
-\frac{2(h_1-2)(m+(h_1-2)+1)}{2(h_1-2)+1}\,
q_{\mathrm{B}}^{h_1-1,h_2, h}(m,r)
\nonu \\
\! & - \! &
\frac{(2(h_1-2)+2)(m+(h_1-2)+1)}{2(h_1-2)+1}\,
q_{\mathrm{F}}^{h_1-1,h_2, h}(m,r) \ ,
\nonu  \\
\bar{q}^{h_1, h_2, h}(m,r) \! & \equiv \! &
\frac{2(h_1-2)(m+(h_1-2)+1)}{2(h_1-2)+1}\,
q_{\mathrm{B}}^{h_1-1,h_2, h}(m,r)
\nonu \\
\! & -\! & \frac{(2(h_1-2)+2)(m+(h_1-2)+1)}{2(h_1-2)+1}\,
q_{\mathrm{F}}^{h_1-1,h_2, h}(m,r) \ .
\label{otherstructure}
\eea
Although we introduce the weight $h_2$ in the structure
constant in (\ref{otherstructure}), their appearance in
(\ref{WQn=1}) takes the form of $(h_2+\frac{1}{2})$
which is the weight of the fermionic current in the left
hand side.
The shift in the weight $h_1$
of the structure constants $(q_B,q_F)$
in the last two of (\ref{otherstructure})
can be understood from the derivative of the bosonic current
in the left hand side.
Moreover, the last three terms in the last equation of
(\ref{WQn=1}) have the weight of the first three terms minus one.
We can combine the last three terms by introducing a new dummy variable
$(h+1)$ with the first three terms in addition to other term.
In this way, we can simplify the last equation of (\ref{WQn=1})
further. 

Then one of our main results with a deformation parameter
$\la$ is summarized by
(\ref{WWn=1}) and (\ref{WQn=1}) together with (\ref{qqtilde}) and
(\ref{otherstructure}).
We will present the realization of this algebra in the supersymmetric
Einstein-Yang-Mills theory.

\subsection{Free field realization }

By combining (\ref{singletadjoint}) and (\ref{WWQQ1}),
we can write down the free field realization
for the singlet current and the adjoint current  of $U(K)$
as follows:
\bea
W_h  \!& = \!&  W_{B,h} +W_{F,h} - \la\, \frac{2(h-2)}{2(h-2)+1}\,
\bar{\pa} \, W_{B,h-1} + \la \, \frac{2(h-2)+2}{2(h-2)+1}\,
\bar{\pa}\, W_{F,h-1} \, ,
\nonu \\
\!& = \!&
 \frac{2^{h-2} (h-1)!}{(2(h-2)+1)!!}\, \la^{h-2}\, 
\sum_{l=0}^{h-2}\sum_{i,\, \bar{\imath}=1}^N 
\delta_{i,\bar{\imath}}\, (-1)^l\,
\left(\begin{array}{c}
h-2 \\  l \\
\end{array}\right)
\left(\begin{array}{c}
h-1 \\  l \\
\end{array}\right)
\,\nonu \\
\!& \times \!& \Bigg(\,\bar{\partial}^{h-l-1}\bar{\phi}^{i,\bar{a}}  \,
\de_{b \bar{a}}\,
\bar{\partial}^{l+1}\phi^{\bar{\imath},b} +
\bar{\partial}^{h-l-1}\bar{\psi}^{i,\bar{\alpha}} \,
\de_{\beta \bar{\al}}\,
\bar{\partial}^l\psi^{\bar{\imath},\beta}
\,\Bigg)\, ,
\nonu \\
W^{\hat{A}}_h  \!& = \!&  W^{\hat{A}}_{B,h} +W^{\hat{A}}_{F,h} - \la\, \frac{2(h-2)}{2(h-2)+1}\,
\bar{\pa} \, W^{\hat{A}}_{B,h-1} + \la \, \frac{2(h-2)+2}{2(h-2)+1}\,
\bar{\pa}\, W^{\hat{A}}_{F,h-1} \,\nonu \\
\!& = \!&
\frac{2^{h-2} (h-1)!}{(2(h-2)+1)!!}\, \la^{h-2}\, 
\sum_{l=0}^{h-2}\sum_{i,\, \bar{\imath}=1}^N 
\delta_{i,\bar{\imath}}\, (-1)^l\,
\left(\begin{array}{c}
h-2 \\  l \\
\end{array}\right)
\left(\begin{array}{c}
h-1 \\  l \\
\end{array}\right)
\,\nonu \\
\!& \times \!& \Bigg(\,\bar{\partial}^{h-l-1}\bar{\phi}^{i,\bar{a}}  \,
t^{\hat{A}}_{b \bar{a}}\,
\bar{\partial}^{l+1}\phi^{\bar{\imath},b} +
\bar{\partial}^{h-l-1}\bar{\psi}^{i,\bar{\alpha}} \,
t^{\hat{A}}_{\beta \bar{\al}}\,
\bar{\partial}^l\psi^{\bar{\imath},\beta}
\,\Bigg)\, .
\label{FREE1}
\eea
There is some difference in the sign when we compare with
the result of \cite{PRSS} because this comes from the fact that
we are using different normalization in the footnote \ref{typos}.

The remaining superpartner currents come from (\ref{WWQQ1})
as follows:
\bea
Q_{h+\frac{1}{2}} 
\!&
=\!&
\frac{2^{h-\frac{1}{2}}h!}{(2h-1)!!}\,\la^{h-1}\,
\sum_{l=0}^{h-1}\sum_{i,\, \bar{\imath}=1}^N 
\delta_{i,\bar{\imath}}(-1)^l
\left(\begin{array}{c}
h-1 \\  l \\
\end{array}\right)
\left(\begin{array}{c}
h \\  l \\
\end{array}\right)
\,(\,\bar{\partial}^{h-l}\bar{\phi}^{i,\bar{a}}\,
\de_{\al \bar{a}}\,
\bar{\partial}^l
\psi^{\bar{\imath},\alpha}\,)\, , \nonu \\
Q^{\hat{A}}_{h+\frac{1}{2}} 
\!&
=\!&
\frac{2^{h-\frac{1}{2}}h!}{(2h-1)!!}\,\la^{h-1}\,
\sum_{l=0}^{h-1}\sum_{i,\, \bar{\imath}=1}^N 
\delta_{i,\bar{\imath}}(-1)^l
\left(\begin{array}{c}
h-1 \\  l \\
\end{array}\right)
\left(\begin{array}{c}
h \\  l \\
\end{array}\right)
\,(\,\bar{\partial}^{h-l}\bar{\phi}^{i,\bar{a}}
\,
t^{\hat{A}}_{\al \bar{a}} \,
\bar{\partial}^l
\psi^{\bar{\imath},\alpha}\,)\, .
\label{FREE2}
\eea
In principle, we can check the previous commutator relations
by using these free field realizations, 
calculating the various OPEs 
and rewriting down these in terms of commutator relations.
The power of deformation parameter is given
by the weight minus $2$ or $\frac{3}{2}$.
As in the abstract, the results of
(\ref{FREE1}) and (\ref{FREE2}) are the matrix
generalization of previous work of \cite{PRSS}.

\subsection{ The
${\cal N}=1$
supersymmetric
$w_{\infty}^{K}$
algebra with $U(K)$
symmetry}

In order to keep the nonzero lowest power of the deformation parameter
in each term of (\ref{WWn=1}) and (\ref{WQn=1}),
we consider the scales for the currents with the deformation parameter
whose power depends on the weights.
According to the following
transformations,
\bea
W_{h} \!& \rightarrow \!& \la^{h-2}\, W_h \ , \qquad W_{h}^{\hat{A}} \rightarrow
\la^{h} \, W_{h}^{\hat{A}}\ , \nonu \\
Q_{h+\frac{1}{2}} \!& \rightarrow \!& \la^{h+\frac{1}{2}-2}\,
Q_{h+\frac{1}{2}} \ , 
\qquad
Q^{\hat{A}}_{h+\frac{1}{2}} \rightarrow \la^{h+\frac{1}{2}}\,
Q^{\hat{A}}_{h+\frac{1}{2}} \ ,
\label{rescale}
\eea
we obtain the following 
the ${\cal N}=1$
supersymmetric
$w_{1+\infty}^{K}$
algebra with $U(K)$
symmetry after redefining as (\ref{rescale})
and taking $\la \rightarrow 0$ limit
with the help of (\ref{qqtilde}) and (\ref{otherstructure})
\bea
\big[(W_{h_1})_m,(W_{h_2})_n\big] 
\!& \rightarrow & \!
q^{h_1, h_2, 0}(m,n)\, (W_{h_1+h_2-2})_{m+n} \nonu \\
\! & = \! &
\Big[m(h_2-1)-n(h_1-1)\Big] \, (W_{h_1+h_2-2})_{m+n} \ ,
\nonu \\
\big[(W_{h_1})_m,(W^{\hat{A}}_{h_2})_n\big] 
\!& \rightarrow & \!
q^{h_1, h_2, 0}(m,n)\, (W^{\hat{A}}_{h_1+h_2-2})_{m+n}
\nonu \\
\! & = \! &
\Big[m(h_2-1)-n(h_1-1)\Big] \, (W^{\hat{A}}_{h_1+h_2-2})_{m+n} \ ,
\nonu \\
\big[(W^{\hat{A}}_{h_1})_m,(W^{\hat{B}}_{h_2})_n\big] 
\!& \rightarrow & \!
-  \tilde{q}^{h_1,h_2, -1}(m,n)
\, \frac{i}{2}\, f^{\hat{A} \hat{B}  \hat{C}} \,
(   W^{\hat{C}}_{h_1+h_2-1} )_{m+n} \nonu \\
\!& = \!&
-  
\frac{i}{4}\, f^{\hat{A} \hat{B}  \hat{C}} \,
(   W^{\hat{C}}_{h_1+h_2-1} )_{m+n}
\ ,
\nonu \\
\big[(W_{h_1})_m,(Q_{h_2+\frac{1}{2}})_r\big] 
\!& \rightarrow & \!
q^{h_1, h_2+1, 0}(m,r-\frac{1}{2})\, (Q_{h_1+h_2-\frac{3}{2}})_{m+r} \nonu \\
\! & = \! &
\Big[m(h_2+1-1)-(r-\frac{1}{2})(h_1-1)\Big] \,
(Q_{h_1+h_2-\frac{3}{2}})_{m+r}\ ,
\nonu \\
\big[(W_{h_1})_m,(Q^{\hat{A}}_{h_2+\frac{1}{2}})_r\big] 
\!& \rightarrow & \!
q^{h_1, h_2+1, 0}(m,r-\frac{1}{2})\, (Q^{\hat{A}}_{h_1+h_2-\frac{3}{2}})_{m+r} \nonu \\
\! & = \! &
\Big[m(h_2+1-1)-(r-\frac{1}{2})(h_1-1)\Big] \,
(Q^{\hat{A}}_{h_1+h_2-\frac{3}{2}})_{m+r}\ ,
\nonu \\
\big[(W^{\hat{A}}_{h_1})_m,(Q_{h_2+\frac{1}{2}})_r\big] 
\!& \rightarrow & \!
q^{h_1, h_2+1, 0}(m,r-\frac{1}{2})\, (Q^{\hat{A}}_{h_1+h_2-\frac{3}{2}})_{m+r} \nonu \\
\! & = \! &
\Big[m(h_2+1-1)-(r-\frac{1}{2})(h_1-1)\Big] \,
(Q^{\hat{A}}_{h_1+h_2-\frac{3}{2}})_{m+r}\ ,
\nonu \\
\big[(W^{\hat{A}}_{h_1})_m,(Q^{\hat{B}}_{h_2+\frac{1}{2}})_r\big] 
\!& \rightarrow & \!
\hat{q}^{h_1, h_2+\frac{1}{2}, -1}(m,r)
\,
\frac{i}{2}\, f^{\hat{A} \hat{B}  \hat{C}} \,
(Q^{\hat{C}}_{h_1+h_2-\frac{1}{2}})_{m+r}
\nonu \\
\!& = \!&  
-\frac{i}{4}\, f^{\hat{A} \hat{B}  \hat{C}} \,
(Q^{\hat{C}}_{h_1+h_2-\frac{1}{2}})_{m+r} \ .
\label{firstalgebra}
\eea
Due to the weight $(h_2+\frac{1}{2})$ for the
fermionic current rather than
$(h_2+ \frac{3}{2})$,
the shift in $h_2$ in the right hand side
appears.
Note that the structure constant $q^{h_1, h_2+1, -1}(m,r-\frac{1}{2})$
vanishes. In other words,
the lowest dummy variable $h$ in the first three equations
of (\ref{WQn=1}) starts with $h=0$.
Moreover, the structure constant
$\check{q}^{h_1,h_2+\frac{1}{2},-1}(m,r)$ in the last equation of
(\ref{WQn=1}) vanishes.
This algebra (\ref{firstalgebra}) with a rescaling of the
structure constant $f^{\hat{A}\hat{B}\hat{C}}$ was
found in \cite{Ahn2111}
previously.
The point here is that the present description is more
transparent because the last three commutator relations of
(\ref{firstalgebra}) in \cite{Ahn2111}
are introduced abstractly but in this paper
we prove that they can be obtained from the
above ${\cal N}=1$ supersymmetric $W_{\infty}^K$ algebra
by taking the vanishing $\la$ limit.
Therefore, at least, the celestial holography
between the above two-dimensional symmetry algebra
and  the OPEs \cite{FSTZ}
from the supersymmetric Einstein-Yang-Mills
theory holds at vanishing $\la$ limit.

\subsection{The seven OPEs }

In order to present the above commutator relations in terms of
OPEs, we need to introduce the following quantity \cite{PRS1990-1}
\bea
M_h^{h_1,h_2}(m,n) \equiv \sum_{k=0}^{h+1}\, (-1)^k \,
\left(\begin{array}{c}
h+1 \\  k \\
\end{array}\right)
(2h_1-h-2)_{k}[2h_2-2-k]_{h+1-k}\, m^{h+1-k}\, n^k \ .
\label{Mdef}
\eea
The degree of this polynomial is given by
$(h+1)$.
Then the above seven commutator relations
can be written in terms of the OPEs (See also \cite{AKK})
\bea
W_{h_1}(\bar{z}) \, W_{h_2}(\bar{w}) \!& = \!&
\sum_{h=0}^{h_1+h_2-4}\, \la^h \,
(-1)^{h-1} \, f^{h_1,h_2,h}(\bar{\pa}_{\bar{z}}, \bar{\pa}_{\bar{w}})\,
\Bigg[ \frac{W_{h_1+h_2-2-h}(\bar{w})}{(\bar{z}-\bar{w})} \Bigg]
+ \cdots \ ,
\nonu \\
W_{h_1}(\bar{z}) \, W^{\hat{A}}_{h_2}(\bar{w}) \!& = \!&
\sum_{h=0}^{h_1+h_2-4}\, \la^h \,
(-1)^{h-1} \, f^{h_1,h_2,h}(\bar{\pa}_{\bar{z}}, \bar{\pa}_{\bar{w}})\,
\Bigg[\frac{W^{\hat{A}}_{h_1+h_2-2-h}(\bar{w})}{(\bar{z}-\bar{w})} \Bigg]
+ \cdots \ ,
\nonu \\
W^{\hat{A}}_{h_1}(\bar{z}) \, W^{\hat{B}}_{h_2}(\bar{w}) \!& = \!&
 \sum_{h=-1}^{h_1+h_2-4}\, \la^h \,
(-1)^{h-1} \, \tilde{f}^{h_1,h_2,h}(\bar{\pa}_{\bar{z}}, \bar{\pa}_{\bar{w}})\,
\Bigg[ \frac{-\frac{i}{2}\,f^{\hat{A} \hat{B} \hat{C}} \,
 W^{\hat{C}}_{h_1+h_2-2-h}(\bar{w})}{(\bar{z}-\bar{w})}
\Bigg]+ \cdots \ ,
\nonu \\
\!& + \!&
\sum_{h=0}^{h_1+h_2-4}\, \la^h \,
(-1)^{h-1} \, f^{h_1,h_2,h}(\bar{\pa}_{\bar{z}}, \bar{\pa}_{\bar{w}})\,
\nonu \\
\!& \times \!& \Bigg[ \frac{\frac{1}{2} \,
d^{\hat{A} \hat{B} \hat{C}} \, W^{\hat{C}}_{h_1+h_2-2-h}(\bar{w})
+ \frac{1}{K} \,
\de^{\hat{A} \hat{B}} \,  W_{h_1+h_2-2-h}(\bar{w})}{(\bar{z}-\bar{w})}\Bigg] \
+ \cdots \ ,
\nonu \\ 
W_{h_1}(\bar{z}) \, Q_{h_2+\frac{1}{2}}(\bar{w}) \!& = \!&
\sum_{h=-1}^{h_1+h_2-3}\, \la^h \,
(-1)^{h-1} \, f^{h_1,h_2+1,h}(\bar{\pa}_{\bar{z}}, \bar{\pa}_{\bar{w}})\,
\Bigg[ \frac{Q_{h_1+h_2-\frac{3}{2}-h}(\bar{w})}{(\bar{z}-\bar{w})}
\Bigg] + \cdots \ ,
\nonu \\
W_{h_1}(\bar{z}) \, Q^{\hat{A}}_{h_2+\frac{1}{2}}(\bar{w}) \!& = \!&
\sum_{h=-1}^{h_1+h_2-3}\, \la^h \,
(-1)^{h-1} \, f^{h_1,h_2+1,h}(\bar{\pa}_{\bar{z}}, \bar{\pa}_{\bar{w}})\,
\Bigg[\frac{Q^{\hat{A}}_{h_1+h_2-\frac{3}{2}-h}(\bar{w})}{(\bar{z}-\bar{w})}
\Bigg]+ \cdots \ ,
\nonu \\
W^{\hat{A}}_{h_1}(\bar{z}) \, Q_{h_2+\frac{1}{2}}(\bar{w}) \!& = \!&
\sum_{h=-1}^{h_1+h_2-3}\, \la^h \,
(-1)^{h-1} \, \Bigg[ f^{h_1,h_2+1,h}(\bar{\pa}_{\bar{z}}, \bar{\pa}_{\bar{w}})\,
\frac{Q^{\hat{A}}_{h_1+h_2-\frac{3}{2}-h}(\bar{w})}{(\bar{z}-\bar{w})}
\Bigg]+ \cdots \ ,
\nonu \\
W^{\hat{A}}_{h_1}(\bar{z}) \, Q^{\hat{B}}_{h_2+\frac{1}{2}}(\bar{w}) \!& = \!&
 \sum_{h=-1}^{h_1+h_2-4}\, \la^h \,
(-1)^{h-1} \, \hat{f}^{h_1,h_2+\frac{1}{2},h}(\bar{\pa}_{\bar{z}}, \bar{\pa}_{\bar{w}})\,
\Bigg[ \frac{\frac{i}{2}\,f^{\hat{A} \hat{B} \hat{C}} \,
Q^{\hat{C}}_{h_1+h_2-\frac{3}{2}-h}(\bar{w})}{(\bar{z}-\bar{w})}
  \Bigg]
\nonu \\
\!& + \!&
\sum_{h=0}^{h_1+h_2-4}\, \la^h \,
(-1)^{h-1} \,
\check{f}^{h_1,h_2+\frac{1}{2},h}(\bar{\pa}_{\bar{z}}, \bar{\pa}_{\bar{w}})\,
\nonu \\
\!& \times \!& \Bigg[ \frac{\frac{1}{2} \,
d^{\hat{A} \hat{B} \hat{C}} \, Q^{\hat{C}}_{h_1+h_2-\frac{3}{2}-h}(\bar{w})
+ \frac{1}{K} \,
\de^{\hat{A} \hat{B}} \,  Q_{h_1+h_2-\frac{3}{2}-h}(\bar{w})}{(\bar{z}-\bar{w})} \
\Bigg] \nonu \\
\!&+ \!&
 \sum_{h=-1}^{h_1+h_2-4}\, \la^h \,
(-1)^{h} \, \dot{f}^{h_1,h_2+\frac{1}{2},h}(\bar{\pa}_{\bar{z}}, \bar{\pa}_{\bar{w}})\,
\Bigg[ \frac{\frac{i}{2}\,f^{\hat{A} \hat{B} \hat{C}} \,
Q^{\hat{C}}_{h_1-1+h_2-\frac{3}{2}-h}(\bar{w})}{(\bar{z}-\bar{w})}
\Bigg] \nonu \\
\!& + \!&
\sum_{h=0}^{h_1+h_2-4}\, \la^h \,
(-1)^h \, \bar{f}^{h_1,h_2+\frac{1}{2},h}(\bar{\pa}_{\bar{z}}, \bar{\pa}_{\bar{w}})\,
\nonu \\
\!& \times \!& \Bigg[ \frac{\frac{1}{2} \,
d^{\hat{A} \hat{B} \hat{C}} \, Q^{\hat{C}}_{h_1-1+h_2-\frac{3}{2}-h}(\bar{w})
+ \frac{1}{K} \,
\de^{\hat{A} \hat{B}} \,  Q_{h_1-1+h_2-\frac{3}{2}-h}(\bar{w})}{(\bar{z}-\bar{w})}
\Bigg] + \cdots \ .
\label{OPES}
\eea
The various differential operators coming from the
structure constants act on the two complex coordinates
$(\bar{z},\bar{w})$.
The currents in the right hand sides do depend on the
coordinate $\bar{w}$.

From the structure constants where
the quantity $N_h^{h_1,h_2}(m,n)$ in (\ref{Nphi}) is replaced by
the quantity $M_h^{h_1,h_2}(m,n)$ in (\ref{Mdef})
\bea
f_{\mathrm{F}}^{h_1,h_2, h}(m,r) 
\! &
\equiv \! &\frac{(-1)^h}{4(h+2)!}\Bigg[
(h_1-1)\,\phi^{h_1,h_2+\frac{1}{2}}_{h+1}(0,0)
\nonu \\
\!& - \!& (h_1-h-3)\,\phi^{h_1,h_2+\frac{1}{2}}_{h+1}(0,\textstyle{-\frac{1}{2}})
\Bigg]
\,M^{h_1,h_2}_{h}(m,r) \ ,
\label{ff}
\\
f_{\mathrm{B}}^{h_1,h_2, h}(m,r) 
\! &
\equiv \! &\frac{-1}{4(h+2)!}\Bigg[
(h_1-h-2)\,\phi^{h_1,h_2+\frac{1}{2}}_{h+1}(0,0)-(h_1)\,\phi^{h_1,h_2+\frac{1}{2}}_{h+1}(0,\textstyle{-\frac{1}{2}})
\Bigg]
\,M^{h_1,h_2}_{h}(m,r) \ ,
\nonu
\eea
the previous structure constants together with (\ref{ff})
can be expressed as follows:
\bea
f^{h_1, h_2, h}(m,n) \!& \equiv \!&
 f_{\mathrm{B}}^{h_1,h_2-\frac{1}{2}, h}(m,n+\frac{1}{2})
\nonu \\
\!& + \!&  \frac{2(h_1-2)(m+(h_1-2)+1)}{2(h_1-2)+1}\,
f_{\mathrm{B}}^{h_1-1,h_2-\frac{1}{2}, h-1}(m,n+\frac{1}{2})
\nonu \\
\!&+\!& f_{\mathrm{F}}^{h_1,h_2-\frac{1}{2}, h}(m,n+\frac{1}{2})
\nonu \\
\! & -\! & 
\frac{(2(h_1-2)+2)(m+(h_1-2)+1)}{2(h_1-2)+1}\,
f_{\mathrm{F}}^{h_1-1,h_2-\frac{1}{2}, h-1}(m,n+\frac{1}{2}) \ ,
\nonu \\
\tilde{f}^{h_1, h_2, h}(m,n) & \equiv &
 f_{\mathrm{B}}^{h_1,h_2-\frac{1}{2}, h}(m,n+\frac{1}{2})
\nonu \\
\!& + \!&  \frac{2(h_1-2)(m+(h_1-2)+1)}{2(h_1-2)+1}\,
f_{\mathrm{B}}^{h_1-1,h_2-\frac{1}{2}, h-1}(m,n+\frac{1} {2})
\nonu \\
\!&-\!& f_{\mathrm{F}}^{h_1,h_2-\frac{1}{2}, h}(m,n+\frac{1}{2})
\nonu \\
\!& + \!& 
\frac{(2(h_1-2)+2)(m+(h_1-2)+1)}{2(h_1-2)+1}\,
f_{\mathrm{F}}^{h_1-1,h_2-\frac{1}{2}, h-1}(m,n+\frac{1}{2}) \ ,
\nonu \\
\hat{f}^{h_1, h_2, h}(m,r) \! & \equiv \! & -f_{\mathrm{B}}^{h_1,h_2, h}(m,r)
+ f_{\mathrm{F}}^{h_1,h_2, h}(m,r) \ ,
\nonu  \\
\check{f}^{h_1, h_2, h}(m,r) \! & \equiv \! & f_{\mathrm{B}}^{h_1,h_2, h}(m,r)
+ f_{\mathrm{F}}^{h_1,h_2, h}(m,r) \ ,
\nonu  \\
\dot{f}^{h_1, h_2, h}(m,r) \! & \equiv \! &
-\frac{2(h_1-2)(m+(h_1-2)+1)}{2(h_1-2)+1}\,
f_{\mathrm{B}}^{h_1-1,h_2, h}(m,r)
\nonu \\
\! & - \! &
\frac{(2(h_1-2)+2)(m+(h_1-2)+1)}{2(h_1-2)+1}\,
f_{\mathrm{F}}^{h_1-1,h_2, h}(m,r) \ ,
\nonu  \\
\bar{f}^{h_1, h_2, h}(m,r) \! & \equiv \! &
 \frac{2(h_1-2)(m+(h_1-2)+1)}{2(h_1-2)+1}\,
f_{\mathrm{B}}^{h_1-1,h_2, h}(m,r)
\nonu \\
\! & -\! & \frac{(2(h_1-2)+2)(m+(h_1-2)+1)}{2(h_1-2)+1}\,
f_{\mathrm{F}}^{h_1-1,h_2, h}(m,r) \ .
\label{struct}
\eea
In Appendix $B$, we present the explicit OPEs between the
currents
$W_4$, $W_4^{\hat{A}}$, $Q_{\frac{7}{2}}$ and
$Q_{\frac{7}{2}}^{\hat{A}}$.
On the one hand, from the construction of
free field realization, the OPEs can be obtained
from either by hands or by Thielemans package \cite{Thielemans}
for fixed weights $h_1$ and $h_2$. On the other hand,
we can write down (\ref{OPES}) explicitly by
substituting the structure constants which are differential
operators into (\ref{OPES}).
The crucial point here is that
from the transformation of (\ref{struct}) to
the ones in (\ref{OPES}) where
$m \rightarrow \bar{\pa}_{\bar{z}}$ and
$n, r \rightarrow \bar{\pa}_{\bar{w}}$, we should keep only
the terms having a degree $(h+1)$ of the polynomial.
In general, the structure constants in (\ref{struct})
do have the terms having a power sum of two variables less than
the above $(h+1)$ of the polynomial, contrary to the case of
(\ref{ff}).
We will provide some details in Appendix $B$.

\subsection{The possible realization in the
${\cal N}=1$ supersymmetric Einstein-Yang-Mills theory}

It is known, in \cite{GHPS},
that the leading OPE for two positive helicity
gluons is given by the simple pole in the holomorphic sector
with Euler beta function whose arguments depend on the
two conformal weights appearing on the left hand side. 
They consider the scattering states in MHV (Maximally
Helicity Violating) tree amplitudes.
The celestial amplitudes
can be written as the massless $n$ particle amplitudes,
which depend on the energies and the points on the
celestial sphere, in the Mellin space.  
These amplitudes can be interpreted as correlation functions of
$n$ primary operators on the celestial sphere.
The simplest nontrivial scattering process occurs
in the $n=4$ gluons.
Then the leading order behavior for the above
massless four particle amplitudes can be obtained
by taking the holomorphic collinear limit for the
positive helicity (outgoing) gluons. 
Then it is possible to relate the full $n=4$ amplitude
to $n=3$ amplitude and the corresponding celestial four point
correlation
function can be expressed as an integral over
some integral parameter where the integrand contains the
three point correlation function.
Moreover the OPE of two positive helicity (outgoing) gluons
can be described as a conformal block including all the
antiholomorphic descendants. Then 
the leading OPE for two positive helicity
gluons we mentioned before can be further simplified after
performing a Taylor expansion.
Finally, the OPE of the conformally soft gluon operators
takes the simple pole in the holomorphic sector
with finite sum of antiholomorphic derivatives acting on the
the second gluon appearing on the left hand side of the OPE.
In this soft limit, the poles appearing in the OPE coefficients
(Euler beta function) disappear and the rescaled
soft gluon operators with specific weights occur in the OPE.

For the supersymmetric Einstein-Yang-Mills theory
which is a generalization of previous paragraph,
from the collinear limit of the respective
Feynman matrix element,
the OPEs can be obtained by performing
the Mellin transforms \cite{FSTZ}. For example,
the OPE of the conformal primary graviton and the
conformal primary gravitino of
arbitrary weights takes the simple pole in the
holomorphic sector and the OPE coefficient
is given by Euler beta function which depends on
the two previous weights on the left hand side of the OPE.
By taking the soft limit, this OPE
between the soft positive helicity graviton and
the soft positive helicity gravitino can be written
in terms of binomial coefficient which depends on
the two weights of the soft operators on the left hand side
and dummy variable for the finite sum for the antiholomorphic
derivatives acting on the soft positive helicity gravitino \cite{Ahn2111}.
The structure of this OPE looks like the one in previous
case between two soft gluon operators in the sense that
the numerical values appearing in the binomial coefficient
and the power of the difference in the antiholomorphic
coordinates are little different.

The MHV gluon amplitudes in previous paragraph
are the simplest amplitudes in Yang-Mills
theory.
The next-to-simplest amplitudes, Next-to-MHV or
NMHV sector is studied in \cite{MRSV} based on the non minimal couplings
of gluons and gravitons by following the work of \cite{HPS}.
From the six-point NMHV analysis, the amplitude is no longer
a finite polynomial in the complex coordinates
(and its complex conjugated ones) of the soft particle.
This leads to the fact that
the lower limits
of the holomorphic and the
antiholomorphic mode expansions
of the soft graviton, gravitino, gluon and gluino
are given by $-\infty$ instead of finite values
in MHV sector. Furthermore, the upper limits
in the dummy variable appearing on the
right hand sides of the OPEs between
these soft operators are not finite values but
$\infty$.
This allows us to obtain the mode dependent function,
which also depends on the three weights, when we express
the OPEs in terms of
the commutators between the soft operators.

According to \cite{HPS,MRSV},
the OPE between the soft positive helicity
graviton, where the weights are
$h_1=\frac{k-2}{2}$
and $h_2=\frac{l-2}{2}$, is given by, after taking the soft limit,
\bea
H^k(z_1,\bar{z}_1)\, H^l(z_2,\bar{z}_2)
\!& = \!& -\frac{\kappa}{2} \,
\frac{1}{z_{12}} \, \sum_{n=0}^{\infty}\,
\left(
\begin{array}{c}
2-2 h-k-l-n \\
1-h-l
\end{array}
\right)\, \frac{\bar{z}_{12}^{n+h+1}}{n!}\, \bar{\pa}^n \,
H^{k+l+h}(z_2, \bar{z}_2) \nonu \\
\!& + \!& \cdots \ .
\label{HH}
\eea
We can check the weights in the
antiholomorphic sector both sides
\footnote{We denote the antiholomorphic
  weights as $h_1, h_2$ and $h$ without barred notation by taking
  the notations in previous sections. For
  the notation of derivative,
  we are using the barred notation as in $\bar{\pa}_{\bar{z}}$.
Taking the holomorphic and antiholomorphic
expansions for the above soft graviton current we obtain:
$
H^k(z,\bar{z})
=  \sum_{n=-\infty}^{\frac{2-k}{2}}\,
\frac{H_{n}^k(z)}{\bar{z}^{n+\frac{k-2}{2}}}
= \sum_{m=-\infty}^{\frac{-2-k}{2}}\,
\sum_{n=-\infty}^{\frac{2-k}{2}}\,
\frac{H_{m,n}^k}{z^{m+\frac{k+2}{2}}\, \bar{z}^{n+\frac{k-2}{2}}}$.
The operator $H_{m,n}^k$ is therefore independent of $z$ and
$\bar{z}$ and 
we focus on the case where the mode $m$ is equal to
$(1-h)$:
$
\hat{H}_{n}^k \equiv H_{m=-\frac{k}{2},n}^k$.}.
The left hand side has the weight $\frac{k-2}{2} + \frac{l-2}{2}$ while
the right hand side has the weight $-(n+h+1)+n +\frac{k+l-2}{2}+h$
from the last three factors.
We observe that for the $n$ and $h$ dependences
in (\ref{HH}), there are no additional weight contributions.
In other words, they are cancelled each other and the inclusion of
$h$ does not change the weight in the form of (\ref{HH}).
Note that there exists
the $h$-dependence on the right hand side of (\ref{HH}) and
for $h=0$, we reproduce the result of \cite{GHPS} with proper
upper limit for the summation variable $n$.
It turns out that by performing the explicit contour integrals
presented in \cite{MRSV},
the following commutator relation from (\ref{HH}) is satisfied
\footnote{\label{for}
In \cite{MRSV}, there are two important identities
around 
their equations $(3.8)$ and $(3.10)$. Their $p$ corresponds to our
$(h+1)$. In their equation $(3.8)$, we see that the summation over
dummy variable with the upper bound $p$ reflects the
mode dependent function
$ N^{h_1, h_2}_{h}(m,n)$ (\ref{Nphi}). They manage to express
the infinite sum over their $\al$ variable in terms of
a product of binomial coefficients. The final result is given by
their equation (3.16) after changing the nontrivial
transformation in the mode expansion \cite{Strominger}.
Note that the general expression for the binomial coefficient is
given by $\left(
\begin{array}{c}
-2h_1-2h_2 -2 (1+h)-n \\
-2h_2-(1+h)
\end{array}
\right)$ and their results will be used later.   }
\bea
\big[(\hat{w}_{h_1})_m,(\hat{w}_{h_2})_n\big] 
\!&=& \!
(-1)^{h+1}\, N^{h_1, h_2}_{h}(m,n)\, (\hat{w}_{h_1+h_2-2-h})_{m+n} \ .
\label{Volexpression}
\eea
Here $\hat{w}_h$ is the rescaled weight-$h$ current
and does not depend on the complex coordinate \cite{MRSV,Ahn2111}.
The mode dependent function
$ N^{h_1, h_2}_{h}(m,n)$ is given by the equation (\ref{Nphi}).

By using the two relations in (\ref{HH}) and (\ref{Volexpression})
with associated other relations for the gravitino, gluon and gluino,
we would like to construct the OPEs
in the supersymmetric Einstein-Yang-Mills theory
related to the previous seven
commutator relations. Note how the weights $(h_1,h_2)$ and
the modes $(m, n)$
of the left hand side appear  
on the right hand side of (\ref{Volexpression}).

\subsubsection{The OPE between the soft positive helicity
  graviton and the soft positive helicity gravitino}

At vanishing deformation parameter,
the OPE of
the conformal primary graviton and
the conformal primary gravitino
of arbitrary weights is given by \cite{FSTZ}
\footnote{ We have
$
{\cal O}_{\Delta_1, +2}(z_1, \bar{z}_1)\,
{\cal O}_{\Delta_2, +\frac{3}{2}}(z_2,\bar{z}_2)=
-\frac{\frac{\kappa}{2}}{z_{12}}\, \sum_{n=0}^{\infty}\,
B(\Delta_1-1+n, \Delta_2-\frac{1}{2})\,
\frac{\bar{z}_{12}^{n+1}}{n!}\, \bar{\pa}^n \, 
{\cal O}_{\Delta_1+\Delta_2, +\frac{3}{2}}(z_2,\bar{z}_2) + \cdots $.
}.
Then the question is how the contributions from the
nonzero deformation parameter occur on the right hand side of this
OPE.

\begin{itemize}
\item[]
i) At least, we should have the OPE structure found by
\cite{HPS} which is the fact that
the OPE contains the $h$ dependence in
the $\bar{z}_{12}$, the binomial coefficient
and the weight of the operator
appearing on the right hand side.
These OPE coefficients are also obtained from the analysis in the
collinear limits.
See also the equation
$(3.2)$ of \cite{HPS} in which the addition of fermions
is also valid. 

ii) As mentioned before (in the non minimal couplings
of gluons and gravitons),
after taking the soft limit for the OPE in i) in order to
absorb the infinite number of poles appearing in the binomial coefficient,
we also require that
there is no restriction on
the lower limits of the holomorphic and antiholomorphic
mode expansions of the soft operators \cite{MRSV}.
See also the equation $(A.17)$ of \cite{MRSV}.

iii) We also should sum over all the contributions
from each fixed $h$ on the right hand side of the OPE.  
Moreover, due to the construction of
${\cal N}=1$ supersymmetric theory in previous section,
we should also consider the contributions
from the antiholomorphic derivatives acting on the
soft currents on the left hand sides of the OPEs.
In some sense, we make the supersymmetric generalization of
the work of \cite{MRSV}.

\end{itemize}

Let us consider the case where one of the soft current
on the left hand side in the OPE contains the fermionic
current.

First of all,
the OPE between the conformally soft positive helicity
gravitons and the  conformally soft positive helicity gravitinos,
where
the weights in the antiholomorphic sector are given by
$h_1=\frac{k-2}{2}$ and $h_2=\frac{l-\frac{3}{2}}{2}$,
can be generalized to the following expression
\bea
\!& \!& H^k(z_1,\bar{z}_1)\, I^l(z_2,\bar{z}_2)
=  -\frac{\kappa}{2} \,
\frac{1}{z_{12}} \,
\sum_{h=0}^{h_1+h_2-3}\, (-1)^{h+1}\, \la^h \,
\Bigg[ q_B^{h_1,h_2+\frac{1}{2},h} +q_F^{h_1,h_2+\frac{1}{2},h} \Bigg] \, 
\nonu \\
\!& \!& \times 
\sum_{n=0}^{\infty}\,
\left(
\begin{array}{c}
\frac{3}{2}-2 h-k-l-n \\
\frac{1}{2}-h-l
\end{array}
\right)\, \frac{\bar{z}_{12}^{n+h+1}}{n!}\, \bar{\pa}^n \,
I^{k+l+h}(z_2, \bar{z}_2) \ 
\nonu \\
\!& \! & -\frac{\kappa}{2} \,
\frac{1}{z_{12}} \,
\sum_{h=0}^{h_1+h_2-3}\, (-1)^{h}\,\la^{h} \,
\Bigg[  \frac{2(h_1-2)}{2(h_1-2)+1} \,
q_B^{h_1-1,h_2+\frac{1}{2},h-1}-
\frac{2(h_1-2)+2}{2(h_1-2)+1} \,q_F^{h_1-1,h_2+\frac{1}{2},h-1}
\Bigg] \, 
\nonu \\
\!& \!& \times 
\sum_{n=0}^{\infty}\,
\left(
\begin{array}{c}
\frac{3}{2}-2(h-1)-(k-2)-l-n \\
\frac{1}{2}-(h-1)-l
\end{array}
\right)\, \frac{\bar{\pa}_{\bar{z_1}}\,
  \bar{z}_{12}^{n+(h-1)+1}}{n!}\, \bar{\pa}^n \,
I^{(k-2)+l+(h-1)}(z_2, \bar{z}_2) \nonu \\
\!& \!& + \cdots \ .
\label{fourthres}
\eea
The numerical factor $\frac{3}{2}$
in the first line of   the
first binomial coefficient
is given by
$(2+\frac{3}{2})$ (which is the
sum of numerical numbers with minus sign in the numerator
of previous weights) minus $2$.
The numerical factor $\frac{1}{2}$
in the second line of the
first binomial coefficient
is given by $\frac{3}{2}$ (which is the
numerical number for the gravitino
with minus sign in the numerator
of previous weights) minus $1$ \footnote{
Or we can use the formula in the footnote \ref{for}.}.
The first binomial coefficient above can be obtained
from (\ref{HH}) by taking the shift
$l \rightarrow (l+\frac{1}{2})$.

In the above, we multiplied the factor $(-1)^{h+1}$
in order to absorb $(-1)^{h+1}$ factor on
the right hand side of (\ref{Volexpression}).
Here the graviton current is replaced
by the gravitino current
on the left hand side of the OPE in (\ref{HH})
\footnote{That is, $
H^k(z_1,\bar{z}_1)\, I^l(z_2,\bar{z}_2)
 =  -\frac{\kappa}{2} \,
\frac{1}{z_{12}} \, \sum_{n=0}^{\infty}\,
\left(
\begin{array}{c}
\frac{3}{2}-2 h-k-l-n \\
\frac{1}{2}-h-l
\end{array}
\right)\, \frac{\bar{z}_{12}^{n+h+1}}{n!}\, \bar{\pa}^n \,
I^{k+l+h}(z_2, \bar{z}_2) +  \cdots \ .
$ For $h=0$ with proper upper limit
of dummy variable $n$,
we observe the result of \cite{Ahn2111}.}.
Now we sum over the dummy variable $h$
from $0$ to $(h_1+h_2-3)$
together with the structure constant, which depends on
$h_1$, $h_2$ and $h$ only without mode-dependent factor in
(\ref{modedependence}),
in order to obtain the first two lines of
(\ref{fourthres}). In other words, 
from the first equation of (\ref{WQn=1}) and (\ref{qqtilde}),
the full structure constant contains 
$ q_B^{h_1,h_2+\frac{1}{2},h}(m,n) +q_F^{h_1,h_2+\frac{1}{2},h}(m,n)$,
which is redefined as $( q_B^{h_1,h_2+\frac{1}{2},h} +
q_F^{h_1,h_2+\frac{1}{2},h})\, N_h^{h_1, h_2+\frac{1}{2}}(m,n)$ together with
(\ref{modedependence}).
Then we can apply the property of (\ref{Volexpression})
to the commutator,  the first equation of (\ref{WQn=1}),
and arrive at the first two lines of above OPE.
Here the mode-independent factor, $( q_B^{h_1,h_2+\frac{1}{2},h} +
q_F^{h_1,h_2+\frac{1}{2},h})$,
can be combined to other $h$-dependent factor,
$(-1)^{h+1}\, \la^h$,
inside the $h$ summation.
We observe that the weights in the antiholomorphic
sector are preserved at both sides because the powers of
the dummy variable $n$ and the weight $h$ are the same as before.


In the third and fourth lines of (\ref{fourthres}), we
use the property between the mode of current and the current itself
in (\ref{WWhat}) and (\ref{FREE1}).
Note that
there exists minus sign between them.
Recall the remaining full structure constant
contains
$ \frac{2(h_1-2)(m+(h_1-2)+1)}{2(h_1-2)+1} \,
q_B^{h_1-1,h_2+\frac{1}{2},h-1}(m,n)-
\frac{(2(h_1-2)+2)(m+(h_1-2)+1)}{
  2(h_1-2)+1} \,q_F^{h_1-1,h_2+\frac{1}{2},h-1}(m,n)$.
As we did before, we factorize the
mode-dependent term $(m+(h_1-2)+1)\,
N^{h_1-1, h_2+\frac{1}{2}}_{h-1}(m,n)$
and the other mode-independent terms.
The partial derivative $\bar{\pa}_{\bar{z}_1}$
of $H^{k-2}(z_1,\bar{z}_1)$ on the left hand side
appears in the form of
$\bar{\pa}_{\bar{z_1}}\,
\bar{z}_{12}^{n+(h-1)+1}$ on the right hand side of the OPE.
Further shifts in the weights $h_1$ and $h$
in the structure constant affect
the shifts in the corresponding the $k$ and the weight $h$
in the OPE. 
Note that the weight in the antiholomorphic sector in the
last line of (\ref{fourthres}) behaves correctly because two additional
weight $2$ in the derivative term $\bar{\pa}_{\bar{z_1}}\,
\bar{z}_{12}^{n+(h-1)+1}$ is cancelled by the corresponding
additional $-2$ coming from the factors
$(k-2)$ and $(h-1)$,
compared to the second line of (\ref{fourthres}).

Then
by performing the contour integrals as in \cite{MRSV} (See also
\cite{GHPS,Ahn2111}),
the above OPE (\ref{fourthres}), from the analysis of
two previous paragraphs, provides the
first equation of (\ref{WQn=1}) precisely
\footnote{
In an expression of $\bar{\pa}_{\bar{z_1}}\,
\bar{z}_{12}^{n+(h-1)+1}$, we can write down this as
$(n+(h-1)+1) \, \bar{z}_{12}^{n+(h-1)}$ after a differentiation.
Compared to 
the one without a derivative, there is an extra factor
$(n+(h-1)+1)$ with different power of
$\bar{z}_{12}$. Now we can combine this with
the binomial coefficient
$\left(
\begin{array}{c}
n+(h-1) \\
-m-h_1
\end{array}
\right)$ appearing in the $\bar{z}_1$ integral
when we calculate the commutator from the OPE.
Then it is easy to check that
the following relation is satisfied $(n+(h-1)+1)
\left(
\begin{array}{c}
n+(h-1) \\
-m-h_1
\end{array}
\right) =-(m+h_1-1) \, \Bigg[ \frac{(n+(h-1)+1)!}{
    (n+(h-1)+m+h_1)!(1-m-h_1)!} \Bigg]
= -(m+h_1-1) \, \Bigg[ \frac{(n+(h-1)+1)!}{
  (n+(h-1)+1+m+h_1)!(-m-h_1)!} \Bigg] \Bigg|_{h_1 \rightarrow
  h_1-1}$. This implies that the remaining calculation is the same as
the one in \cite{MRSV} with an overall factor $ -(m+h_1-1)=
-(m+(h_1-2)+1)$ which appears in the second term of
the first structure constant in (\ref{qqtilde}) as
we expected. Of course, the factor $(-\bar{z}_2)^{n+(h-1)+m+h_1}$
coming from the $\bar{z}_1$ integral
can be written as $(-\bar{z}_2)^{n+(h-1)+1+m+(h_1-1)}=
(-\bar{z}_2)^{n+(h-1)+1+m+h_1}\Bigg|_{h_1 \rightarrow h_1-1}$.
Moreover, the binomial coefficient is $\left(
\begin{array}{c}
\frac{3}{2}-2(h-1)-(k-2)-l-n \\
\frac{1}{2}-(h-1)-l
\end{array}
\right)=\left(
\begin{array}{c}
\frac{3}{2}-2(h-1)-k-l-n \\
\frac{1}{2}-(h-1)-l
\end{array}
\right)\Bigg|_{h_1 \rightarrow h_1-1}$.
Of course, $I^{(k-2)+l+(h-1)}=I^{k+l+(h-1)}\Bigg|_{h_1 \rightarrow h_1-1}$.
\label{firstder} }.
Therefore, we have determined the
OPE between the soft positive helicity graviton and
the soft positive helicity
gravitino which can be obtained from (or
leads to) the corresponding commutator
in two dimensions studied in previous section.
We present the details for other OPEs including the
soft positive helicity gravitinos or
soft positive helicity
gluinos
in Appendix $C$ explicitly where other structure constants
(\ref{otherstructure}) can be used appropriately.

\subsubsection{The OPE between the soft positive helicity gravitons}

Let us consider the second example, which is more nontrivial,
for the appearance of the two dimensional symmetry algebra
in the four dimensional Einstein-Yang-Mills theory.
Based on the
three features i), ii) and iii), in previous
subsection,
we can write down the corresponding
OPE for soft positive helicity graviton
in the supersymmetric Einstein-Yang-Mills theory,
by comparing (\ref{Volexpression}) with the first equation of
(\ref{WWn=1}),
as follows:
\bea
\!& \!& H^k(z_1,\bar{z}_1)\, H^l(z_2,\bar{z}_2)
=
 -\frac{\kappa}{2} \,
\frac{1}{z_{12}} \,
\sum_{h=1,\mbox{\footnotesize odd}}^{h_1+h_2-4}\, (-1)^{h}\,\la^{h} \,
\Bigg[ f_1^{h_1-1,h_2,h-1}
\nonu \\
\!& \!& \times 
\sum_{n=0}^{\infty}\,
\left(
\begin{array}{c}
2-2(h-1)-(k-2)-l-n \\
1-(h-1)-l
\end{array}
\right)\, \frac{\bar{\pa}_{\bar{z_1}}\,
  \bar{z}_{12}^{n+(h-1)+1}}{n!}\, \bar{\pa}^n \,
H^{(k-2)+l+(h-1)}(z_2, \bar{z}_2) \nonu \\
\!& \! & +
f_2^{h_1,h_2-1,h-1}
\,
\sum_{n=0}^{\infty}\,
\left(
\begin{array}{c}
2-2(h-1)-k-(l-2)-n \\
1-(h-1)-(l-2)
\end{array}
\right)\,
\nonu \\
\!& \!&  \times 
\frac{1}{n!}\, \bar{\pa}_{\bar{z_2}}\,
 \big[ \bar{z}_{12}^{n+(h-1)+1}\, \bar{\pa}^n \,
   H^{k+(l-2)+(h-1)}(z_2, \bar{z}_2) \big]
 \Bigg]  -\frac{\kappa}{2} \,
\frac{1}{z_{12}} \,
\sum_{h=0,\mbox{\footnotesize even}}^{h_1+h_2-4}\, (-1)^{h+1}\, \la^h \,
\Bigg[
  f_3^{h_1,h_2,h}
  \nonu \\
\!& \!& \times \, \sum_{n=0}^{\infty}\,
\left(
\begin{array}{c}
2-2 h-k-l-n \\
1-h-l
\end{array}
\right)\, \frac{\bar{z}_{12}^{n+h+1}}{n!}\, \bar{\pa}^n \,
 H^{k+l+h}(z_2, \bar{z}_2) 
 \nonu \\
 \!& \!&
+ f_4^{h_1-1,h_2-1,h-2}
\, \sum_{n=0}^{\infty}\,
\left(
\begin{array}{c}
2-2(h-2)-(k-2)-(l-2)-n \\
1-(h-2)-(l-2)
\end{array}
\right)\, \nonu \\
\!& \!& \times \frac{1}{n!}\,\bar{\pa}_{\bar{z_1}}\, \bar{\pa}_{\bar{z_2}}\,
 \big[ \bar{z}_{12}^{n+(h-2)+1}\, \bar{\pa}^n \,
   H^{(k-2)+(l-2)+(h-2)}(z_2, \bar{z}_2) \big] \Bigg]
 + \cdots \ .
\label{HH1}
\eea
Here we introduce 
the following quantities
\bea
 f_1^{h_1-1,h_2,h-1} & \equiv & 
\Bigg[  \frac{2(h_1-2)}{2(h_1-2)+1} \,
  \frac{(h_1+h_2-2-h)}{2(h_1+h_2-2-h)+1}\,
  p_B^{h_1-1,h_2,h-1}
  \label{fourf}
  \\
  \!& \!& -
  \frac{(h_1+h_2-2-h-1)}{2(h_1+h_2-2-h-1)+1} \,
 \frac{2(h_1-2)+2}{2(h_1-2)+1}  \, 
p_F^{h_1-1,h_2,h-1}
\Bigg], 
\nonu \\
f_2^{h_1,h_2-1,h-1}
& \equiv &
\Bigg[
  \frac{2(h_2-2)}{2(h_2-2)+1}\,
\frac{(h_1+h_2-2-h)}{2(h_1+h_2-2-h-1)+1}
  \,p_B^{h_1,h_2-1,h-1}
  \nonu \\
  \!& \!& -
  \frac{(h_1+h_2-2-h-1)}{2(h_1+h_2-2-h-1)+1}\,
  \frac{2(h_2-2)+2}{2(h_2-2)+1}\,
p_F^{h_1,h_2-1,h-1}  
\Bigg],
\nonu \\
f_3^{h_1,h_2,h} & \equiv &
\Bigg[ \frac{(h_1+h_2-2-h)}{2(h_1+h_2-2-h-1)+1}  p_B^{h_1,h_2,h}
  +\frac{(h_1+h_2-2-h-1)}{2(h_1+h_2-2-h-1)+1}
  p_F^{h_1,h_2,h}\Bigg],
\nonu \\
f_4^{h_1-1,h_2-1,h-2} & \equiv &
\Bigg[
  \frac{2(h_1-2)}{2(h_1-2)+1}\,
\frac{2(h_2-2)}{2(h_2-2)+1}
  \,  \frac{(h_1+h_2-2-h)}{2(h_1+h_2-2-h-1)+1}\, p_B^{h_1-1,h_2-1,h-2}
  \nonu \\
  \!& \!& + \frac{2(h_1-2)+2}{2(h_1-2)+1}\,
\frac{2(h_2-2)+2}{2(h_2-2)+1}
\,  \frac{(h_1+h_2-2-h-1)}{2(h_1+h_2-2-h-1)+1}\, p_F^{h_1-1,h_2-1,h-2}
\Bigg].
\nonu
\eea
Let us describe how we obtain this result.
%
Now we return to the OPE (\ref{HH1}) for the soft
positive helicity gravitons.
We multiply the factor $(-1)^{h+1}$ into the relation (\ref{HH})
and sum over the variable $h$ from $0$ to $(h_1+h_2-4)$
together with the structure constant which depends on
$h_1$, $h_2$ and $h$ without mode-dependent factor in
(\ref{modedependence}) as before.
For the mode-dependent part, we want to use the previous relation
(\ref{Volexpression}).
In the first three lines of (\ref{HH1}), we again
use the property between the mode of current and the current itself
in (\ref{WWhat}) and (\ref{FREE1}).
The partial derivative $\bar{\pa}_{\bar{z}_1}$
of $H^{k-2}(z_1,\bar{z}_1)$ appears in the form of
$\bar{\pa}_{\bar{z_1}}\,
\bar{z}_{12}^{n+(h-1)+1}$ on the right hand side of the OPE.
The mode-independent coefficients come from
the second and the fourth terms of the first equation in
(\ref{qotherexp}) with (\ref{fourf}).
Further shifts in the weights
$h_1$ and $h$ in the structure constant affect
the shifts in the corresponding $k$ and $h$. 
Note that the weight in the antiholomorphic sector here
behaves correctly. See also the footnote \ref{firstder}.

In the next line,
the $\bar{\pa}_{\bar{z}_2}$
acts on both
the coordinate $\bar{z}_{12}$ and $H^{k+(l-2)+(h-1)}(z_2,\bar{z}_2)$
(corresponding to  $\bar{\pa}_{\bar{z}_2}$
of $H^{l-2}(z_2,\bar{z}_2)$ in the left hand side)
\footnote{From the contribution of  $\bar{\pa}_{\bar{z}_2}\,
  \bar{z}_{12}^{n'+(h-1)+1}$, there is an extra factor
  $-(n'+(h-1)+1)$ with $ \bar{z}_{12}^{n'+(h-1)}$
  where $n'$ is a previous dummy variable $n$.
  By recalling the $\bar{z}_1$ contour integral, we have
$-(n'+(h-1)+1)\, \left(
\begin{array}{c}
n'+(h-1) \\
-m-h_1
\end{array}
\right)\, (-1)^{n'+(h-1)+m+h_1}$ where $h_1=\frac{k-2}{2}$.
Then this can be rewritten as
$ (n'+(h-1)+1+m+h_1) \, \left(
\begin{array}{c}
n'+(h-1)+1 \\
-m-h_1
\end{array}
\right)\, (-1)^{n'+(h-1)+1+m+h_1}$.
Note that the power of $\bar{z}_{12}$ is $n'+(h-1)$.
This implies that
there exists  a factor $(n'+(h-1)+1+m+h_1)$ in the presence of
$\bar{\pa}_{\bar{z}_2}$.
Of course, the factor $(\bar{z}_2)^{n'+(h-1)+m+h_1+n+h_2-1}$
coming from the $\bar{z}_1$ integral
and the exponent of $\bar{z}_2$ in the contour integral
can be written as $(\bar{z}_2)^{n'+(h-1)+1+m+h_1+n+(h_2-1)-1}$
which can be obtained by taking $h_2 \rightarrow (h_2-1)$
from the expression without a derivative.

Furthermore, from the
contribution of
$\bar{\pa}_{\bar{z}_2}\, H^{k+(l-2)+(h-1)}(z_2,\bar{z}_2)$,
we have the extra factor $(-m-n-h_1-(h_2-1)-(h-1)-1-n')$
(with $h_2=\frac{l-2}{2}$)
which can be obtained by taking $h_2 \rightarrow (h_2-1)$
from the expression without a derivative factor.
The exponent of $\bar{z}_2$ is given by 
$(\bar{z}_2)^{n'+(h-1)+1+m+h_1+n+h_2-1}$ originally and
from the above derivative $\bar{\pa}_{\bar{z}_2}$
we have an additional $\bar{z}_2^{-1}$, compared to the case without
this derivative.
Note that the power of $\bar{z}_{12}$ is $n'+(h-1)+1$.
Therefore, the above power
can be written as $n'+(h-1)+1+m+h_1+n+(h_2-1)-1$ which is obtained
by taking $h_2 \rightarrow h_2-1$. The binomial coefficient
can be shifted similarly as in the footnote \ref{firstder}.
Then finally
we are left with the factor $[n'+(h-1)+1+m+h_1]+[-m-n-h_1-(h_2-1)-(h-1)-1-n'
]=-n-(h_2-1)$
by adding these two contributions.
\label{firstder1}}.
In this case,
the mode-independent coefficients come from
the first and the third terms of the first equation in
(\ref{qotherexp}). 

For the remaining lines, we describe similarly
by applying the first and the third terms in the second equation of
(\ref{qotherexp}) without any derivatives in the summation over
even weight $h$. Finally,
the second and the fourth terms in the second equation of
(\ref{qotherexp}) with two derivatives play the role in
the second summation over
even weight $h$ \footnote{We can analyze
the action of $\bar{\pa}_{\bar{z}_1}\,\bar{\pa}_{\bar{z}_2}$
appearing at the end of (\ref{HH1}) by taking the procedures in
the footnote \ref{firstder} and the footnote \ref{firstder1}.}.
The additional weight-$2$ from the two derivatives
is cancelled by those from the power of $(k-2)$
and $(l-2)$ in the soft current.

Then
by performing the contour integrals as in \cite{MRSV} similarly
(See also \cite{GHPS,Ahn2111}),
the above OPEs (\ref{fourthres}) and (\ref{HH1}) provide the  
first equations of (\ref{WQn=1}) and (\ref{WWn=1}) respectively.
We can extract other five OPEs similarly by taking into
account of the weights for the soft
graviton, gravitino, gluon and
gluino in the binomial coefficient above. We present the details
in Appendix $C$ explicitly. In particular, the
useful structure constants $\tilde{q}^{h_1,h_2,h}(m,n)$ we are using,
which appear in the third equation of (\ref{WWn=1}),
are given at the end of  Appendix $C$.
Therefore, we have found the precise correspondence between
the OPEs between the conformally soft operators in
the ${\cal N}=1$ supersymmetric Einstein-Yang-Mills theory
and the two dimensional symmetry algebra.

\section{ Conclusions and outlook}

The main result of this paper can be summarized by (\ref{WWn=1})
and (\ref{WQn=1}) together with (\ref{qqtilde}) and (\ref{otherstructure}).
That is,
we have determined the
three commutator relations by combining the
the corresponding commutator relations from the
bosonic currents made of the bosonic free fields
with
the corresponding commutator relations from the
bosonic currents made of the fermionic free fields. 
Similarly, the
remaining
four commutator relations can be obtained
from two different kinds of commutator relations. 
Their OPE version can be found in (\ref{OPES}) with (\ref{ff}) and
(\ref{struct}).
We observe that there are also the nontrivial
singular terms whose poles are greater than two
and they play the role of the contributions from the deformation.

By using the celestial holography, we
have obtained the results, summarized by (\ref{fourthres}),
(\ref{HH1}), Appendices 
(\ref{fifthres}), (\ref{sixthres}),
(\ref{seventhres}), (\ref{secondres}), and
(\ref{thirdres})
in the supersymmetric Einstein-Yang-Mills theory
at nonzero deformation parameter.
The common behavior is as follows.
There exist a simple pole in the holomorphic
sector, the nontrivial structure constants which depend on
the three weights, binomial coefficients containing the
dummy variable also as well as the three weights and the descendant
fields associated with the second soft currents on the left hand sides.
Furthermore, after calculating the various commutator relations
by using these seven OPEs between the soft currents,
we have checked the above seven commutators in
two dimensions discussed in previous paragraph
\footnote{So far, we have focused on the soft currents as mentioned
  in the footnote \ref{firstfootnote}.
  According to the findings of \cite{HPS,Jiang:2021ovh},
  the leading tree level celestial OPEs from the cubic vertices
  of three spinning massless particles contains the
  OPE coefficients given by Euler beta function
  whose arguments are  $(\Delta_1 +s_2-s_3-1)$ and $(\Delta_2+s_1
  -s_3-1)$ where $s_i$ is a helicity or spin.
  In particular, in \cite{Jiang:2021ovh},
  it is found that 
  all the previous known seven nontrivial celestial OPEs
  are obtained from this general formula. We can ask
  whether the OPEs between the hard operators, which do not
  satisfy the conditions for the conformal dimensions in the footnote
  \ref{firstfootnote}, provide any two dimensional symmetry algebra
  similar to the ones in this paper or not.
  It seems that the infinite number of poles appearing in the above
  Euler beta function can appear only for the corresponding
  conformal dimensions of the soft operators. In other words,
  for the conformal dimensions of the hard operators,
  we cannot cover the infinite number of poles and therefore we cannot remove
  all of them.}.

We have not discussed about the implications of
the ${\cal N}=2$ supersymmetric $W_{\infty}^{K,K}$ algebra
in the section
$2$. It would be interesting to
find out whether the possibility of
the ${\cal N}=2$ supersymmetric $w_{\infty}^{K,K}$ algebra
occurs or not by further examination.
In the context of $AdS_3/CFT_2$ correspondence,
the previous algebra in (\ref{seven}) is related to the
case of vanishing 't Hooft-like coupling constant.
Therefore, it is an open question how the another deformed case with
nonvanishing 't Hooft-like coupling constant \cite{AK2009}
will arise in the context of the present paper.
We expect that the currents from the free field realization 
will depend on this nonzero coupling constant explicitly
and they will become the currents we have described in this paper
by taking this coupling constant to be zero.

\vspace{.7cm}

\centerline{\bf Acknowledgments}

We
would like to
thank
M.H. Kim
for discussions.
This work was supported by
the National Research Foundation of Korea(NRF) grant
funded by the Korea government(MSIT)(No. 2020R1F1A1066893).

\newpage

\appendix

\renewcommand{\theequation}{\Alph{section}\mbox{.}\arabic{equation}}

\section{ The remaining (anti)commutator relations
in the ${\cal N}=2$ supersymmetric
$W_{\infty}^{K,K}$ algebra
with $U(K)\times U(K)$ symmetry
}

In this Appendix, we present some details which are related to
the contents in section $2$.

\subsection{The commutators between the bosonic and the other fermionic
currents}

We can multiply the generators into the fourth equation of
(\ref{seven})
and obtain the following commutator relations
\bea
\big[(W_{\mathrm{F},h_1})_m,(\bar{Q}_{h_2+\frac{1}{2}})_r\big] 
\!&= &\!
\sum^{h_1+h_2-3}_{h=-1}\, \la^h \, (-1)^h\, q_{\mathrm{F}}^{h_1,h_2+\frac{1}{2}, h}(m,r) 
\, (\bar{Q}_{h_1+h_2-\frac{3}{2}-h})_{m+r}\ ,
\nonu \\
\big[(W_{\mathrm{F},h_1})_m,(\bar{Q}^{\hat{A}}_{h_2+\frac{1}{2}})_r\big] 
\!&= &\!
\sum^{h_1+h_2-3}_{h=-1}\, \la^h \,  (-1)^h\, q_{\mathrm{F}}^{h_1,h_2+\frac{1}{2}, h}(m,r) 
\, (\bar{Q}^{\hat{A}}_{h_1+h_2-\frac{3}{2}-h})_{m+r}\ ,
\nonu\\
\big[(W^{\hat{A}}_{\mathrm{F},h_1})_m,(\bar{Q}_{h_2+\frac{1}{2}})_r\big] 
\!&= &\!
\sum^{h_1+h_2-3}_{h=-1}\, \la^h \,  (-1)^h\, q_{\mathrm{F}}^{h_1,h_2+\frac{1}{2}, h}(m,r) 
\, (\bar{Q}^{\hat{A}}_{h_1+h_2-\frac{3}{2}-h})_{m+r}\ ,
\nonu \\
\big[(W^{\hat{A}}_{\mathrm{F},h_1})_m,(\bar{Q}^{\hat{B}}_{h_2+\frac{1}{2}})_r\big] 
\!&=& \!
-\sum^{h_1+h_2-3}_{h=-1} \, \la^h\,  (-1)^h\, q_{\mathrm{F}}^{h_1,h_2+\frac{1}{2}, h}(m,r)
\, \frac{i}{2}\, f^{\hat{A} \hat{B}  \hat{C}} \,
( \bar{Q}^{\hat{C}}_{h_1+h_2-\frac{3}{2}-h} )_{m+r}
\nonu \\
\!&+\!& \sum^{h_1+h_2-3}_{h= -1} \, \la^h\,  (-1)^h\,
q_{\mathrm{F}}^{h_1,h_2+\frac{1}{2}, h}(m,r)
\, \Bigg[ \frac{1}{2}\, d^{\hat{A} \hat{B} \hat{C}} \,
(   \bar{Q}^{\hat{C}}_{h_1+h_2-\frac{3}{2}-h} )_{m+r} \nonu \\
\!& + \!& \frac{1}{K}\, \delta^{\hat{A} \hat{B}}\,
(   \bar{Q}_{h_1+h_2-\frac{3}{2}-h} )_{m+r} \Bigg] \ .
\label{app1}
\eea

It is easy to check the following
reduced commutator relations by taking
(\ref{rescaling})
\bea
\big[(W_{\mathrm{F},h_1})_m,(\bar{Q}_{h_2+\frac{1}{2}})_r\big] 
\!&= &\!
\frac{1}{4}
\, (\bar{Q}_{h_1+h_2-\frac{1}{2}})_{m+r}\ ,
\nonu \\
\big[(W_{\mathrm{F},h_1})_m,(\bar{Q}^{\hat{A}}_{h_2+\frac{1}{2}})_r\big] 
\!&= &\!
\frac{1}{4}
\, (\bar{Q}^{\hat{A}}_{h_1+h_2-\frac{1}{2}})_{m+r}\ ,
\nonu\\
\big[(W^{\hat{A}}_{\mathrm{F},h_1})_m,(\bar{Q}_{h_2+\frac{1}{2}})_r\big] 
\!&= &\!
\frac{1}{4}
\, (\bar{Q}^{\hat{A}}_{h_1+h_2-\frac{1}{2}})_{m+r}\ ,
\nonu \\
\big[(W^{\hat{A}}_{\mathrm{F},h_1})_m,(\bar{Q}^{\hat{B}}_{h_2+\frac{1}{2}})_r\big] 
\!&=& \!
- \frac{i}{8}\, f^{\hat{A} \hat{B}  \hat{C}} \, (
\bar{Q}^{\hat{C}}_{h_1+h_2-\frac{1}{2}} )_{m+r}
+ \frac{1}{4}
\, \Bigg[ \frac{1}{2}\, d^{\hat{A} \hat{B} \hat{C}} \,
(   \bar{Q}^{\hat{C}}_{h_1+h_2-\frac{1}{2}} )_{m+r} \nonu \\
\!& + \!& \frac{1}{K}\, \delta^{\hat{A} \hat{B}}\,
(   \bar{Q}_{h_1+h_2-\frac{1}{2}} )_{m+r} \Bigg] \ . 
\label{n2reduced1}
\eea

\subsection{The commutators between the other bosonic and the other
  fermionic
currents}

The sixth equation of (\ref{seven}) with the addition of
generators leads to the following
commutator relations
\bea
\big[(W_{\mathrm{B},h_1})_m,(\bar{Q}_{h_2+\frac{1}{2}})_r\big] 
\!&= &\!
\sum^{h_1+h_2-3}_{h=-1}\, \la^h \, (-1)^h \, q_{\mathrm{B}}^{h_1,h_2+\frac{1}{2}, h}(m,r) 
\, (\bar{Q}_{h_1+h_2-\frac{3}{2}-h})_{m+r}\ ,
\nonu \\
\big[(W_{\mathrm{B},h_1})_m,(\bar{Q}^{\hat{A}}_{h_2+\frac{1}{2}})_r\big] 
\!&= &\!
\sum^{h_1+h_2-3}_{h=-1}\, \la^h \,  (-1)^h \,q_{\mathrm{B}}^{h_1,h_2+\frac{1}{2}, h}(m,r) 
\, (\bar{Q}^{\hat{A}}_{h_1+h_2-\frac{3}{2}-h})_{m+r}\ ,
\nonu\\
\big[(W^{\hat{A}}_{\mathrm{B},h_1})_m,(\bar{Q}_{h_2+\frac{1}{2}})_r\big] 
\!&= &\!
\sum^{h_1+h_2-3}_{h=-1}\, \la^h \,  (-1)^h \,q_{\mathrm{B}}^{h_1,h_2+\frac{1}{2}, h}(m,r) 
\, (\bar{Q}^{\hat{A}}_{h_1+h_2-\frac{3}{2}-h})_{m+r}\ ,
\nonu \\
\big[(W^{\hat{A}}_{\mathrm{B},h_1})_m,(\bar{Q}^{\hat{B}}_{h_2+\frac{1}{2}})_r\big] 
\!&=& \!
\sum^{h_1+h_2-3}_{h= -1} \, \la^h\,  (-1)^h \,q_{\mathrm{B}}^{h_1,h_2+\frac{1}{2}, h}(m,r)
\, \frac{i}{2}\, f^{\hat{A} \hat{B}  \hat{C}} \,
(   \bar{Q}^{\hat{C}}_{h_1+h_2-\frac{3}{2}-h} )_{m+r}
\nonu \\
\!&+\!& \sum^{h_1+h_2-3}_{h= -1} \, \la^h\,  (-1)^h \,q_{\mathrm{B}}^{h_1,h_2+\frac{1}{2}, h}(m,r)
\, \Bigg[ \frac{1}{2}\, d^{\hat{A} \hat{B} \hat{C}} \,
(   \bar{Q}^{\hat{C}}_{h_1+h_2-\frac{3}{2}-h} )_{m+r} \nonu \\
\!& +\! & \frac{1}{K}\, \delta^{\hat{A} \hat{B}}\,
(   \bar{Q}_{h_1+h_2-\frac{3}{2}-h} )_{m+r} \Bigg] \ .
\label{app2}
\eea

By taking
(\ref{rescaling}),
the following
reduced commutator relations hold
\bea
\big[(W_{\mathrm{B},h_1})_m,(\bar{Q}_{h_2+\frac{1}{2}})_r\big] 
\!&= &\!
-\frac{1}{4}
\, (\bar{Q}_{h_1+h_2-\frac{1}{2}})_{m+r}\ ,
\nonu \\
\big[(W_{\mathrm{B},h_1})_m,(Q^{\hat{A}}_{h_2+\frac{1}{2}})_r\big] 
\!&= &\!
-\frac{1}{4}
\, (\bar{Q}^{\hat{A}}_{h_1+h_2-\frac{1}{2}})_{m+r}\ ,
\nonu \\
\big[(W^{\hat{A}}_{\mathrm{B},h_1})_m,(\bar{Q}_{h_2+\frac{1}{2}})_r\big] 
\!&= &\!
-\frac{1}{4}
\, (\bar{Q}^{\hat{A}}_{h_1+h_2-\frac{1}{2}})_{m+r}\ ,
\nonu \\
\big[(W^{\hat{A}}_{\mathrm{B},h_1})_m,(Q^{\hat{B}}_{h_2+\frac{1}{2}})_r\big] 
\!&=& \!
 -\frac{i}{8}\, f^{\hat{A} \hat{B}  \hat{C}} \, (
\bar{Q}^{\hat{C}}_{h_1+h_2-\frac{1}{2}} )_{m+r}
- \frac{1}{4}
\, \Bigg[ \frac{1}{2}\, d^{\hat{A} \hat{B} \hat{C}} \,
(   \bar{Q}^{\hat{C}}_{h_1+h_2-\frac{1}{2}} )_{m+r} \nonu \\
\!& + \!& \frac{1}{K}\, \delta^{\hat{A} \hat{B}}\,
(   \bar{Q}_{h_1+h_2-\frac{1}{2}} )_{m+r} \Bigg] \ .
\label{n2reduced2}
\eea

\subsection{The anticommutators between the fermionic
currents}

Finally, the seventh equation of (\ref{seven})
provides the following anticommutator relations
\bea
\{(Q_{h_1+\frac{1}{2}})_r,(\bar{Q}_{h_2+\frac{1}{2}})_s\} 
\!&=&\!
\sum^{h_1+h_2-1}_{h=0} \,
\la^h 
\, o_{\mathrm{F}}^{h_1+\frac{1}{2},h_2+\frac{1}{2},h}(r,s) \,
(W_{\mathrm{F},h_1+h_2-h})_{r+s}
\nonu \\
\!& + \!& \sum^{h_1+h_2-2}_{h=0} \,
\la^h \,
o_{\mathrm{B}}^{h_1+\frac{1}{2},h_2+\frac{1}{2},h}(r,s) \,
(W_{\mathrm{B},h_1+h_2-h})_{r+s} \ ,
\nonu \\
\!&+\!&  K\, c_{Q\bar{Q}_{h_1+\frac{1}{2}}}
\,  \delta^{h_1 h_2} \, \la^{2(h_1+\frac{1}{2}-1)}
\delta_{r+s} \ ,
\nonu \\
\{(Q_{h_1+\frac{1}{2}})_r,(\bar{Q}^{\hat{A}}_{h_2+\frac{1}{2}})_s\} 
\!&=&\!
\sum^{h_1+h_2-1}_{h=0}\,
\la^h  \,
\, o_{\mathrm{F}}^{h_1+\frac{1}{2},h_2+\frac{1}{2},h}(r,s) \,
(W^{\hat{A}}_{\mathrm{F},h_1+h_2-h})_{r+s}
\nonu \\
\!& + \!&
\sum^{h_1+h_2-2}_{h=0}\,
\la^h  \,
o_{\mathrm{B}}^{h_1+\frac{1}{2},h_2+\frac{1}{2},h}(r,s) \,
(W^{\hat{A}}_{\mathrm{B},h_1+h_2-h})_{r+s}  \ ,
\nonu \\
\{(Q^{\hat{A}}_{h_1+\frac{1}{2}})_r,(\bar{Q}_{h_2+\frac{1}{2}})_s\} 
\!&=&\!
\sum^{h_1+h_2-1}_{h=0} \,
\la^h 
\, o_{\mathrm{F}}^{h_1+\frac{1}{2},h_2+\frac{1}{2},h}(r,s) \,
(W^{\hat{A}}_{\mathrm{F},h_1+h_2-h})_{r+s}
\nonu \\
\!& + \!& 
\,\sum^{h_1+h_2-2}_{h=0} \,
\la^h 
o_{\mathrm{B}}^{h_1+\frac{1}{2},h_2+\frac{1}{2},h}(r,s) \,
(W^{\hat{A}}_{\mathrm{B},h_1+h_2-h})_{r+s}  \ ,
\nonu \\
\{(Q^{\hat{A}}_{h_1+\frac{1}{2}})_r,(\bar{Q}^{\hat{B}}_{h_2+\frac{1}{2}})_s\} 
\!&=& \!
\sum^{h_1+h_2-1}_{h= 0} \, \la^h \,o_{\mathrm{F}}^{h_1+\frac{1}{2},h_2+
\frac{1}{2}, h}(r,s)
\, \frac{i}{2}\, f^{\hat{A} \hat{B}  \hat{C}} \,
(   W^{\hat{C}}_{F,h_1+h_2-h} )_{r+s}
\nonu \\
\!&+\!& \sum^{h_1+h_2-1}_{h= 0} \, \la^h \,o_{\mathrm{F}}^{h_1+\frac{1}{2},
h_2+\frac{1}{2}, h}(r,s)
\, \Bigg[ \frac{1}{2}\, d^{\hat{A} \hat{B} \hat{C}} \,
(   W^{\hat{C}}_{F,h_1+h_2-h} )_{r+s} \nonu \\
\!& + \!& \frac{1}{K}\, \delta^{\hat{A} \hat{B}}\,
(   W_{F,h_1+h_2-h} )_{r+s} \Bigg]
\nonu \\
\!&-\!& \sum^{h_1+h_2-2}_{h= 0} \, \la^h \,
o_{\mathrm{B}}^{h_1+\frac{1}{2},h_2+\frac{1}{2}, h}(r,s)
\, \frac{i}{2}\, f^{\hat{A} \hat{B}  \hat{C}} \,
(   W^{\hat{C}}_{B,h_1+h_2-h} )_{r+s}
\nonu \\
\!&+\!& \sum^{h_1+h_2-2}_{h= 0} \, \la^h \,
o_{\mathrm{B}}^{h_1+\frac{1}{2},h_2+\frac{1}{2}, h}(r,s)
\, \Bigg[ \frac{1}{2}\, d^{\hat{A} \hat{B} \hat{C}} \,
(   W^{\hat{C}}_{B,h_1+h_2-h} )_{r+s} \nonu \\
\!& +\! & \frac{1}{K}\, \delta^{\hat{A} \hat{B}}\,
(   W_{B,h_1+h_2-h} )_{r+s} \Bigg]
\nonu \\
\!&+\!&  c_{Q\bar{Q}_{h_1+\frac{1}{2}}}  \, \delta^{\hat{A} \hat{B}}
\, \delta^{h_1 h_2} \, \la^{2(h_1+\frac{1}{2}-1)}
\delta_{r+s} \ .
\label{app3}
\eea

Then the complete algebra
consists of
(\ref{OSalgebra}), (\ref{BKalgebra}),
(\ref{WFQ}), (\ref{WBQ}), Appendix (\ref{app1}), Appendix
(\ref{app2}) and
Appendix (\ref{app3}).

Note that
there are no nontrivial reduced anticommutator relations
after taking (\ref{rescaling}).
Then we have the complete results, (\ref{n2reduced}),
Appendix (\ref{n2reduced1}) and Appendix (\ref{n2reduced2}).

\section{ The operator product expansions
in the ${\cal N}=1$ supersymmetric
$W_{\infty}^{K}$ algebra with $U(K)$ symmetry
}

In this Appendix, we present some details which are related to
the contents in section $3$.

\subsection{ The seven OPEs for fixed $h_1$ and $h_2$}

It is straightforward to calculate the following OPEs
by using the Thielemans package \cite{Thielemans}
inside the mathematica \cite{mathematica}.
We use the equations (\ref{OPE}), (\ref{FREE1}) and (\ref{FREE2}).
For fixed $h_1$ and $h_2$, we perform each pole starting from the
highest order pole. For each pole, we can consider the possible terms
(descendant terms and new higher spin current).
The higher spin currents are, in general,  not quasiprimary.
Although the current $W_h(\bar{w})$ for fixed $h$
does not appear in the particular pole, its derivative term
$\bar{\pa} \, W_h(\bar{w})$ appears at the next order pole.
This derivative term plays the role of the quasiprimary current
of the weight-$(h+1)$. Once we rearrange each OPE in terms of
quasiprimary currents then the standard expressions in the right hand
sides appear \footnote{Note that
  $W_4,W_4^{\hat{A}},Q_{\frac{7}{2}}$ and $Q^{\hat{A}}_{\frac{7}{2}}$
  are not quasiprimary under the
  $W_2$ stress energy tensor which has zero central charge.
  We can make them to be quasiprimary
  by adding the derivatives of currents having lower weights.
  The weights for $Q_{\frac{7}{2}}$ and $Q^{\hat{A}}_{\frac{7}{2}}$
are $4$ not $\frac{7}{2}$ after topological twisting.}.
We have the following OPEs for fixed $h_1$ and $h_2$
\bea
&& W_4(\bar{z})\, W_4(\bar{w}) = 
\frac{1}{(\bar{z}-\bar{w})^6}\, \frac{9216 \la^4}{5}\,
W_2(\bar{w}) + \frac{1}{(\bar{z}-\bar{w})^5}\,
\frac{4608 \la^4}{5}\,
\bar{\pa}\, W_2(\bar{w})
\nonu \\
&& + \frac{1}{(\bar{z}-\bar{w})^4}\,
\Bigg[\frac{1536 \la^4}{5}\,
  \bar{\pa}^2 \, W_2 +
  \frac{2304 \la^3}{125}\,  \bar{\pa} \, W_3
  + \frac{8256 \la^2}{25}\,
 W_4
  \Bigg](\bar{w})
\nonu \\
&& + \frac{1}{(\bar{z}-\bar{w})^3}\,
\Bigg[\frac{384 \la^4}{5}\,
  \bar{\pa}^3 \, W_2 +
  \frac{1152 \la^3}{125}\,  \bar{\pa}^2 \, W_3
  + \frac{4128 \la^2}{25}\,
 \bar{\pa} \, W_4
  \Bigg](\bar{w})
\nonu \\
&& + \frac{1}{(\bar{z}-\bar{w})^2}\,
\Bigg[\frac{384 \la^4}{25}\,
  \bar{\pa}^4 \, W_2 +
  \frac{2304 \la^3}{875}\,  \bar{\pa}^3 \, W_3
  + \frac{8256 \la^2}{175}\,
  \bar{\pa}^2 \, W_4
  +  \frac{8 \la}{15}\,
  \bar{\pa} \, W_5
  + 6 \, W_6
  \Bigg](\bar{w})
\nonu \\
&& + \frac{1}{(\bar{z}-\bar{w})}\,
\Bigg[\frac{64 \la^4}{25}\,
  \bar{\pa}^5 \, W_2 +
  \frac{96 \la^3}{175}\,  \bar{\pa}^4 \, W_3
  + \frac{344 \la^2}{35}\,
  \bar{\pa}^3 \, W_4
  +  \frac{4 \la}{15}\,
  \bar{\pa}^2 \, W_5
  + 3 \, \bar{\pa} \, W_6
  \Bigg](\bar{w})
\nonu \\
&& + 
\cdots \ ,
\nonu \\
&& W_4(\bar{z})\, W^{\hat{A}}_4(\bar{w})  = 
\frac{1}{(\bar{z}-\bar{w})^6}\, \frac{9216 \la^4}{5}\,
W^{\hat{A}}_2(\bar{w}) + \frac{1}{(\bar{z}-\bar{w})^5}\,
\frac{4608 \la^4}{5}\,
\bar{\pa}\, W^{\hat{A}}_2(\bar{w})
\nonu \\
&& + \frac{1}{(\bar{z}-\bar{w})^4}\,
\Bigg[\frac{1536 \la^4}{5}\,
  \bar{\pa}^2 \, W^{\hat{A}}_2 +
  \frac{2304 \la^3}{125}\,  \bar{\pa} \, W^{\hat{A}}_3
  + \frac{8256 \la^2}{25}\,
 W^{\hat{A}}_4
  \Bigg](\bar{w})
\nonu \\
&& + \frac{1}{(\bar{z}-\bar{w})^3}\,
\Bigg[\frac{384 \la^4}{5}\,
  \bar{\pa}^3 \, W^{\hat{A}}_2 +
  \frac{1152 \la^3}{125}\,  \bar{\pa}^2 \, W^{\hat{A}}_3
  + \frac{4128 \la^2}{25}\,
 \bar{\pa} \, W^{\hat{A}}_4
  \Bigg](\bar{w})
\nonu \\
&& + \frac{1}{(\bar{z}-\bar{w})^2}\,
\Bigg[\frac{384 \la^4}{25}\,
  \bar{\pa}^4 \, W^{\hat{A}}_2 +
  \frac{2304 \la^3}{875}\,  \bar{\pa}^3 \, W^{\hat{A}}_3
  + \frac{8256 \la^2}{175}\,
  \bar{\pa}^2 \, W^{\hat{A}}_4
  +  \frac{8 \la}{15}\,
  \bar{\pa} \, W^{\hat{A}}_5
  + 6 \, W^{\hat{A}}_6
  \Bigg](\bar{w})
\nonu \\
&& + \frac{1}{(\bar{z}-\bar{w})}\,
\Bigg[\frac{64 \la^4}{25}\,
  \bar{\pa}^5 \, W^{\hat{A}}_2 +
  \frac{96 \la^3}{175}\,  \bar{\pa}^4 \, W^{\hat{A}}_3
  + \frac{344 \la^2}{35}\,
  \bar{\pa}^3 \, W^{\hat{A}}_4
  +  \frac{4 \la}{15}\,
  \bar{\pa}^2 \, W^{\hat{A}}_5
  + 3 \, \bar{\pa} \, W^{\hat{A}}_6
  \Bigg](\bar{w})
\nonu \\
& &+ 
\cdots \ ,
\nonu \\
&& W^{\hat{A}}_4(\bar{z})\, W^{\hat{B}}_4(\bar{w})  = 
\frac{1}{(\bar{z}-\bar{w})^6}\, \frac{9216 \la^4}{5}\,
\Bigg[ \frac{1}{K}\, \de^{\hat{A} \hat{B}} W_2 + \frac{1}{2}\, d^{\hat{A}
  \hat{B} \hat{C}} \, W^{\hat{C}}_2 \, \Bigg](\bar{w})
\nonu \\
&& +
\frac{1}{(\bar{z}-\bar{w})^5}\, \Bigg[
\frac{4608 \la^4}{5}\,\Bigg(
  \frac{1}{K}\, \de^{\hat{A} \hat{B}}\, \bar{\pa} \, W_2+
 \frac{1}{2}\, d^{\hat{A}
  \hat{B} \hat{C}} \, \bar{\pa} \, W^{\hat{C}}_2 \Bigg) \nonu \\
&& +
\frac{1536 \la^4}{25}\, (-\frac{i}{2})\, f^{\hat{A}
  \hat{B} \hat{C}} \, \bar{\pa} \, W^{\hat{C}}_2
+\frac{29952 \la^3}{25}\, (-\frac{i}{2})\, f^{\hat{A}
  \hat{B} \hat{C}} \, W^{\hat{C}}_3
\Bigg](\bar{w})
\nonu \\
&& +
\frac{1}{(\bar{z}-\bar{w})^4}\, \Bigg[
\frac{1536 \la^4}{5}\,\Bigg(
  \frac{1}{K}\, \de^{\hat{A} \hat{B}}\, \bar{\pa}^2 \, W_2+
 \frac{1}{2}\, d^{\hat{A}
   \hat{B} \hat{C}} \, \bar{\pa}^2 \, W^{\hat{C}}_2 \Bigg) \nonu \\
&& +
 \frac{2304 \la^3}{125}\,\Bigg(
  \frac{1}{K}\, \de^{\hat{A} \hat{B}}\, \bar{\pa} \, W_3+
 \frac{1}{2}\, d^{\hat{A}
   \hat{B} \hat{C}} \, \bar{\pa} \, W^{\hat{C}}_3 \Bigg)
  +  \frac{8256 \la^2}{25}\,\Bigg(
  \frac{1}{K}\, \de^{\hat{A} \hat{B}} \, W_4+
 \frac{1}{2}\, d^{\hat{A}
   \hat{B} \hat{C}}  \, W^{\hat{C}}_4 \Bigg)
\nonu \\ 
&& +
\frac{768 \la^4}{25}\, (-\frac{i}{2})\, f^{\hat{A}
  \hat{B} \hat{C}} \, \bar{\pa}^2 \, W^{\hat{C}}_2
+\frac{14976 \la^3}{25}\, (-\frac{i}{2})\, f^{\hat{A}
  \hat{B} \hat{C}} \, \bar{\pa} \, W^{\hat{C}}_3
\Bigg](\bar{w})
\nonu \\
&& +
\frac{1}{(\bar{z}-\bar{w})^3}\, \Bigg[
\frac{384 \la^4}{5}\,\Bigg(
  \frac{1}{K}\, \de^{\hat{A} \hat{B}}\, \bar{\pa}^3 \, W_2+
 \frac{1}{2}\, d^{\hat{A}
   \hat{B} \hat{C}} \, \bar{\pa}^3 \, W^{\hat{C}}_2 \Bigg) \nonu \\
&& +
 \frac{1152 \la^3}{125}\,\Bigg(
  \frac{1}{K}\, \de^{\hat{A} \hat{B}}\, \bar{\pa}^2 \, W_3+
 \frac{1}{2}\, d^{\hat{A}
   \hat{B} \hat{C}} \, \bar{\pa}^2 \, W^{\hat{C}}_3 \Bigg)
 +  \frac{4128 \la^2}{25}\,\Bigg(
  \frac{1}{K}\, \de^{\hat{A} \hat{B}} \, \bar{\pa}\, W_4+
 \frac{1}{2}\, d^{\hat{A}
   \hat{B} \hat{C}}  \, \bar{\pa} \, W^{\hat{C}}_4 \Bigg)
\nonu \\ 
&& +
\frac{1152 \la^4}{125}\, (-\frac{i}{2})\, f^{\hat{A}
  \hat{B} \hat{C}} \, \bar{\pa}^3 \, W^{\hat{C}}_2
+\frac{22464\la^3}{125}\, (-\frac{i}{2})\, f^{\hat{A}
  \hat{B} \hat{C}} \, \bar{\pa}^2 \, W^{\hat{C}}_3
\nonu \\
&& +
\frac{96 \la^2}{25}\, (-\frac{i}{2})\, f^{\hat{A}
  \hat{B} \hat{C}} \, \bar{\pa} \, W^{\hat{C}}_4
+\frac{1344 \la}{25}\, (-\frac{i}{2})\, f^{\hat{A}
  \hat{B} \hat{C}}  \, W^{\hat{C}}_5
\Bigg](\bar{w})
  \nonu \\
&& +
\frac{1}{(\bar{z}-\bar{w})^2}\, \Bigg[
\frac{384 \la^4}{25}\,\Bigg(
  \frac{1}{K}\, \de^{\hat{A} \hat{B}}\, \bar{\pa}^4 \, W_2+
 \frac{1}{2}\, d^{\hat{A}
   \hat{B} \hat{C}} \, \bar{\pa}^4 \, W^{\hat{C}}_2 \Bigg) \nonu \\
&& +
 \frac{2304 \la^3}{875}\,\Bigg(
  \frac{1}{K}\, \de^{\hat{A} \hat{B}}\, \bar{\pa}^3 \, W_3+
 \frac{1}{2}\, d^{\hat{A}
   \hat{B} \hat{C}} \, \bar{\pa}^3 \, W^{\hat{C}}_3 \Bigg)
  +  \frac{8256 \la^2}{175}\,\Bigg(
  \frac{1}{K}\, \de^{\hat{A} \hat{B}} \, \bar{\pa}^2\, W_4+
 \frac{1}{2}\, d^{\hat{A}
   \hat{B} \hat{C}}  \, \bar{\pa}^2 \, W^{\hat{C}}_4 \Bigg)
 \nonu \\
&& +  \frac{8 \la}{15}\,\Bigg(
  \frac{1}{K}\, \de^{\hat{A} \hat{B}} \, \bar{\pa}\, W_5+
 \frac{1}{2}\, d^{\hat{A}
   \hat{B} \hat{C}}   \, \bar{\pa}\, W^{\hat{C}}_5 \Bigg)
 +  6 \,\Bigg(
  \frac{1}{K}\, \de^{\hat{A} \hat{B}} \, W_6+
 \frac{1}{2}\, d^{\hat{A}
   \hat{B} \hat{C}}   \,  W^{\hat{C}}_6 \Bigg)
\nonu \\ 
&& +
\frac{256 \la^4}{125}\, (-\frac{i}{2})\, f^{\hat{A}
  \hat{B} \hat{C}} \, \bar{\pa}^4 \, W^{\hat{C}}_2
+\frac{4992\la^3}{125}\, (-\frac{i}{2})\, f^{\hat{A}
  \hat{B} \hat{C}} \, \bar{\pa}^3 \, W^{\hat{C}}_3
\nonu \\
&& +
\frac{48 \la^2}{25}\, (-\frac{i}{2})\, f^{\hat{A}
  \hat{B} \hat{C}} \, \bar{\pa}^2 \, W^{\hat{C}}_4
+\frac{672 \la}{25}\, (-\frac{i}{2})\, f^{\hat{A}
  \hat{B} \hat{C}}  \, \bar{\pa}\, W^{\hat{C}}_5
\Bigg](\bar{w})
\nonu \\
&& +
\frac{1}{(\bar{z}-\bar{w})}\, \Bigg[
\frac{64 \la^4}{25}\,\Bigg(
  \frac{1}{K}\, \de^{\hat{A} \hat{B}}\, \bar{\pa}^5 \, W_2+
 \frac{1}{2}\, d^{\hat{A}
   \hat{B} \hat{C}} \, \bar{\pa}^5 \, W^{\hat{C}}_2 \Bigg) \nonu \\
&& +
 \frac{96 \la^3}{175}\,\Bigg(
  \frac{1}{K}\, \de^{\hat{A} \hat{B}}\, \bar{\pa}^4 \, W_3+
 \frac{1}{2}\, d^{\hat{A}
   \hat{B} \hat{C}} \, \bar{\pa}^4 \, W^{\hat{C}}_3 \Bigg)
 +  \frac{344 \la^2}{35}\,\Bigg(
  \frac{1}{K}\, \de^{\hat{A} \hat{B}} \, \bar{\pa}^3\, W_4+
 \frac{1}{2}\, d^{\hat{A}
   \hat{B} \hat{C}}  \, \bar{\pa}^3 \, W^{\hat{C}}_4 \Bigg)
 \nonu \\
&& +  \frac{4 \la}{15}\,\Bigg(
  \frac{1}{K}\, \de^{\hat{A} \hat{B}} \, \bar{\pa}^2 \, W_5+
 \frac{1}{2}\, d^{\hat{A}
   \hat{B} \hat{C}}   \, \bar{\pa}^2 \, W^{\hat{C}}_5 \Bigg)
 +  3 \,\Bigg(
  \frac{1}{K}\, \de^{\hat{A} \hat{B}} \, \bar{\pa} \, W_6+
 \frac{1}{2}\, d^{\hat{A}
   \hat{B} \hat{C}}   \,  \bar{\pa} \, W^{\hat{C}}_6 \Bigg)
\nonu \\ 
&& +
\frac{64 \la^4}{175}\, (-\frac{i}{2})\, f^{\hat{A}
  \hat{B} \hat{C}} \, \bar{\pa}^5 \, W^{\hat{C}}_2
+\frac{1248\la^3}{175}\, (-\frac{i}{2})\, f^{\hat{A}
  \hat{B} \hat{C}} \, \bar{\pa}^4 \, W^{\hat{C}}_3
\nonu \\
&& +
\frac{8 \la^2}{15}\, (-\frac{i}{2})\, f^{\hat{A}
  \hat{B} \hat{C}} \, \bar{\pa}^3 \, W^{\hat{C}}_4
+\frac{112 \la}{15}\, (-\frac{i}{2})\, f^{\hat{A}
  \hat{B} \hat{C}}  \, \bar{\pa}^2 \, W^{\hat{C}}_5
\nonu \\
&& +
\frac{3 }{55}\, (-\frac{i}{2})\, f^{\hat{A}
  \hat{B} \hat{C}} \, \bar{\pa} \, W^{\hat{C}}_6
+\frac{1 }{2 \, \la}\, (-\frac{i}{2})\, f^{\hat{A}
  \hat{B} \hat{C}}   \, W^{\hat{C}}_7
\Bigg](\bar{w}) + \cdots \ ,
\nonu \\
&& W_4(\bar{z})\, Q_{\frac{7}{2}}(\bar{w})  =
\frac{1}{(\bar{z}-\bar{w})^6}\, \frac{9216 \la^4}{5}\,
Q_{\frac{3}{2}}(\bar{w}) + \frac{1}{(\bar{z}-\bar{w})^5}\,
\frac{4608 \la^4}{5}\,
\bar{\pa}\, Q_{\frac{3}{2}}(\bar{w})
\nonu \\
& & + \frac{1}{(\bar{z}-\bar{w})^4}\,
\Bigg[\frac{1536 \la^4}{5}\,
  \bar{\pa}^2 \, Q_{\frac{3}{2}} +
  \frac{2304 \la^3}{125}\,  \bar{\pa} \, Q_{\frac{5}{2}}
  + \frac{8256 \la^2}{25}\,
 Q_{\frac{7}{2}}
  \Bigg](\bar{w})
\nonu \\
&&+ \frac{1}{(\bar{z}-\bar{w})^3}\,
\Bigg[\frac{384 \la^4}{5}\,
  \bar{\pa}^3 \, Q_{\frac{3}{2}} +
  \frac{1152 \la^3}{125}\,  \bar{\pa}^2 \, Q_{\frac{5}{2}}
  + \frac{4128 \la^2}{25}\,
 \bar{\pa} \, Q_{\frac{7}{2}}
  \Bigg](\bar{w})
\nonu \\
&&+ \frac{1}{(\bar{z}-\bar{w})^2}\,
\Bigg[\frac{384 \la^4}{25}\,
  \bar{\pa}^4 \, Q_{\frac{3}{2}} +
  \frac{2304 \la^3}{875}\,  \bar{\pa}^3 \, Q_{\frac{5}{2}}
  + \frac{8256 \la^2}{175}\,
  \bar{\pa}^2 \, Q_{\frac{7}{2}}
  +  \frac{8 \la}{15}\,
  \bar{\pa} \, Q_{\frac{9}{2}}
  + 6 \, Q_{\frac{11}{2}}
  \Bigg](\bar{w})
\nonu \\
&&+ \frac{1}{(\bar{z}-\bar{w})}\,
\Bigg[\frac{64 \la^4}{25}\,
  \bar{\pa}^5 \, Q_{\frac{3}{2}} +
  \frac{96 \la^3}{175}\,  \bar{\pa}^4 \, Q_{\frac{5}{2}}
  + \frac{344 \la^2}{35}\,
  \bar{\pa}^3 \, Q_{\frac{7}{2}}
  +  \frac{4 \la}{15}\,
  \bar{\pa}^2 \, Q_{\frac{9}{2}}
  + 3 \, \bar{\pa} \, Q_{\frac{11}{2}}
  \Bigg](\bar{w})
\nonu \\
&&+ 
\cdots \ ,
\nonu  \\
&& W_4(\bar{z})\, Q^{\hat{A}}_{\frac{7}{2}}(\bar{w})  = 
\frac{1}{(\bar{z}-\bar{w})^6}\, \frac{9216 \la^4}{5}\,
Q^{\hat{A}}_{\frac{3}{2}}(\bar{w}) + \frac{1}{(\bar{z}-\bar{w})^5}\,
\frac{4608 \la^4}{5}\,
\bar{\pa}\, Q^{\hat{A}}_{\frac{3}{2}}(\bar{w})
\nonu \\
&& + \frac{1}{(\bar{z}-\bar{w})^4}\,
\Bigg[\frac{1536 \la^4}{5}\,
  \bar{\pa}^2 \, Q^{\hat{A}}_{\frac{3}{2}} +
  \frac{2304 \la^3}{125}\,  \bar{\pa} \, Q^{\hat{A}}_{\frac{5}{2}}
  + \frac{8256 \la^2}{25}\,
 Q^{\hat{A}}_{\frac{7}{2}}
  \Bigg](\bar{w})
\nonu \\
& &+ \frac{1}{(\bar{z}-\bar{w})^3}\,
\Bigg[\frac{384 \la^4}{5}\,
  \bar{\pa}^3 \, Q^{\hat{A}}_{\frac{3}{2}} +
  \frac{1152 \la^3}{125}\,  \bar{\pa}^2 \, Q^{\hat{A}}_{\frac{5}{2}}
  + \frac{4128 \la^2}{25}\,
 \bar{\pa} \, Q^{\hat{A}}_{\frac{7}{2}}
  \Bigg](\bar{w})
\nonu \\
&&+ \frac{1}{(\bar{z}-\bar{w})^2}\,
\Bigg[\frac{384 \la^4}{25}\,
  \bar{\pa}^4 \, Q^{\hat{A}}_{\frac{3}{2}} +
  \frac{2304 \la^3}{875}\,  \bar{\pa}^3 \, Q^{\hat{A}}_{\frac{5}{2}}
  + \frac{8256 \la^2}{175}\,
  \bar{\pa}^2 \, Q^{\hat{A}}_{\frac{7}{2}}
  +  \frac{8 \la}{15}\,
  \bar{\pa} \, Q^{\hat{A}}_{\frac{9}{2}}
  + 6 \, Q^{\hat{A}}_{\frac{11}{2}}
  \Bigg](\bar{w})
\nonu \\
&&+ \frac{1}{(\bar{z}-\bar{w})}\,
\Bigg[\frac{64 \la^4}{25}\,
  \bar{\pa}^5 \, Q^{\hat{A}}_{\frac{3}{2}} +
  \frac{96 \la^3}{175}\,  \bar{\pa}^4 \, Q^{\hat{A}}_{\frac{5}{2}}
  + \frac{344 \la^2}{35}\,
  \bar{\pa}^3 \, Q^{\hat{A}}_{\frac{7}{2}}
  +  \frac{4 \la}{15}\,
  \bar{\pa}^2 \, Q^{\hat{A}}_{\frac{9}{2}}
  + 3 \, \bar{\pa} \, Q^{\hat{A}}_{\frac{11}{2}}
  \Bigg](\bar{w})
\nonu \\
&&+ 
\cdots \ ,
\nonu \\
&& W^{\hat{A}}_4(\bar{z})\, Q_{\frac{7}{2}}(\bar{w})  = 
\frac{1}{(\bar{z}-\bar{w})^6}\, \frac{9216 \la^4}{5}\,
Q^{\hat{A}}_{\frac{3}{2}}(\bar{w}) + \frac{1}{(\bar{z}-\bar{w})^5}\,
\frac{4608 \la^4}{5}\,
\bar{\pa}\, Q^{\hat{A}}_{\frac{3}{2}}(\bar{w})
\nonu \\
&& + \frac{1}{(\bar{z}-\bar{w})^4}\,
\Bigg[\frac{1536 \la^4}{5}\,
  \bar{\pa}^2 \, Q^{\hat{A}}_{\frac{3}{2}} +
  \frac{2304 \la^3}{125}\,  \bar{\pa} \, Q^{\hat{A}}_{\frac{5}{2}}
  + \frac{8256 \la^2}{25}\,
 Q^{\hat{A}}_{\frac{7}{2}}
  \Bigg](\bar{w})
\nonu \\
&&+ \frac{1}{(\bar{z}-\bar{w})^3}\,
\Bigg[\frac{384 \la^4}{5}\,
  \bar{\pa}^3 \, Q^{\hat{A}}_{\frac{3}{2}} +
  \frac{1152 \la^3}{125}\,  \bar{\pa}^2 \, Q^{\hat{A}}_{\frac{5}{2}}
  + \frac{4128 \la^2}{25}\,
 \bar{\pa} \, Q^{\hat{A}}_{\frac{7}{2}}
  \Bigg](\bar{w})
\nonu \\
&&+ \frac{1}{(\bar{z}-\bar{w})^2}\,
\Bigg[\frac{384 \la^4}{25}\,
  \bar{\pa}^4 \, Q^{\hat{A}}_{\frac{3}{2}} +
  \frac{2304 \la^3}{875}\,  \bar{\pa}^3 \, Q^{\hat{A}}_{\frac{5}{2}}
  + \frac{8256 \la^2}{175}\,
  \bar{\pa}^2 \, Q^{\hat{A}}_{\frac{7}{2}}
  +  \frac{8 \la}{15}\,
  \bar{\pa} \, Q^{\hat{A}}_{\frac{9}{2}}
  + 6 \, Q^{\hat{A}}_{\frac{11}{2}}
  \Bigg](\bar{w})
\nonu \\
&&+ \frac{1}{(\bar{z}-\bar{w})}\,
\Bigg[\frac{64 \la^4}{25}\,
  \bar{\pa}^5 \, Q^{\hat{A}}_{\frac{3}{2}} +
  \frac{96 \la^3}{175}\,  \bar{\pa}^4 \, Q^{\hat{A}}_{\frac{5}{2}}
  + \frac{344 \la^2}{35}\,
  \bar{\pa}^3 \, Q^{\hat{A}}_{\frac{7}{2}}
  +  \frac{4 \la}{15}\,
  \bar{\pa}^2 \, Q^{\hat{A}}_{\frac{9}{2}}
  + 3 \, \bar{\pa} \, Q^{\hat{A}}_{\frac{11}{2}}
  \Bigg](\bar{w})
\nonu \\
&&+ 
\cdots \ ,
\nonu \\
&& W^{\hat{A}}_4(\bar{z})\, Q^{\hat{B}}_{\frac{7}{2}}(\bar{w})  = 
\frac{1}{(\bar{z}-\bar{w})^6}\, \frac{9216 \la^4}{5}\,
\Bigg[ \frac{1}{K}\, \de^{\hat{A} \hat{B}} Q_{\frac{3}{2}} +
  \frac{1}{2}\, d^{\hat{A}
  \hat{B} \hat{C}} \, Q^{\hat{C}}_{\frac{3}{2}} \, \Bigg](\bar{w})
\nonu \\
&& + 
\frac{1}{(\bar{z}-\bar{w})^5}\, \Bigg[
\frac{4608 \la^4}{5}\,\Bigg(
  \frac{1}{K}\, \de^{\hat{A} \hat{B}}\, \bar{\pa} \, Q_\frac{3}{2}+
 \frac{1}{2}\, d^{\hat{A}
  \hat{B} \hat{C}} \, \bar{\pa} \, Q^{\hat{C}}_{\frac{3}{2}} \Bigg) \nonu \\
&& - 
\frac{1536 \la^4}{25}\, (\frac{i}{2})\, f^{\hat{A}
  \hat{B} \hat{C}} \, \bar{\pa} \, Q^{\hat{C}}_{\frac{3}{2}}
-\frac{29952 \la^3}{25}\, (\frac{i}{2})\, f^{\hat{A}
  \hat{B} \hat{C}} \, W^{\hat{C}}_{\frac{5}{2}}
\Bigg](\bar{w})
\nonu \\
&& + 
\frac{1}{(\bar{z}-\bar{w})^4}\, \Bigg[
\frac{1536 \la^4}{5}\,\Bigg(
  \frac{1}{K}\, \de^{\hat{A} \hat{B}}\, \bar{\pa}^2 \, Q_{\frac{3}{2}}+
 \frac{1}{2}\, d^{\hat{A}
   \hat{B} \hat{C}} \, \bar{\pa}^2 \, Q^{\hat{C}}_{\frac{3}{2}} \Bigg)
 \nonu \\
&&+
 \frac{2304 \la^3}{125}\,\Bigg(
  \frac{1}{K}\, \de^{\hat{A} \hat{B}}\, \bar{\pa} \, Q_{\frac{5}{2}}+
 \frac{1}{2}\, d^{\hat{A}
   \hat{B} \hat{C}} \, \bar{\pa} \, Q^{\hat{C}}_{\frac{5}{2}} \Bigg)
 +   \frac{8256 \la^2}{25}\,\Bigg(
  \frac{1}{K}\, \de^{\hat{A} \hat{B}} \, Q_{\frac{7}{2}}+
 \frac{1}{2}\, d^{\hat{A}
   \hat{B} \hat{C}}  \, Q^{\hat{C}}_{\frac{7}{2}} \Bigg)
\nonu \\ 
&& - 
\frac{768 \la^4}{25}\, (\frac{i}{2})\, f^{\hat{A}
  \hat{B} \hat{C}} \, \bar{\pa}^2 \, Q^{\hat{C}}_{\frac{3}{2}}
-\frac{14976 \la^3}{25}\, (\frac{i}{2})\, f^{\hat{A}
  \hat{B} \hat{C}} \, \bar{\pa} \, Q^{\hat{C}}_{\frac{5}{2}}
\Bigg](\bar{w})
\nonu \\
&& + 
\frac{1}{(\bar{z}-\bar{w})^3}\, \Bigg[
\frac{384 \la^4}{5}\,\Bigg(
  \frac{1}{K}\, \de^{\hat{A} \hat{B}}\, \bar{\pa}^3 \, Q_{\frac{3}{2}}+
 \frac{1}{2}\, d^{\hat{A}
   \hat{B} \hat{C}} \, \bar{\pa}^3 \, Q^{\hat{C}}_{\frac{3}{2}}
 \Bigg) \nonu \\
&&+
 \frac{1152 \la^3}{125}\,\Bigg(
  \frac{1}{K}\, \de^{\hat{A} \hat{B}}\, \bar{\pa}^2 \, Q_{\frac{5}{2}}+
 \frac{1}{2}\, d^{\hat{A}
   \hat{B} \hat{C}} \, \bar{\pa}^2 \, Q^{\hat{C}}_{\frac{5}{2}} \Bigg)
  + \frac{4128 \la^2}{25}\,\Bigg(
  \frac{1}{K}\, \de^{\hat{A} \hat{B}} \, \bar{\pa}\, Q_{\frac{7}{2}}+
 \frac{1}{2}\, d^{\hat{A}
   \hat{B} \hat{C}}  \, \bar{\pa} \, Q^{\hat{C}}_{\frac{7}{2}} \Bigg)
\nonu \\ 
&& - 
\frac{1152 \la^4}{125}\, (\frac{i}{2})\, f^{\hat{A}
  \hat{B} \hat{C}} \, \bar{\pa}^3 \, Q^{\hat{C}}_{\frac{3}{2}}
-\frac{22464\la^3}{125}\, (\frac{i}{2})\, f^{\hat{A}
  \hat{B} \hat{C}} \, \bar{\pa}^2 \, Q^{\hat{C}}_{\frac{5}{2}}
\nonu \\
&& - 
\frac{96 \la^2}{25}\, (\frac{i}{2})\, f^{\hat{A}
  \hat{B} \hat{C}} \, \bar{\pa} \, Q^{\hat{C}}_{\frac{7}{2}}
-\frac{1344 \la}{25}\, (\frac{i}{2})\, f^{\hat{A}
  \hat{B} \hat{C}}  \, Q^{\hat{C}}_{\frac{9}{2}}
\Bigg](\bar{w})
  \nonu \\
&& + 
\frac{1}{(\bar{z}-\bar{w})^2}\, \Bigg[
\frac{384 \la^4}{25}\,\Bigg(
  \frac{1}{K}\, \de^{\hat{A} \hat{B}}\, \bar{\pa}^4 \, Q_{\frac{3}{2}}+
 \frac{1}{2}\, d^{\hat{A}
   \hat{B} \hat{C}} \, \bar{\pa}^4 \, Q^{\hat{C}}_{\frac{3}{2}}
 \Bigg) \nonu \\
&&+
 \frac{2304 \la^3}{875}\,\Bigg(
  \frac{1}{K}\, \de^{\hat{A} \hat{B}}\, \bar{\pa}^3 \, Q_{\frac{5}{2}}+
 \frac{1}{2}\, d^{\hat{A}
   \hat{B} \hat{C}} \, \bar{\pa}^3 \, Q^{\hat{C}}_{\frac{5}{2}} \Bigg)
 +   \frac{8256 \la^2}{175}\,\Bigg(
  \frac{1}{K}\, \de^{\hat{A} \hat{B}} \, \bar{\pa}^2\, Q_{\frac{7}{2}}+
 \frac{1}{2}\, d^{\hat{A}
   \hat{B} \hat{C}}  \, \bar{\pa}^2 \, Q^{\hat{C}}_{\frac{7}{2}} \Bigg)
 \nonu \\
&& +   \frac{8 \la}{15}\,\Bigg(
  \frac{1}{K}\, \de^{\hat{A} \hat{B}} \, \bar{\pa}\, Q_{\frac{9}{2}}+
 \frac{1}{2}\, d^{\hat{A}
   \hat{B} \hat{C}}   \, \bar{\pa}\, Q^{\hat{C}}_{\frac{9}{2}} \Bigg)
 +   6 \,\Bigg(
  \frac{1}{K}\, \de^{\hat{A} \hat{B}} \, Q_{\frac{11}{2}}+
 \frac{1}{2}\, d^{\hat{A}
   \hat{B} \hat{C}}   \,  Q^{\hat{C}}_{\frac{11}{2}} \Bigg)
\nonu \\ 
&& - 
\frac{256 \la^4}{125}\, (\frac{i}{2})\, f^{\hat{A}
  \hat{B} \hat{C}} \, \bar{\pa}^4 \, Q^{\hat{C}}_{\frac{3}{2}}
-\frac{4992\la^3}{125}\, (\frac{i}{2})\, f^{\hat{A}
  \hat{B} \hat{C}} \, \bar{\pa}^3 \, Q^{\hat{C}}_{\frac{5}{2}}
\nonu \\
&& - 
\frac{48 \la^2}{25}\, (\frac{i}{2})\, f^{\hat{A}
  \hat{B} \hat{C}} \, \bar{\pa}^2 \, Q^{\hat{C}}_{\frac{7}{2}}
-\frac{672 \la}{25}\, (\frac{i}{2})\, f^{\hat{A}
  \hat{B} \hat{C}}  \, \bar{\pa}\, Q^{\hat{C}}_{\frac{9}{2}}
\Bigg](\bar{w})
\nonu \\
&& + 
\frac{1}{(\bar{z}-\bar{w})}\, \Bigg[
\frac{64 \la^4}{25}\,\Bigg(
  \frac{1}{K}\, \de^{\hat{A} \hat{B}}\, \bar{\pa}^5 \, Q_{\frac{3}{2}}+
 \frac{1}{2}\, d^{\hat{A}
   \hat{B} \hat{C}} \, \bar{\pa}^5 \, Q^{\hat{C}}_{\frac{3}{2}} \Bigg)
 \nonu \\
&&+
 \frac{96 \la^3}{175}\,\Bigg(
  \frac{1}{K}\, \de^{\hat{A} \hat{B}}\, \bar{\pa}^4 \, Q_{\frac{5}{2}}+
 \frac{1}{2}\, d^{\hat{A}
   \hat{B} \hat{C}} \, \bar{\pa}^4 \, Q^{\hat{C}}_{\frac{5}{2}} \Bigg)
 +   \frac{344 \la^2}{35}\,\Bigg(
  \frac{1}{K}\, \de^{\hat{A} \hat{B}} \, \bar{\pa}^3\, Q_{\frac{7}{2}}+
 \frac{1}{2}\, d^{\hat{A}
   \hat{B} \hat{C}}  \, \bar{\pa}^3 \, Q^{\hat{C}}_{\frac{7}{2}} \Bigg)
 \nonu \\
&& +   \frac{4 \la}{15}\,\Bigg(
  \frac{1}{K}\, \de^{\hat{A} \hat{B}} \, \bar{\pa}^2 \, Q_{\frac{9}{2}}+
 \frac{1}{2}\, d^{\hat{A}
   \hat{B} \hat{C}}   \, \bar{\pa}^2 \, Q^{\hat{C}}_{\frac{9}{2}} \Bigg)
 +   3 \,\Bigg(
  \frac{1}{K}\, \de^{\hat{A} \hat{B}} \, \bar{\pa} \, Q_{\frac{11}{2}}+
 \frac{1}{2}\, d^{\hat{A}
   \hat{B} \hat{C}}   \,  \bar{\pa} \, Q^{\hat{C}}_{\frac{11}{2}} \Bigg)
\nonu \\ 
&& - 
\frac{64 \la^4}{175}\, (\frac{i}{2})\, f^{\hat{A}
  \hat{B} \hat{C}} \, \bar{\pa}^5 \, Q^{\hat{C}}_{\frac{3}{2}}
-\frac{1248\la^3}{175}\, (\frac{i}{2})\, f^{\hat{A}
  \hat{B} \hat{C}} \, \bar{\pa}^4 \, Q^{\hat{C}}_{\frac{5}{2}}
\nonu \\
&& - 
\frac{8 \la^2}{15}\, (\frac{i}{2})\, f^{\hat{A}
  \hat{B} \hat{C}} \, \bar{\pa}^3 \, Q^{\hat{C}}_{\frac{7}{2}}
-\frac{112 \la}{15}\, (\frac{i}{2})\, f^{\hat{A}
  \hat{B} \hat{C}}  \, \bar{\pa}^2 \, Q^{\hat{C}}_{\frac{9}{2}}
\nonu \\
&& - 
\frac{3 }{55}\, (\frac{i}{2})\, f^{\hat{A}
  \hat{B} \hat{C}} \, \bar{\pa} \, Q^{\hat{C}}_{\frac{11}{2}}
-\frac{1 }{2 \, \la}\, (\frac{i}{2})\, f^{\hat{A}
  \hat{B} \hat{C}}   \, Q^{\hat{C}}_{\frac{13}{2}}
\Bigg](\bar{w}) + \cdots \ .
\label{sevenope}
\eea
Note that the structure constants appearing in the third
and last equations in Appendix (\ref{sevenope})
are common. Of course, the corresponding currents in the
right hand sides are different from each other.
We will see in next subsection
that the sum of structure constants
$\hat{q}^{h_1,h_2+\frac{1}{2},h}(m,n)$
and  $\dot{q}^{h_1,h_2+\frac{1}{2},h-1}(m,n)$ is equal to
$-\tilde{q}^{h_1,h_2+1,h}(m,n)$ up to the degree $(h+1)$ of the polynomial
and
the sum of structure constants $\check{q}^{h_1,h_2+\frac{1}{2},h}(m,n)$
and  $\bar{q}^{h_1,h_2+\frac{1}{2},h-1}(m,n)$ is equal to
the structure constant
$q^{h_1,h_2+1,h}(m,n)$ up to the degree $(h+1)$ of the polynomial.
This implies that in the OPE language, the independent
structure constants
are given by (\ref{qqtilde}).

\subsection{The structure constants for fixed
$h_1$, $h_2$}

Let us check whether the equations (\ref{OPES}) are consistent with
Appendix (\ref{sevenope}) obtained
from the free field realizations for fixed $h_1$ and $h_2$.
First of all, we need to obtain the following possible
polynomials explicitly
\bea
q^{4,4,4}(m,n) \!& = \!&
\frac{4}{25}
\Big(16 m^5-16 m^4 n+32 m^4+16 m^3 n^2-16 m^3 n-12 m^3-16 m^2 n^3
+12 m^2 n\nonu \\
\!& + \!&
4 m^2+16 m n^4+16 m n^3-4 m n-m-16 n^5-32 n^4-24 n^3-8 n^2-n \Big)
\ ,
\nonu \\
q^{4,4,3}(m,n) \!& = \!&
-\frac{6}{875}
\Big(80 m^4-64 m^3 n+288 m^3-384 m^2 n-192 m^2+64 m n^3+384 m n^2\nonu \\
\!& + \!& 336 m n+80 m-80 n^4-288 n^3-312 n^2-136 n-21
\Big)\ ,
\nonu \\
q^{4,4,2}(m,n) \!& = \!&
\frac{86}{175}
\Big(20 m^3-36 m^2 n+12 m^2+36 m n^2-9 m-20 n^3-12 n^2+3 n+2\Big)
\ ,
\nonu \\
q^{4,4,1}(m,n) \!& = \!&
\frac{1}{15} \Big(-4 m^2-20 m+4 n^2+20 n+9\Big) \ , \qquad
q^{4,4,0}(m,n)  = 
3 (m-n) \ ,
\nonu \\
\tilde{q}^{4,4,4}(m,n) \!& = \!&
-\frac{4}{875}  \Big(80 m^5-48 m^4 n+256 m^4+16 m^3 n^2-208 m^3 n-108 m^3+16 m^2 n^3\nonu \\
\!& + \!&
192 m^2 n^2+180 m^2 n+44 m^2-48 m n^4-208 m n^3-240 m n^2-108 m n
-17 m \nonu \\
\!& + \!& 80 n^5+256 n^4+312 n^3+184 n^2+53 n+6\Big)
\ ,
\nonu \\
\tilde{q}^{4,4,3}(m,n) \!& = \!&
\frac{78}{875} \Big(80 m^4-128 m^3 n+96 m^3+144 m^2 n^2-48 m^2 n-60 m^2-128 m n^3\nonu \\
\!& - \!& 48 m n^2+48 m n+20 m+80 n^4+96 n^3+24 n^2-8 n-3\Big)
\ ,
\nonu \\
\tilde{q}^{4,4,2}(m,n) \!& = \!&
-\frac{2}{75}  \Big(20 m^3-12 m^2 n+84 m^2-12 m n^2-120 m n-57 m+20 n^3+84 n^2\nonu \\
\!& + \!& 69 n+16\Big) \ ,
\nonu \\
\tilde{q}^{4,4,1}(m,n) \!& = \!&
\frac{28}{75} \Big(20 m^2-32 m n+4 m+20 n^2+4 n-3\Big)\ ,
\nonu \\
\tilde{q}^{4,4,0}(m,n)  \!& = \!& 
-\frac{3}{55}  (m+n+6) \ ,
\qquad
\tilde{q}^{4,4,-1}(m,n)  =  \frac{1}{2} \ ,
\nonu \\
\hat{q}^{4,\frac{7}{2},4}(m,n)
\!& = \!& -\frac{32}{525}  \Big(m^5-2 m^4 n+3 m^3 n^2-4 m^2 n^3+5 m n^4-6 n^5\Big)\ ,
\nonu \\
\hat{q}^{4,\frac{7}{2},3}(m,n)
\!& = \!& -\frac{416}{875} 
\Big(5 m^4-12 m^3 n+18 m^2 n^2-20 m n^3+15 n^4\Big) \ , 
\nonu \\
\hat{q}^{4,\frac{7}{2},2}(m,n)
\!& = \!&
-\frac{4}{75}  \Big(5 m^3-12 m^2 n+15 m n^2-10 n^3\Big) \ ,
\nonu \\
\hat{q}^{4,\frac{7}{2},1}(m,n)
\!& = \!& -\frac{112}{45}  \Big(2 m^2-4 m n+3 n^2\Big) \ ,
\qquad
\hat{q}^{4,\frac{7}{2},0}(m,n)
 =  \frac{1}{110} (6 n-5 m) \ ,
\nonu \\
\hat{q}^{4,\frac{7}{2},-1}(m,n)
\!& = \!& -\frac{1}{2}\ ,
\nonu \\
\dot{q}^{4,\frac{7}{2},3}(m,n) \!& = \!&
\frac{32}{375} (m+3) \Big(5 m^4-4 m^3 n+3 m^2 n^2-2 m n^3+n^4
\Big) \ ,
\nonu \\
\dot{q}^{4,\frac{7}{2},2}(m,n) \!& = \!&
-\frac{416}{875}  (m+3) \Big(10 m^3-12 m^2 n+9 m n^2-4 n^3
\Big) \ ,
\nonu \\
\dot{q}^{4,\frac{7}{2},1}(m,n) \!& = \!&
\frac{4}{25} (m+3) \Big(5 m^2-6 m n+3 n^2\Big) \ ,
\nonu \\
\dot{q}^{4,\frac{7}{2},0}(m,n) \!& = \!&
-\frac{112}{225}  (m+3) \Big(5 m-4 n\Big) \ ,
\qquad
\dot{q}^{4,\frac{7}{2},-1}(m,n)  = 
\frac{1}{10} (m+3) \ ,
\nonu \\
\dot{q}^{4,\frac{7}{2},-2}(m,n) \!& = \!& 0 \ ,
\nonu \\
\check{q}^{4,\frac{7}{2},4}(m,n) \!& = \!&
\frac{32}{75} \Big(m^5-2 m^4 n+3 m^3 n^2-4 m^2 n^3+5 m n^4-6 n^5
\Big)\ ,
\nonu \\
\check{q}^{4,\frac{7}{2},3}(m,n) \!& = \!&
\frac{32}{875} \Big(5 m^4-12 m^3 n+18 m^2 n^2-20 m n^3+15 n^4
\Big) \ ,
\nonu \\
\check{q}^{4,\frac{7}{2},2}(m,n) \!& = \!&
\frac{172}{175} \Big(5 m^3-12 m^2 n+15 m n^2-10 n^3\Big) \ ,
\nonu \\
\check{q}^{4,\frac{7}{2},1}(m,n) \!& = \!&
\frac{4}{45} \Big(2 m^2-4 m n+3 n^2\Big) \ ,
\qquad
\check{q}^{4,\frac{7}{2},0}(m,n)  = 
\frac{1}{2} (5 m-6 n) \ ,
\nonu \\
\check{q}^{4,\frac{7}{2},-1}(m,n) \!& = \!& 0 \ ,
\nonu \\
\bar{q}^{4,\frac{7}{2},3}(m,n) \!& = \!&
\frac{32}{75} (m+3) \Big(5 m^4-4 m^3 n+3 m^2 n^2-2 m n^3+n^4\Big)\ ,
\nonu \\
\bar{q}^{4,\frac{7}{2},2}(m,n) \!& = \!&
-\frac{64}{875}  (m+3) \Big(10 m^3-12 m^2 n+9 m n^2-4 n^3\Big)\ ,
\nonu \\
\bar{q}^{4,\frac{7}{2},1}(m,n) \!& = \!&
\frac{172}{175} (m+3) \Big(5 m^2-6 m n+3 n^2\Big) \ ,
\nonu \\
\bar{q}^{4,\frac{7}{2},0}(m,n) \!& = \!&
-\frac{4}{45}  (m+3) (5 m-4 n) \ ,
\qquad
\bar{q}^{4,\frac{7}{2},-1}(m,n)  = 
\frac{1}{2} (m+3) \ ,
\nonu \\
\bar{q}^{4,\frac{7}{2},-2}(m,n) \!& = \!& 0 \ .
\label{mnexp}
\eea
Note that all the terms in
some of these structure constants have the degree
$(h+1)$ while some terms in other structure constants
have the degree $(h+1)$.

Let us focus on the $W_2(\bar{w})$ in right hand side
of the first OPE of
Appendix (\ref{sevenope}).
Then from the explicit form of the first OPE,
we should calculate $\la^4 \, (-1)^{3}\, f^{4,4,4}(\bar{\pa}_{\bar{z}},
\bar{\pa}_{\bar{w}}) \, \Bigg[\frac{W_2(\bar{w})}{(\bar{z}-\bar{w})}
  \Bigg]$. Here $ f^{4,4,4}(\bar{\pa}_{\bar{z}},
\bar{\pa}_{\bar{w}})$ can be obtained from
$q^{4,4,4}(m,n)$ in Appendix (\ref{mnexp}) under the constraint by taking
the terms having a degree $(h+1)=5$ and $m$ and $n$ are replaced by
$\bar{\pa}_{\bar{z}}$ and 
$\bar{\pa}_{\bar{w}}$
respectively.
Then we have
\bea
q^{4,4,4}(m, n) \rightarrow \frac{4}{25}
\Big(16 m^5-16 m^4 n+16 m^3 n^2-16 m^2 n^3
+16 m n^4-16 n^5 \Big).
\label{reducedq}
\eea
This is due to the fact that
$q^{h_1,h_2,h}(m, n)$ has the second and the fourth terms in
(\ref{qqtilde}) from the derivative terms in the current.
The corresponding
$f^{h_1,h_2,h}(m, n)$ has similar terms in (\ref{struct}).

The corresponding differential operator
$ f^{4,4,4}(\bar{\pa}_{\bar{z}},
\bar{\pa}_{\bar{w}})$ from Appendix (\ref{reducedq}) is given by
\bea
f^{4,4,4}(\bar{\pa}_{\bar{z}},
\bar{\pa}_{\bar{w}}) \rightarrow \frac{4}{25}
\Big(16 \bar{\pa}_{\bar{z}}^5-16 \bar{\pa}_{\bar{z}}^4 \bar{\pa}_{\bar{w}}+16 \bar{\pa}_{\bar{z}}^3 \bar{\pa}_{\bar{w}}^2-16 \bar{\pa}_{\bar{z}}^2 \bar{\pa}_{\bar{w}}^3
+16 \bar{\pa}_{\bar{z}} \bar{\pa}_{\bar{w}}^4-16 \bar{\pa}_{\bar{w}}^5 \Big).
\label{expf}
\eea
The next thing is
to calculate
$-\la^4 \, f^{4,4,4}(\bar{\pa}_{\bar{z}},
\bar{\pa}_{\bar{w}}) \, \Bigg[\frac{W_2(\bar{w})}{(\bar{z}-\bar{w})}
\Bigg]$ from Appendix (\ref{expf}). It turns out that
\bea
\!& \!& \frac{1}{(\bar{z}-\bar{w})^6}\,
\frac{9216 \la^4}{5}\,
W_2(\bar{w}) + \frac{1}{(\bar{z}-\bar{w})^5}\,
\frac{4608 \la^4}{5}\,
\bar{\pa}\, W_2(\bar{w})
+ \frac{1}{(\bar{z}-\bar{w})^4}\,
\frac{1536 \la^4}{5}\,
\bar{\pa}^2 \, W_2 (\bar{w})
\label{finalexp}
  \\
\!& \!& + \frac{1}{(\bar{z}-\bar{w})^3}\,
\frac{384 \la^4}{5}\,
  \bar{\pa}^3 \, W_2 (\bar{w})
 + \frac{1}{(\bar{z}-\bar{w})^2}\,
\frac{384 \la^4}{25}\,
  \bar{\pa}^4 \, W_2 (\bar{w})
 + \frac{1}{(\bar{z}-\bar{w})}\,
\frac{64 \la^4}{25}\,
  \bar{\pa}^5 \, W_2(\bar{w}) \ .
\nonu
\eea
This Appendix
(\ref{finalexp}) are exactly the terms with $h_1+h_2-2-h=2$ appearing
in the first OPE of Appendix (\ref{sevenope}).

In this way, we can check that the equations (\ref{OPES})
are right seven OPEs.

\section{ Other OPEs for soft currents
in the supersymmetric Einstein-Yang-Mills theory
}

In this Appendix, we present some details which are related to
the contents in section $3$.

According to (\ref{WWn=1}) and (\ref{qqtilde}),
there is a shift in the weight $h_2$
coming from the quantity $(h_2-\frac{1}{2})$ in the
structure constant, 
therefore
we cannot use  the previous result in (\ref{Volexpression})
directly.
This is due to the fact that
the corresponding structure constant is written in terms of
the one in previous commutator in the first equation of
(\ref{WQn=1}). They are equivalent to each other \cite{PRSS}.
We should find out the right basis where
the structure constant can be obtained from the previous
relation like as (\ref{Volexpression}).
Therefore, we return to the first equation of (\ref{WWn=1})
and read off the structure constant in terms of $p_B$ and $p_F$
rather than $q_B$ and $q_F$. By substituting
the current in the first equation of
(\ref{WWhat}) into the first equation of
(\ref{WWn=1}), then there are eight commutators in the left hand side
and there are
four current terms in the right hand side. We can substitute
the commutators in the equations (\ref{OSalgebra}) and (\ref{BKalgebra})
into the above expression and 
collect each independent term.
Then we obtain
the structure constants in terms of $p_B$ or $p_F$ explicitly.
That is,
\bea
\!& \!& q^{h_1,h_2,h}(m,n)\Bigg|_{h, \mbox{ \footnotesize odd}}
 =   \Bigg[ \frac{2(h_1-2)(m+(h_1-2)+1)}{2(h_1-2)+1} \Bigg]\,
p_B^{h_1-1,h_2,h-1}(m,n) \nonu \\
\!&  \!& + \Bigg[ \frac{2(h_2-2)(n+(h_2-2)+1)}{2(h_2-2)+1}\Bigg] \,
p_B^{h_1,h_2-1,h-1}(m,n)
\nonu \\
\!&  \!& - \Bigg[ \frac{2(h_1+h_2-2-h-1)(m+n+h_1+h_2-2-h-1+1)}{
  2(h_1+h_2-2-h-1)+1} \Bigg] \,
q^{h_1,h_2,h-1}(m,n) \ ,
\nonu \\
\!& \!& 
 =   -\Bigg[\frac{(2(h_1-2)+2)(m+(h_1-2)+1)}{2(h_1-2)+1}\Bigg] \,
p_F^{h_1-1,h_2,h-1}(m,n) \nonu \\
\!&  \!& - \Bigg[ \frac{(2(h_2-2)+2)(n+(h_2-2)+1)}{2(h_2-2)+1}\Bigg]
\,
p_F^{h_1,h_2-1,h-1}(m,n)
\nonu \\
\!&  \!& + \Bigg[ \frac{(2(h_1+h_2-2-h-1)+2)(m+n+h_1+h_2-2-h-1+1)}{
  2(h_1+h_2-2-h-1)+1} \Bigg] \,
q^{h_1,h_2,h-1}(m,n) \ ,
\nonu \\
\!& \!& q^{h_1,h_2,h}(m,n)\Bigg|_{h, \mbox{ \footnotesize even}}
 =  
p_B^{h_1,h_2,h}(m,n) \nonu \\
\!&  \!& +  \Bigg[  \frac{2(h_1-2)(m+(h_1-2)+1)}{2(h_1-2)+1}\,
\frac{2(h_2-2)(n+(h_2-2)+1)}{2(h_2-2)+1} \Bigg] \, 
p_B^{h_1-1,h_2-1,h-2}(m,n)
\nonu \\
\!&  \!& - \Bigg[
  \frac{2(h_1+h_2-2-h-1)(m+n+h_1+h_2-2-h-1+1)}{2(h_1+h_2-2-h-1)+1}
  \Bigg] \,
q^{h_1,h_2,h-1}(m,n)\nonu \\
\!& \!&
= p_F^{h_1,h_2,h}(m,n) +
\Bigg[  \frac{(2(h_1-2)+2)(m+(h_1-2)+1)}{2(h_1-2)+1}
  \nonu \\
\!& \!&  \times
\frac{(2(h_2-2)+2)(n+(h_2-2)+1)}{2(h_2-2)+1} \Bigg]\, 
p_F^{h_1-1,h_2-1,h-2}(m,n)
\label{STRUCT}
\\
\!&  \!& +\Bigg[
  \frac{(2(h_1+h_2-2-h-1)+2)(m+n+h_1+h_2-2-h-1+1)}{2(h_1+h_2-2-h-1)+1}
  \Bigg] \,
q^{h_1,h_2,h-1}(m,n) \ .
\nonu
\eea
Note that $q^{h_1,h_2,0}(m,n)= p_B^{h_1,h_2,0}(m,n)= p_F^{h_1,h_2,0}(m,n)$,
which appears in (\ref{firstalgebra}), because
the other terms in (\ref{STRUCT}) vanish.
Although there appear the unwanted terms
$(m+n+h_1+h_2-2-h)\, q^{h_1,h_2,h-1}(m,n)$
in the right hand side of (\ref{STRUCT}) because
it is not obvious how we can deal with the mode dependent piece
with a factor $(m+n+h_1+h_2-2-h+)$ and others,  
we can express this as the linear combination of
$p_B$ and $p_F$ by realizing that
the relative coefficients of these unwanted terms are
different from each other and they have common behavior of above
$(m+n+h_1+h_2-2-h)$-dependent factor.
Then we can write down it in terms of other wanted terms
by solving each two equations in (\ref{STRUCT}).
After substituting $(m+n+h_1+h_2-2-h)\,
q^{h_1,h_2,h-1}(m,n)$ (for odd and even cases) 
written in terms of the structure constants $p_B$ and $p_F$
(Note that these structure constants terms contain only $m$ or
$n$ dependence in their coefficients)
into the above equations back
then we determine the following
relations \footnote{Compared to the first equation of (\ref{OSalgebra})
  and the first equation (\ref{BKalgebra}), as described before, the
  $h=0$ case leads to the result
  of  $q^{h_1,h_2,0}(m,n)= p_B^{h_1,h_2,0}(m,n)= p_F^{h_1,h_2,0}(m,n)$.
  But for general $h$, they are different from each other. The range
  for $h$ is also different. Note that the sum of coefficients of
  $p_B^{h_1,h_2,h}(m,n)$ and $p_F^{h_1,h_2,h}(m,n)$ in the second equation
of (\ref{qotherexp}) is equal to $1$. }
\bea
\!& \!& q^{h_1,h_2,h}(m,n)\Bigg|_{h, \mbox{ \footnotesize odd}}
= \nonu \\
\!& \!& 
\Bigg[ \frac{2(h_2-2)(n+(h_2-2)+1)}{2(h_2-2)+1}\,
\frac{(h_1+h_2-2-h)}{2(h_1+h_2-2-h-1)+1}
 \Bigg] \,p_B^{h_1,h_2-1,h-1}(m,n)
  \nonu \\
\!& \!&\Bigg[ \frac{2(h_1-2)(m+(h_1-2)+1)}{2(h_1-2)+1} \,
  \frac{(h_1+h_2-2-h)}{2(h_1+h_2-2-h-1)+1} \Bigg] \,
  p_B^{h_1-1,h_2,h-1}(m,n)
  \nonu \\
  \!& \!& -
  \Bigg[ \frac{(h_1+h_2-2-h-1)}{2(h_1+h_2-2-h-1)+1}\,
  \frac{(2(h_2-2)+2)(n+(h_2-2)+1)}{2(h_2-2)+1} \Bigg] \,
  p_F^{h_1,h_2-1,h-1}(m,n)
  \nonu \\
   \!& \!& -\Bigg[
  \frac{(h_1+h_2-2-h-1)}{2(h_1+h_2-2-h-1)+1} \,
 \frac{(2(h_1-2)+2)(m+(h_1-2)+1)}{2(h_1-2)+1}\Bigg]  \, 
p_F^{h_1-1,h_2,h-1}(m,n) \ ,
\nonu \\
\!& \!& q^{h_1,h_2,h}(m,n)\Bigg|_{h, \mbox{ \footnotesize even}}
= \nonu \\
\!& \!& \Bigg[
\frac{(h_1+h_2-2-h)}{2(h_1+h_2-2-h-1)+1} \Bigg] \,  p_B^{h_1,h_2,h}(m,n)
 +   \Bigg[ \frac{2(h_1-2)(m+(h_1-2)+1)}{2(h_1-2)+1}\,
 \nonu \\
 \!& \! & \times \frac{2(h_2-2)(n+(h_2-2)+1)}{2(h_2-2)+1}\, 
\frac{(h_1+h_2-2-h)}{2(h_1+h_2-2-h-1)+1}\Bigg]\, p_B^{h_1-1,h_2-1,h-2}(m,n)
\label{qotherexp}
 \\
  \!& \!& +\Bigg[ \frac{(h_1+h_2-2-h-1)}{2(h_1+h_2-2-h-1)+1}
    \Bigg] \,
  p_F^{h_1,h_2,h}(m,n)
+ \Bigg[ \frac{(2(h_1-2)+2)(m+(h_1-2)+1)}{2(h_1-2)+1}\,
\nonu
\\
\!& \!& \times \frac{(2(h_2-2)+2)(n+(h_2-2)+1)}{2(h_2-2)+1}
\,
\frac{(h_1+h_2-2-h-1)}{2(h_1+h_2-2-h-1)+1}\Bigg]\,
p_F^{h_1-1,h_2-1,h-2}(m,n)
\ .
\nonu
\eea
In (\ref{qotherexp}),
all the mode dependent terms are given by
either $(m+(h_1-2)+1)$ or $(n+(h_2-2)+1)$. We remove the previous
unwanted $(m+n+h_1+h_2-2-h)$ dependence completely.
Instead of using the previous
relations for the structure constants (\ref{qqtilde})
which is appropriate for the first example or the OPEs including
the gravitino and gluino, we use
the above basis (\ref{qotherexp}) for the structure constants
which is more appropriate for the OPEs including the graviton
and gluon.
Therefore, the structure constant consists of both $p_B$ and $p_F$
coming from (\ref{BKalgebra}) and (\ref{OSalgebra}) respectively
and the corresponding weights for the first current
are either $h_1$ or $(h_1-1)$
while the corresponding weights for the second current
are either $h_2$ or $(h_2-1)$.
This is reasonable because from (\ref{FREE1})
we allow to have the first derivative terms.
For the dummy variable $h$, the corresponding weight
is given by $h$, $(h-1)$ or $(h-2)$.
The last one occurs when we consider the commutators
where the corresponding
two currents contain each derivative term.

\subsection{The OPE between the graviton and the gluino}

The OPE between the conformally soft gravitons and the gluinos
where
the weights in the antiholomorphic sector are given by
$h_1=\frac{k-2}{2}$ and $h_2=\frac{l-\frac{1}{2}}{2}$
can be expressed as 
\bea
\!& \!& H^k(z_1,\bar{z}_1)\, L^{l,\hat{A}}(z_2,\bar{z}_2)
=  -\frac{\kappa}{2} \,
\frac{1}{z_{12}} \,
\sum_{h=0}^{h_1+h_2-3}\, (-1)^{h+1}\, \la^h \,
\Bigg[ q_B^{h_1,h_2+\frac{1}{2},h} +q_F^{h_1,h_2+\frac{1}{2},h} \Bigg] \, 
\nonu \\
\!& \!& \times 
\sum_{n=0}^{\infty}\,
\left(
\begin{array}{c}
\frac{1}{2}-2 h-k-l-n \\
-\frac{1}{2}-h-l
\end{array}
\right)\, \frac{\bar{z}_{12}^{n+h+1}}{n!}\, \bar{\pa}^n \,
L^{k+l+h,\hat{A}}(z_2, \bar{z}_2) \ 
\nonu \\
\!& \! & -\frac{\kappa}{2} \,
\frac{1}{z_{12}} \,
\sum_{h=0}^{h_1+h_2-3}\, (-1)^{h}\,\la^{h} \,
\Bigg[  \frac{2(h_1-2)}{2(h_1-2)+1} \,
q_B^{h_1-1,h_2+\frac{1}{2},h-1}-
\frac{2(h_1-2)+2}{2(h_1-2)+1} \,q_F^{h_1-1,h_2+\frac{1}{2},h-1}
\Bigg] \, 
\nonu \\
\!& \!& \times 
\sum_{n=0}^{\infty}\,
\left(
\begin{array}{c}
\frac{1}{2}-2(h-1)-(k-2)-l-n \\
-\frac{1}{2}-(h-1)-l
\end{array}
\right)\, \frac{\bar{\pa}_{\bar{z_1}}\,
  \bar{z}_{12}^{n+(h-1)+1}}{n!}\, \bar{\pa}^n \,
L^{(k-2)+l+(h-1),\hat{A}}(z_2, \bar{z}_2) \nonu \\
\!& \!& + \cdots \ .
\label{fifthres}
\eea
The numerical factor $\frac{1}{2}$
of the
first binomial coefficient is given by
$(2+\frac{1}{2})$
minus $2$.
The numerical factor $-\frac{1}{2}$
in the second line of the
first binomial coefficient
is given by $\frac{1}{2}$
minus $1$ \footnote{As mentioned before,
  the general form for the element of
  binomial coefficient in the footnote \ref{for} can be used.}.
The first binomial coefficient above can be obtained
from (\ref{HH1}) by taking $l \rightarrow (l+\frac{3}{2})$.

\subsection{The OPE between the gluon and the gravitino}

The OPE between the conformally soft gluons and the gravitinos
where
the weights in the antiholomorphic sector are given by
$h_1=\frac{k-1}{2}$ and $h_2=\frac{l-\frac{3}{2}}{2}$
can be described as
\bea
\!& \!& R^{k,\hat{A}}(z_1,\bar{z}_1)\, I^l(z_2,\bar{z}_2)
=  -\frac{\kappa}{2} \,
\frac{1}{z_{12}} \,
\sum_{h=0}^{h_1+h_2-3}\, (-1)^{h+1}\, \la^h \,
\Bigg[ q_B^{h_1,h_2+\frac{1}{2},h} +q_F^{h_1,h_2+\frac{1}{2},h} \Bigg] \, 
\nonu \\
\!& \!& \times 
\sum_{n=0}^{\infty}\,
\left(
\begin{array}{c}
\frac{1}{2}-2 h-k-l-n \\
\frac{1}{2}-h-l
\end{array}
\right)\, \frac{\bar{z}_{12}^{n+h+1}}{n!}\, \bar{\pa}^n \,
L^{k+l+h,\hat{A}}(z_2, \bar{z}_2) \ 
\nonu \\
\!& \! & -\frac{\kappa}{2} \,
\frac{1}{z_{12}} \,
\sum_{h=0}^{h_1+h_2-3}\, (-1)^{h}\,\la^{h} \,
\Bigg[  \frac{2(h_1-2)}{2(h_1-2)+1} \,
q_B^{h_1-1,h_2+\frac{1}{2},h-1}-
\frac{2(h_1-2)+2}{2(h_1-2)+1} \,q_F^{h_1-1,h_2+\frac{1}{2},h-1}
\Bigg] \, 
\nonu \\
\!& \!& \times 
\sum_{n=0}^{\infty}\,
\left(
\begin{array}{c}
\frac{1}{2}-2(h-1)-(k-2)-l-n \\
\frac{1}{2}-(h-1)-l
\end{array}
\right)\, \frac{\bar{\pa}_{\bar{z_1}}\,
  \bar{z}_{12}^{n+(h-1)+1}}{n!}\, \bar{\pa}^n \,
L^{(k-2)+l+(h-1),\hat{A}}(z_2, \bar{z}_2) \nonu \\
\!& \!& + \cdots \ .
\label{sixthres}
\eea
The numerical factor $\frac{1}{2}$
in the first line of the
first binomial coefficient
is given by
$(1+\frac{3}{2})$
minus $2$.
The numerical factor $\frac{1}{2}$
in the second line of the
first binomial coefficient
is given by $\frac{3}{2}$
minus $1$.
The first binomial coefficient above can be obtained
from (\ref{HH1}) by taking $k \rightarrow (k+1)$ and
$l \rightarrow (l+\frac{1}{2})$.

\subsection{The OPE between the gluon and the gluino}

The OPE between the conformally soft gluons and the gluinos
where
the weights in the antiholomorphic sector are given by
$h_1=\frac{k-1}{2}$ and $h_2=\frac{l-\frac{1}{2}}{2}$
can be written as
\bea
\!& \!& R^{k,\hat{A}}(z_1,\bar{z}_1)\, L^{l,\hat{B}}(z_2,\bar{z}_2)
=  -\frac{\kappa}{2} \, (\frac{i}{2})\, f^{\hat{A} \hat{B} \hat{C}}\, 
\frac{1}{z_{12}} \,
\sum_{h=-1}^{h_1+h_2-3}\, (-1)^{h+1}\, \la^h \,
\Bigg[ -q_B^{h_1,h_2+\frac{1}{2},h} +q_F^{h_1,h_2+\frac{1}{2},h} \Bigg] \, 
\nonu \\
\!& \!& \times 
\sum_{n=0}^{\infty}\,
\left(
\begin{array}{c}
-\frac{1}{2}-2 h-k-l-n \\
-\frac{1}{2}-h-l
\end{array}
\right)\, \frac{\bar{z}_{12}^{n+h+1}}{n!}\, \bar{\pa}^n \,
L^{k+l+h,\hat{C}}(z_2, \bar{z}_2) \
\nonu \\
\!& \!&-\frac{\kappa}{2} \, (\frac{1}{2})\, d^{\hat{A} \hat{B} \hat{C}}\, 
\frac{1}{z_{12}} \,
\sum_{h=-1}^{h_1+h_2-3}\, (-1)^{h+1}\, \la^h \,
\Bigg[ q_B^{h_1,h_2+\frac{1}{2},h} +q_F^{h_1,h_2+\frac{1}{2},h} \Bigg] \, 
\nonu \\
\!& \!& \times 
\sum_{n=0}^{\infty}\,
\left(
\begin{array}{c}
-\frac{1}{2}-2 h-k-l-n \\
-\frac{1}{2}-h-l
\end{array}
\right)\, \frac{\bar{z}_{12}^{n+h+1}}{n!}\, \bar{\pa}^n \,
L^{k+l+h,\hat{C}}(z_2, \bar{z}_2) \
\nonu \\
\!& \!&-\frac{\kappa}{2} \, (\frac{1}{K})\, \de^{\hat{A} \hat{B}        }\, 
\frac{1}{z_{12}} \,
\sum_{h=-1}^{h_1+h_2-3}\, (-1)^{h+1}\, \la^h \,
\Bigg[ q_B^{h_1,h_2+\frac{1}{2},h} +q_F^{h_1,h_2+\frac{1}{2},h} \Bigg] \, 
\nonu \\
\!& \!& \times 
\sum_{n=0}^{\infty}\,
\left(
\begin{array}{c}
-\frac{1}{2}-2 h-k-l-n \\
-\frac{1}{2}-h-l
\end{array}
\right)\, \frac{\bar{z}_{12}^{n+h+1}}{n!}\, \bar{\pa}^n \,
I^{k+l+h}(z_2, \bar{z}_2) \
 -\frac{\kappa}{2} \,
(\frac{i}{2})\, f^{\hat{A} \hat{B} \hat{C}}\, 
\frac{1}{z_{12}} \,
\nonu \\
\!& \!& \times  \sum_{h=-1}^{h_1+h_2-3}\, (-1)^{h}\,\la^{h} \,
\Bigg[  \frac{2(h_1-2)}{2(h_1-2)+1} \,
q_B^{h_1-1,h_2+\frac{1}{2},h}+
\frac{2(h_1-2)+2}{2(h_1-2)+1} \,q_F^{h_1-1,h_2+\frac{1}{2},h}
\Bigg] \, 
\nonu \\
\!&   \!& \times
\sum_{n=0}^{\infty}\,
\left(
\begin{array}{c}
-\frac{1}{2}-2h-(k-2)-l   -n \\
-\frac{1}{2}-h-l
\end{array}
\right)\, \frac{\bar{\pa}_{\bar{z_1}}\,
  \bar{z}_{12}^{n+h+1}}{n!}\, \bar{\pa}^n \,
L^{(k-2)+l+h,\hat{C}}(z_2, \bar{z}_2) 
 -\frac{\kappa}{2} \,
(\frac{1}{2})\, d^{\hat{A} \hat{B} \hat{C}}\, 
\frac{1}{z_{12}} \,
\nonu \\
\!& \!& \times  \sum_{h=-1}^{h_1+h_2-3}\, (-1)^{h}\,\la^{h} \,
\Bigg[  -\frac{2(h_1-2)}{2(h_1-2)+1} \,
q_B^{h_1-1,h_2+\frac{1}{2},h}+
\frac{2(h_1-2)+2}{2(h_1-2)+1} \,q_F^{h_1-1,h_2+\frac{1}{2},h}
\Bigg] \, 
\nonu \\
\!&   \!& \times
\sum_{n=0}^{\infty}\,
\left(
\begin{array}{c}
-\frac{1}{2}-2h-(k-2)-l   -n \\
-\frac{1}{2}-h-l
\end{array}
\right)\, \frac{\bar{\pa}_{\bar{z_1}}\,
  \bar{z}_{12}^{n+h+1}}{n!}\, \bar{\pa}^n \,
L^{(k-2)+l+h,\hat{C}}(z_2, \bar{z}_2)
-\frac{\kappa}{2} \,
(\frac{1}{K})\, \de^{\hat{A} \hat{B} }\, 
\frac{1}{z_{12}} \,
\nonu \\
\!& \!& \times 
\sum_{h=-1}^{h_1+h_2-3}\, (-1)^{h}\,\la^{h} \,
\Bigg[  -\frac{2(h_1-2)}{2(h_1-2)+1} \,
q_B^{h_1-1,h_2+\frac{1}{2},h}+
\frac{2(h_1-2)+2}{2(h_1-2)+1} \,q_F^{h_1-1,h_2+\frac{1}{2},h}
\Bigg] \, 
\nonu \\
\!&   \!& \times
\sum_{n=0}^{\infty}\,
\left(
\begin{array}{c}
-\frac{1}{2}-2h-(k-2)-l   -n \\
-\frac{1}{2}-h-l
\end{array}
\right)\, \frac{\bar{\pa}_{\bar{z_1}}\,
  \bar{z}_{12}^{n+h+1}}{n!}\, \bar{\pa}^n \,
I^{(k-2)+l+h}(z_2, \bar{z}_2) \ .
\label{seventhres}
\eea
The various structure constants in (\ref{otherstructure})
are used.
The numerical factor $-\frac{1}{2}$ is given by
$(1+\frac{1}{2})$
minus $2$.
The numerical factor $-\frac{1}{2}$
in the second line of the
first binomial coefficient
is given by $\frac{1}{2}$
minus $1$.
The first binomial coefficient above can be obtained
from (\ref{HH1}) by taking $k \rightarrow (k+1)$ and
$l \rightarrow (l+\frac{3}{2})$.

\subsection{ The OPE between the graviton and the gluon}

The OPE between the conformally soft gravitons and gluons
where
the weights in the antiholomorphic sector are given by
$h_1=\frac{k-2}{2}$ and $h_2=\frac{l-1}{2}$
can be summarized by
\bea
\!& \!& H^k(z_1,\bar{z}_1)\, R^{l,\hat{A}}(z_2,\bar{z}_2)
=
 -\frac{\kappa}{2} \,
\frac{1}{z_{12}} \,
\sum_{h=1,\mbox{\footnotesize odd}}^{h_1+h_2-4}\, (-1)^{h}\,\la^{h} \,
\nonu \\
\!& \!& \times \Bigg[  \frac{2(h_1-2)}{2(h_1-2)+1} \,
  \frac{(h_1+h_2-2-h)}{2(h_1+h_2-2-h)+1}\,
  p_B^{h_1-1,h_2,h-1}
  \nonu \\
  \!& \!& -
  \frac{(h_1+h_2-2-h-1)}{2(h_1+h_2-2-h-1)+1} \,
 \frac{2(h_1-2)+2}{2(h_1-2)+1}  \, 
p_F^{h_1-1,h_2,h-1}
\Bigg] \, 
\nonu \\
\!& \!& \times 
\sum_{n=0}^{\infty}\,
\left(
\begin{array}{c}
1-2(h-1)-(k-2)-l-n \\
-(h-1)-l
\end{array}
\right)\, \frac{\bar{\pa}_{\bar{z_1}}\,
  \bar{z}_{12}^{n+(h-1)+1}}{n!}\, \bar{\pa}^n \,
R^{(k-2)+l+(h-1),\hat{A}}(z_2, \bar{z}_2) \nonu \\
\!& \! & -\frac{\kappa}{2} \,
\frac{1}{z_{12}} \,
\sum_{h=1,\mbox{\footnotesize odd}}^{h_1+h_2-4}\, (-1)^{h}\,\la^{h} \,
 \Bigg[
  \frac{2(h_2-2)}{2(h_2-2)+1}\,
\frac{(h_1+h_2-2-h)}{2(h_1+h_2-2-h-1)+1}
  \,p_B^{h_1,h_2-1,h-1}
  \nonu \\
  \!& \!& -
  \frac{(h_1+h_2-2-h-1)}{2(h_1+h_2-2-h-1)+1}\,
  \frac{2(h_2-2)+2}{2(h_2-2)+1}\,
p_F^{h_1,h_2-1,h-1}  
  \Bigg] \, 
\nonu \\
\!& \!& \times 
\sum_{n=0}^{\infty}\,
\left(
\begin{array}{c}
1-2(h-1)-k-(l-2)-n \\
-(h-1)-(l-2)
\end{array}
\right)\, \frac{1}{n!}\, \bar{\pa}_{\bar{z_2}}\,
 \big[ \bar{z}_{12}^{n+(h-1)+1}\, \bar{\pa}^n \,
   R^{k+(l-2)+(h-1),\hat{A}}(z_2, \bar{z}_2) \big] \nonu \\
\!& \!& -\frac{\kappa}{2} \,
\frac{1}{z_{12}} \,
\sum_{h=0,\mbox{\footnotesize even}}^{h_1+h_2-4}\, (-1)^{h+1}\, \la^h \,
\Bigg[ \frac{(h_1+h_2-2-h)}{2(h_1+h_2-2-h-1)+1}\,  p_B^{h_1,h_2,h}
  \nonu \\
  \!& \!& +\frac{(h_1+h_2-2-h-1)}{2(h_1+h_2-2-h-1)+1}\,
  p_F^{h_1,h_2,h}\Bigg]  \, 
\nonu \\
\!& \!& \times \sum_{n=0}^{\infty}\,
\left(
\begin{array}{c}
1-2 h-k-l-n \\
-h-l
\end{array}
\right)\, \frac{\bar{z}_{12}^{n+h+1}}{n!}\, \bar{\pa}^n \,
 R^{k+l+h,\hat{A}}(z_2, \bar{z}_2) 
 -\frac{\kappa}{2} \,
\frac{1}{z_{12}} \,
\sum_{h=0,\mbox{\footnotesize even}}^{h_1+h_2-4}\, (-1)^{h+1}\,\la^{h} \,
\nonu \\
\!& \!& \times \Bigg[
  \frac{2(h_1-2)}{2(h_1-2)+1}\,
\frac{2(h_2-2)}{2(h_2-2)+1}
  \,  \frac{(h_1+h_2-2-h)}{2(h_1+h_2-2-h-1)+1}\, p_B^{h_1-1,h_2-1,h-2}
  \nonu \\
  \!& \!& + \frac{2(h_1-2)+2}{2(h_1-2)+1}\,
\frac{2(h_2-2)+2}{2(h_2-2)+1}
  \,  \frac{(h_1+h_2-2-h-1)}{2(h_1+h_2-2-h-1)+1}\, p_F^{h_1-1,h_2-1,h-2} \Bigg] \, 
\nonu \\
\!& \!& \times 
\sum_{n=0}^{\infty}\,
\left(
\begin{array}{c}
1-2(h-2)-(k-2)-(l-2)-n \\
-(h-2)-(l-2)
\end{array}
\right)\, \nonu \\
\!& \!& \times \frac{1}{n!}\,\bar{\pa}_{\bar{z_1}}\, \bar{\pa}_{\bar{z_2}}\,
 \big[ \bar{z}_{12}^{n+(h-2)+1}\, \bar{\pa}^n \,
   R^{(k-2)+(l-2)+(h-2),\hat{A}}(z_2, \bar{z}_2) \big]
 + \cdots \ .
\label{secondres}
\eea
According to the second equation of (\ref{WWn=1}),
the structure constants are the same as the first equation of
(\ref{WWn=1}). So we can repeat what we have done in (\ref{HH1}).
The difference appears in the first
binomial coefficient in Appendix
(\ref{secondres}) in the sense that
the numerical factor $1$ is given by $2+1$
minus $2$.
Similarly,
the numerical factor $0$
in the second line of the
first binomial coefficient
is given by $1$
minus $1$.
The above binomial coefficients can be obtained
from (\ref{HH1}) by taking $l \rightarrow (l+1)$.

\subsection{The OPE between the gluons}

Finally, the OPE between the conformally soft gluons
where
the weights in the antiholomorphic sector are given by
$h_1=\frac{k-1}{2}$ and $h_2=\frac{l-1}{2}$
can be expressed as
\bea
\!& \!& R^{k,\hat{A}}(z_1,\bar{z}_1)\, R^{l,\hat{B}}(z_2,\bar{z}_2)
=
 (-\frac{i}{2})\, f^{\hat{A} \hat{B} \hat{C}}\,  \Bigg\{ -\frac{\kappa}{2} \,
\frac{1}{z_{12}} \,
\sum_{h=0,\mbox{\footnotesize even}}^{h_1+h_2-4}\, (-1)^{h}\,\la^{h} \nonu \\
\!& \!& \times 
\Bigg[  \frac{2(h_1-2)}{2(h_1-2)+1} \,
\frac{(h_1+h_2-2-h)}{2(h_1+h_2-2-h-1)+1}\,   p_B^{h_1-1,h_2,h-1}
\nonu \\
\!& \!& -  \frac{2(h_1-2)+2}{2(h_1-2)+1} \,
\frac{(h_1+h_2-2-h-1)}{2(h_1+h_2-2-h-1)+1}\,   p_F^{h_1-1,h_2,h-1}
\Bigg] \, 
\nonu \\
\!& \!& \times 
\sum_{n=0}^{\infty}\,
\left(
\begin{array}{c}
-2(h-1)-(k-2)-l-n \\
-(h-1)-l
\end{array}
\right)\, \frac{\bar{\pa}_{\bar{z_1}}\,
  \bar{z}_{12}^{n+(h-1)+1}}{n!}\, \bar{\pa}^n \,
R^{(k-2)+l+(h-1),\hat{C}}(z_2, \bar{z}_2) \nonu \\
\!& \! & -\frac{\kappa}{2} \,
\frac{1}{z_{12}} \,
\sum_{h=0,\mbox{\footnotesize even}}^{h_1+h_2-4}\, (-1)^{h}\,\la^{h} \,
\Bigg[
  \frac{2(h_2-2)}{2(h_2-2)+1}\,
\frac{(h_1+h_2-2-h)}{2(h_1+h_2-2-h-1)+1}\,
  \,p_B^{h_1,h_2-1,h-1}
  \nonu \\
  \!& \!& -
 \frac{2(h_2-2)+2}{2(h_2-2)+1}\,
\frac{(h_1+h_2-2-h-1)}{2(h_1+h_2-2-h-1)+1}\,
  \,p_F^{h_1,h_2-1,h-1}
  \Bigg] \, 
\nonu \\
\!& \!& \times 
\sum_{n=0}^{\infty}\,
\left(
\begin{array}{c}
-2(h-1)-k-(l-2)-n \\
-(h-1)-(l-2)
\end{array}
\right)\, \frac{1}{n!}\, \bar{\pa}_{\bar{z_2}}\,
 \big[ \bar{z}_{12}^{n+(h-1)+1}\, \bar{\pa}^n \,
   R^{k+(l-2)+(h-1),\hat{C}}(z_2, \bar{z}_2) \big] \nonu \\
\!& \!& -\frac{\kappa}{2} \,
\frac{1}{z_{12}} \,
\sum_{h=-1,\mbox{\footnotesize odd}}^{h_1+h_2-4}\, (-1)^{h+1}\, \la^h \,
\nonu \\
\!& \!& \times \Bigg[\frac{(h_1+h_2-2-h)}{2(h_1+h_2-2-h-1)+1}
  \, p_B^{h_1,h_2,h} +
\frac{(h_1+h_2-2-h-1)}{2(h_1+h_2-2-h-1)+1}
  \, p_F^{h_1,h_2,h}
  \Bigg]
  \nonu \\
  \!& \!& \times 
\sum_{n=0}^{\infty}\,
\left(
\begin{array}{c}
-2 h-k-l-n \\
-h-l
\end{array}
\right)\, \frac{\bar{z}_{12}^{n+h+1}}{n!}\, \bar{\pa}^n \,
 R^{k+l+h,\hat{C}}(z_2, \bar{z}_2) 
 -\frac{\kappa}{2} \,
\frac{1}{z_{12}} \,
\sum_{h=-1,\mbox{\footnotesize odd}}^{h_1+h_2-4}\, (-1)^{h+1}\,\la^{h} \,
\nonu \\
\!& \!& \times \Bigg[
  \frac{2(h_1-2)}{2(h_1-2)+1}\,
\frac{2(h_2-2)}{2(h_2-2)+1}
\, \frac{(h_1+h_2-2-h)}{2(h_1+h_2-2-h-1)+1}  \,p_B^{h_1-1,h_2-1,h-2}
\nonu \\
\!& \!& + \frac{2(h_1-2)+2}{2(h_1-2)+1}\,
\frac{2(h_2-2)+2}{2(h_2-2)+1}
\, \frac{(h_1+h_2-2-h-1)}{2(h_1+h_2-2-h-1)+1}  \,p_F^{h_1-1,h_2-1,h-2}
\Bigg] \, 
\nonu \\
\!& \!& \times 
\sum_{n=0}^{\infty}\,
\left(
\begin{array}{c}
-2(h-2)-(k-2)-(l-2)-n \\
-(h-2)-(l-2)
\end{array}
\right)\, \nonu \\
\!& \!& \times \frac{1}{n!}\,\bar{\pa}_{\bar{z_1}}\, \bar{\pa}_{\bar{z_2}}\,
 \big[ \bar{z}_{12}^{n+(h-2)+1}\, \bar{\pa}^n \,
   R^{(k-2)+(l-2)+(h-2),\hat{C}}(z_2, \bar{z}_2) \big]
  \Bigg\} \nonu \\
 \!& \!& +
 (\frac{1}{2})\, d^{\hat{A} \hat{B} \hat{C}}\, 
 \Bigg\{
 -\frac{\kappa}{2} \,
\frac{1}{z_{12}} \,
\sum_{h=1,\mbox{\footnotesize odd}}^{h_1+h_2-4}\, (-1)^{h}\,\la^{h} \,
\nonu \\
\!& \!& \times \Bigg[  \frac{2(h_1-2)}{2(h_1-2)+1} \,
  \frac{(h_1+h_2-2-h)}{2(h_1+h_2-2-h)+1}\,
  p_B^{h_1-1,h_2,h-1}
  \nonu \\
  \!& \!& -
  \frac{(h_1+h_2-2-h-1)}{2(h_1+h_2-2-h-1)+1} \,
 \frac{2(h_1-2)+2}{2(h_1-2)+1}  \, 
p_F^{h_1-1,h_2,h-1}
\Bigg] \, 
\nonu \\
\!& \!& \times 
\sum_{n=0}^{\infty}\,
\left(
\begin{array}{c}
-2(h-1)-(k-2)-l-n \\
-(h-1)-l
\end{array}
\right)\, \frac{\bar{\pa}_{\bar{z_1}}\,
  \bar{z}_{12}^{n+(h-1)+1}}{n!}\, \bar{\pa}^n \,
R^{(k-2)+l+(h-1),\hat{C}}(z_2, \bar{z}_2) \nonu \\
\!& \! & -\frac{\kappa}{2} \,
\frac{1}{z_{12}} \,
\sum_{h=1,\mbox{\footnotesize odd}}^{h_1+h_2-4}\, (-1)^{h}\,\la^{h} \,
 \Bigg[
  \frac{2(h_2-2)}{2(h_2-2)+1}\,
\frac{(h_1+h_2-2-h)}{2(h_1+h_2-2-h-1)+1}
  \,p_B^{h_1,h_2-1,h-1}
  \nonu \\
  \!& \!& -
  \frac{(h_1+h_2-2-h-1)}{2(h_1+h_2-2-h-1)+1}\,
  \frac{2(h_2-2)+2}{2(h_2-2)+1}\,
p_F^{h_1,h_2-1,h-1}  
  \Bigg] \, 
\nonu \\
\!& \!& \times 
\sum_{n=0}^{\infty}\,
\left(
\begin{array}{c}
-2(h-1)-k-(l-2)-n \\
-(h-1)-(l-2)
\end{array}
\right)\, \frac{1}{n!}\, \bar{\pa}_{\bar{z_2}}\,
 \big[ \bar{z}_{12}^{n+(h-1)+1}\, \bar{\pa}^n \,
   R^{k+(l-2)+(h-1),\hat{C}}(z_2, \bar{z}_2) \big] \nonu \\
\!& \!& -\frac{\kappa}{2} \,
\frac{1}{z_{12}} \,
\sum_{h=0,\mbox{\footnotesize even}}^{h_1+h_2-4}\, (-1)^{h+1}\, \la^h \,
\Bigg[ \frac{(h_1+h_2-2-h)}{2(h_1+h_2-2-h-1)+1}\,  p_B^{h_1,h_2,h}
  \nonu \\
  \!& \!& +\frac{(h_1+h_2-2-h-1)}{2(h_1+h_2-2-h-1)+1}\,
  p_F^{h_1,h_2,h}\Bigg]  \, 
\nonu \\
\!& \!& \times \sum_{n=0}^{\infty}\,
\left(
\begin{array}{c}
-2 h-k-l-n \\
-h-l
\end{array}
\right)\, \frac{\bar{z}_{12}^{n+h+1}}{n!}\, \bar{\pa}^n \,
 R^{k+l+h,\hat{C}}(z_2, \bar{z}_2) 
 -\frac{\kappa}{2} \,
\frac{1}{z_{12}} \,
\sum_{h=0,\mbox{\footnotesize even}}^{h_1+h_2-4}\, (-1)^{h+1}\,\la^{h} \,
\nonu \\
\!& \!& \times \Bigg[
  \frac{2(h_1-2)}{2(h_1-2)+1}\,
\frac{2(h_2-2)}{2(h_2-2)+1}
  \,  \frac{(h_1+h_2-2-h)}{2(h_1+h_2-2-h-1)+1}\, p_B^{h_1-1,h_2-1,h-2}
  \nonu \\
  \!& \!& + \frac{2(h_1-2)+2}{2(h_1-2)+1}\,
\frac{2(h_2-2)+2}{2(h_2-2)+1}
  \,  \frac{(h_1+h_2-2-h-1)}{2(h_1+h_2-2-h-1)+1}\, p_F^{h_1-1,h_2-1,h-2} \Bigg] \, 
\nonu \\
\!& \!& \times 
\sum_{n=0}^{\infty}\,
\left(
\begin{array}{c}
-2(h-2)-(k-2)-(l-2)-n \\
-(h-2)-(l-2)
\end{array}
\right)\, \nonu \\
\!& \!& \times \frac{1}{n!}\,\bar{\pa}_{\bar{z_1}}\, \bar{\pa}_{\bar{z_2}}\,
 \big[ \bar{z}_{12}^{n+(h-2)+1}\, \bar{\pa}^n \,
   R^{(k-2)+(l-2)+(h-2),\hat{C}}(z_2, \bar{z}_2) \big]
 \Bigg\}
 \nonu \\
 \!& \!& +
 (\frac{1}{K})\, \de^{\hat{A} \hat{B} }\, 
 \Bigg\{
 -\frac{\kappa}{2} \,
\frac{1}{z_{12}} \,
\sum_{h=1,\mbox{\footnotesize odd}}^{h_1+h_2-4}\, (-1)^{h}\,\la^{h} \,
\nonu \\
\!& \!& \times \Bigg[  \frac{2(h_1-2)}{2(h_1-2)+1} \,
  \frac{(h_1+h_2-2-h)}{2(h_1+h_2-2-h)+1}\,
  p_B^{h_1-1,h_2,h-1}
  \nonu \\
  \!& \!& -
  \frac{(h_1+h_2-2-h-1)}{2(h_1+h_2-2-h-1)+1} \,
 \frac{2(h_1-2)+2}{2(h_1-2)+1}  \, 
p_F^{h_1-1,h_2,h-1}
\Bigg] \, 
\nonu \\
\!& \!& \times 
\sum_{n=0}^{\infty}\,
\left(
\begin{array}{c}
-2(h-1)-(k-2)-l-n \\
-(h-1)-l
\end{array}
\right)\, \frac{\bar{\pa}_{\bar{z_1}}\,
  \bar{z}_{12}^{n+(h-1)+1}}{n!}\, \bar{\pa}^n \,
R^{(k-2)+l+(h-1)}(z_2, \bar{z}_2) \nonu \\
\!& \! & -\frac{\kappa}{2} \,
\frac{1}{z_{12}} \,
\sum_{h=1,\mbox{\footnotesize odd}}^{h_1+h_2-4}\, (-1)^{h}\,\la^{h} \,
 \Bigg[
  \frac{2(h_2-2)}{2(h_2-2)+1}\,
\frac{(h_1+h_2-2-h)}{2(h_1+h_2-2-h-1)+1}
  \,p_B^{h_1,h_2-1,h-1}
  \nonu \\
  \!& \!& -
  \frac{(h_1+h_2-2-h-1)}{2(h_1+h_2-2-h-1)+1}\,
  \frac{2(h_2-2)+2}{2(h_2-2)+1}\,
p_F^{h_1,h_2-1,h-1}  
  \Bigg] \, 
\nonu \\
\!& \!& \times 
\sum_{n=0}^{\infty}\,
\left(
\begin{array}{c}
-2(h-1)-k-(l-2)-n \\
-(h-1)-(l-2)
\end{array}
\right)\, \frac{1}{n!}\, \bar{\pa}_{\bar{z_2}}\,
 \big[ \bar{z}_{12}^{n+(h-1)+1}\, \bar{\pa}^n \,
   R^{k+(l-2)+(h-1)}(z_2, \bar{z}_2) \big] \nonu \\
\!& \!& -\frac{\kappa}{2} \,
\frac{1}{z_{12}} \,
\sum_{h=0,\mbox{\footnotesize even}}^{h_1+h_2-4}\, (-1)^{h+1}\, \la^h \,
\Bigg[ \frac{(h_1+h_2-2-h)}{2(h_1+h_2-2-h-1)+1}\,  p_B^{h_1,h_2,h}
  \nonu \\
  \!& \!& +\frac{(h_1+h_2-2-h-1)}{2(h_1+h_2-2-h-1)+1}\,
  p_F^{h_1,h_2,h}\Bigg]  \, 
\nonu \\
\!& \!& \times \sum_{n=0}^{\infty}\,
\left(
\begin{array}{c}
-2 h-k-l-n \\
-h-l
\end{array}
\right)\, \frac{\bar{z}_{12}^{n+h+1}}{n!}\, \bar{\pa}^n \,
 R^{k+l+h}(z_2, \bar{z}_2) 
 -\frac{\kappa}{2} \,
\frac{1}{z_{12}} \,
\sum_{h=0,\mbox{\footnotesize even}}^{h_1+h_2-4}\, (-1)^{h+1}\,\la^{h} \,
\nonu \\
\!& \!& \times \Bigg[
  \frac{2(h_1-2)}{2(h_1-2)+1}\,
\frac{2(h_2-2)}{2(h_2-2)+1}
  \,  \frac{(h_1+h_2-2-h)}{2(h_1+h_2-2-h-1)+1}\, p_B^{h_1-1,h_2-1,h-2}
  \nonu \\
  \!& \!& + \frac{2(h_1-2)+2}{2(h_1-2)+1}\,
\frac{2(h_2-2)+2}{2(h_2-2)+1}
  \,  \frac{(h_1+h_2-2-h-1)}{2(h_1+h_2-2-h-1)+1}\, p_F^{h_1-1,h_2-1,h-2} \Bigg] \, 
\nonu \\
\!& \!& \times 
\sum_{n=0}^{\infty}\,
\left(
\begin{array}{c}
-2(h-2)-(k-2)-(l-2)-n \\
-(h-2)-(l-2)
\end{array}
\right)\, \nonu \\
\!& \!& \times \frac{1}{n!}\,\bar{\pa}_{\bar{z_1}}\, \bar{\pa}_{\bar{z_2}}\,
 \big[ \bar{z}_{12}^{n+(h-2)+1}\, \bar{\pa}^n \,
   R^{(k-2)+(l-2)+(h-2)}(z_2, \bar{z}_2) \big]
 \Bigg\}+ \cdots \ .
\label{thirdres}
\eea
The numerical factor $0$ is given by $1+1$
minus $2$.
The numerical factor $0$
in the second line of the
first binomial coefficient
is given by $1$
minus $1$.
The first binomial coefficient above can be obtained
from (\ref{HH1}) by taking $k \rightarrow (k+1)$
and $l \rightarrow (l+1)$.
We use the following relations between the structure constants 
\bea
\!& \!& \tilde{q}^{h_1,h_2,h}(m,n)\Bigg|_{h, \mbox{ \footnotesize odd}}
 =  
p_B^{h_1,h_2,h}(m,n) \nonu \\
\!&  \!& +  \Bigg[  \frac{2(h_1-2)(m+(h_1-2)+1)}{2(h_1-2)+1}\,
\frac{2(h_2-2)(n+(h_2-2)+1)}{2(h_2-2)+1}\Bigg]\, 
p_B^{h_1-1,h_2-1,h-2}(m,n)
\nonu \\
\!&  \!& - \Bigg[
  \frac{2(h_1+h_2-2-h-1)(m+n+h_1+h_2-2-h-1+1)}{2(h_1+h_2-2-h-1)+1}
  \Bigg]\,
\tilde{q}^{h_1,h_2,h-1}(m,n) 
\nonu\\
\!&  \!& = p_F^{h_1,h_2,h}(m,n) +
\Bigg[  \frac{(2(h_1-2)+2)(m+(h_1-2)+1)}{2(h_1-2)+1}
  \nonu \\
\!& \!& \times \frac{(2(h_2-2)+2)(n+(h_2-2)+1)}{2(h_2-2)+1}\Bigg]\, 
p_F^{h_1-1,h_2-1,h-2}(m,n)
\nonu \\
\!&  \!& + 
\Bigg[\frac{(2(h_1+h_2-2-h-1)+2)(m+n+h_1+h_2-2-h-1+1)}{2(h_1+h_2-2-h-1)+1}
  \Bigg]
\tilde{q}^{h_1,h_2,h-1}(m,n) \ ,
\nonu\\
\!& \!& \tilde{q}^{h_1,h_2,h}(m,n)\Bigg|_{h, \mbox{ \footnotesize even}}
 =   \Bigg[ \frac{2(h_1-2)(m+(h_1-2)+1)}{2(h_1-2)+1} \Bigg]\,
p_B^{h_1-1,h_2,h-1}(m,n) \nonu \\
\!&  \!& + \Bigg[ \frac{2(h_2-2)(n+(h_2-2)+1)}{2(h_2-2)+1} \Bigg]\,
p_B^{h_1,h_2-1,h-1}(m,n)
\nonu \\
\!&  \!& - \Bigg[
  \frac{2(h_1+h_2-2-h-1)(m+n+h_1+h_2-2-h-1+1)}{2(h_1+h_2-2-h-1)+1}\Bigg]
\tilde{q}^{h_1,h_2,h-1}(m,n)
\nonu \\
\!&  \!& = - \Bigg[ \frac{(2(h_1-2)+2)(m+(h_1-2)+1)}{2(h_1-2)+1} \Bigg]\,
p_F^{h_1-1,h_2,h-1}(m,n) \nonu \\
\!&  \!& - \Bigg[\frac{(2(h_2-2)+2)(n+(h_2-2)+1)}{2(h_2-2)+1} \Bigg]\,
p_F^{h_1,h_2-1,h-1}(m,n)
\label{STRUCT1}
\\
\!&  \!& + \Bigg[
  \frac{(2(h_1+h_2-2-h-1)+2)(m+n+h_1+h_2-2-h-1+1)}{2(h_1+h_2-2-h-1)+1}\Bigg]
\tilde{q}^{h_1,h_2,h-1}(m,n)\ .
\nonu
\eea
Note that $\tilde{q}^{h_1,h_2,-1}(m,n)= p_B^{h_1,h_2,h-1}(m,n)=\frac{1}{2}$.
From Appendix (\ref{STRUCT1}), finally we obtain
\bea
\!& \!& \tilde{q}^{h_1,h_2,h}(m,n)\Bigg|_{h, \mbox{ \footnotesize odd}}
=
\nonu \\
\!& \!&
\frac{(h_1+h_2-2-h)}{2(h_1+h_2-2-h-1)+1}
  \, p_B^{h_1,h_2,h}
 + \Bigg[ \frac{2(h_1-2)(m+(h_1-2)+1)}{2(h_1-2)+1}\,
 \nonu \\
 \!& \!& \times \frac{2(h_2-2)(n+(h_2-2)+1)}{2(h_2-2)+1}
\frac{(h_1+h_2-2-h)}{2(h_1+h_2-2-h-1)+1}\Bigg]  \,p_B^{h_1-1,h_2-1,h-2}
\nonu \\
\!& \!& +
\frac{(h_1+h_2-2-h-1)}{2(h_1+h_2-2-h-1)+1}
  \, p_F^{h_1,h_2,h}
 + \Bigg[ \frac{(2(h_1-2)+2)(m+(h_1-2)+1)}{2(h_1-2)+1}\,
 \nonu \\
 \!& \!& \times
 \frac{(2(h_2-2)+2)(n+(h_2-2)+1)}{2(h_2-2)+1}
 \frac{(h_1+h_2-2-h-1)}{2(h_1+h_2-2-h-1)+1} \Bigg] \,p_F^{h_1-1,h_2-1,h-2}
\ ,
  \nonu \\
  \!& \!& \tilde{q}^{h_1,h_2,h}(m,n)\Bigg|_{h, \mbox{ \footnotesize even}}
= \nonu \\
\!& \!& \Bigg[ \frac{2(h_2-2)(n+(h_2-2)+1)}{2(h_2-2)+1}\,
\frac{(h_1+h_2-2-h)}{2(h_1+h_2-2-h-1)+1} \Bigg]\,
  \,p_B^{h_1,h_2-1,h-1}
  \nonu \\
  \!& \!& \Bigg[ \frac{2(h_1-2)(m+(h_1-2)+1)}{2(h_1-2)+1} \,
\frac{(h_1+h_2-2-h)}{2(h_1+h_2-2-h-1)+1}\Bigg]\,   p_B^{h_1-1,h_2,h-1}
\nonu \\
  \!& \!& -\Bigg[
 \frac{(2(h_2-2)+2)(n+(h_2-2)+1)}{2(h_2-2)+1}\,
\frac{(h_1+h_2-2-h-1)}{2(h_1+h_2-2-h-1)+1}\Bigg]
\,p_F^{h_1,h_2-1,h-1} 
\nonu \\
 \!& \!& -  \Bigg[ \frac{(2(h_1-2)+2)(m+(h_1-2)+1)}{2(h_1-2)+1} \,
\frac{(h_1+h_2-2-h-1)}{2(h_1+h_2-2-h-1)+1}\Bigg]\,   p_F^{h_1-1,h_2,h-1}
\ .
\label{STRUCT2}
\eea
By considering Appendix (\ref{STRUCT2}), we
can write down the OPE in Appendix (\ref{thirdres}).
The first one of Appendix (\ref{STRUCT2})
is the same as the functional form of the second one of (\ref{qotherexp})
while the second one of Appendix (\ref{STRUCT2})
is the same as the functional form of the first one of (\ref{qotherexp}).


\end{document}